\documentclass{article}

\usepackage{amsmath}
\usepackage{amssymb}
\usepackage{graphicx}
\usepackage[super,numbers,sort&compress]{natbib}
\usepackage{xcolor}
\usepackage{siunitx}
\usepackage{amsfonts}
\usepackage{amsmath}
\usepackage{float}
\usepackage{hyperref}
\usepackage{tikz}
\usepackage{caption}
\usepackage{subcaption}
\usepackage{tocloft}                   
\usetikzlibrary{arrows.meta,positioning,calc,fit,backgrounds,shapes.geometric}

\DeclareUnicodeCharacter{2212}{\textminus}
\DeclareSIUnit\fps{fps}

\newcommand{\SIrang}[3]{\qtyrange[range-phrase=~--~]{#1}{#2}{#3}}

\usepackage{longtable}   

\begin{document}

\begin{titlepage}
\title{\LARGE{\textbf{Biting fly vision and zebra stripes}} }

\vspace{0.1 cm}
\date{}
\author{ \sc{Krispin M. Dettlaff}\\
        \textit{krispind@ethz.ch} \\
        \small Department of Chemistry and Applied Biosciences\\
        \small ETH Zürich \\
        \small Vladimir-Prelog-Weg 2, CH-8093 Zürich, Switzerland
}

\maketitle
\pagenumbering{gobble}

\begin{abstract} 
The function of the zebra's striped coat has been discussed since
\citet{darwin1871descent} and \citet{wallace1867mimicry}. Increasing amount
of comparative \citep{caro2014function,larison2015zebra} and
experimental evidence
\citep{waage1981zebra,gibson1992tsetse,brady1988landing,egri2012polarotactic,caro2019benefits}
supports the hypothesis that the stripes mainly serve as protection
against visually orienting biting Diptera, in particular tabanids
(horse flies), glossinids (tsetse flies), and culicids (mosquitoes). The proposed ways this protection works include polarotactic
disruption \citep{egri2012polarotactic,horvath2010whiteness} and
silhouette break-up \citep{britten2016zebras}, as well as motion based illusions in motion based detectors of the insect visual system \citep{snyder1979physics,landnilsson2002eyes, how2014motion, mouy2025aliasing}. In this work a complementary 
purely optical mechanism is presented to explain why the zebra’s stripe pattern provides a measurable advantage in warding off biting flies. 
We have quantified the optical illusion caused by the interaction between the zebra’s striped coat and the fly’s compound eye with a linear, shift-invariant Fourier model based on published optical data of the diurnal Culicidae \citep{land1997mosquito,land1999fundamental,kawada2006eye}. 
The model predicts that, within an approach range of about \SIrang{1}{5}{\meter}, the interaction of
the stripe pattern with the ommatidial sampling produces parasitic spatial frequencies that are absent from the physical stimulus. These frequencies fall
within the range most important for host fixation and
controlling their landing \citep{borst2014fly,vanbreugel2012visual,baird2013universal}.
A post-retinal motion-detector shows that these
parasitic frequencies create misleading motion signals. The magnitude of this parasitic energy predicts the specific loss of the radial-expansion (looming) cue that gates landing, linking the optical disturbance quantitatively to the motion computation that controls the fly's final approach. By rigorously modeling visual processes and quantifying disturbances, this study makes a decisive contribution to clarifying the role of zebra stripes in defense against biting flies. Furthermore, this first-principles model of sampling-induced parasitic image content and its propagation into motion signals provides a framework that may extend to other aliasing-prone systems, from compound-eyed insects to machine-vision and other spatially sampled sensors.

\end{abstract}
\vfill
\vfill
Zürich, \small \today
\end{titlepage}
\newpage

{
\setlength{\cftbeforesecskip}{2pt}
\setlength{\cftbeforesubsecskip}{1pt}
\setlength{\cftbeforetoctitleskip}{-20pt}

\tableofcontents

\newpage
}

\pagenumbering{arabic}
\setcounter{page}{1}
\section{Introduction}
\subsection{The riddle of the zebra's stripes}

The black-and-white coat of plains, mountain and Grevy's zebras
(\textit{Equus quagga}, \textit{E.\ zebra} and \textit{E.\ grevyi})
is one of the most noticeable coat patterns in the animal kingdom. It has attracted scientific interest for more than 150~years
\citep{wallace1867mimicry,darwin1871descent,kipling1908justso,thayer1909concealing,cott1940adaptive}.
At least five non-mutually exclusive families of hypotheses have been
formulated\citep{ruxton2002fitness,caro2009contrasting,caro2014function}:
crypsis or disruptive camouflage against large mammalian predators,
predator confusion (often called motion dazzle)
\citep{stevens2008dazzle,stevens2011motion,scott2011dazzle}, 
intra-specific signaling, including individual identification and group cohesion \citep{kingdon1984cohesion}, thermoregulation through different heating of the black and white
stripes \citep{cobb2019zebrastripes,horvath2018experimental} and
protection against hematophagous biting Diptera
\citep{waage1981zebra,gibson1992tsetse,egri2012polarotactic,caro2014function}.
Evidence from phylogenetics and geography shifted 
the focus away from the mammalian predator hypothesis. \citet{caro2014function}
mapped the intensity of the striping in the equid subspecies against the geographic
distribution of suspected selective agents. They found a strong, repeated
link between body, leg, belly, and shadow striping with proxies for
tabanid and tsetse fly activity. However, there was no consistent connection with predator distribution, woodland cover, group size, or maximum temperature. \citet{larison2015zebra} reached somewhat different conclusions
about plains-zebra populations using random-forest models. They found the
strongest correlation with temperature, but the association with leg-stripes
does not support a thermoregulatory explanation. Furthermore, direct
thermographic measurements indicate that stripes actually do not lower 
the surface temperature of equid models exposed to sunlight
\citep{horvath2018experimental}. The hypothesis regarding byting flies is the only one backed by both comparative and direct experimental evidence
\citep{waage1981zebra,brady1988landing,gibson1992tsetse,egri2012polarotactic,blaho2013stripes,britten2016zebras,caro2019benefits}.

\subsection{Biting flies}

Equids in areas rich in tabanid and glossinid flies can suffer severe blood
loss, reduced grazing time, and exposure to vector-borne pathogens
including \textit{Trypanosoma} spp.\ \citep{waage1981zebra,caro2014function}.
The selective pressure from biting fly attack is significant and, as Waage noted, comparable to the pressure from predation\citep{waage1981zebra}. Diurnal biting flies find their hosts using a mix of sensory cues. At long distances, they are primarily based on olfactory cues such as $\mathrm{CO_2}$, ammonia,
short-chain fatty acids and species-specific kairomones dominate
\citep{coutinho2022human,mcmeniman2014multimodal,vanbreugel2015mosquitoes}.
At closer distances, vision becomes more important. Many tabanids
exploit positive polarotaxis to detect dark polarizing surfaces such
as wet skin or water, and use the same cue to home on dark hosts
\citep{egri2012polarotactic,horvath2010whiteness,britten2016zebras}.
When they get very close, they control their landing maneuver using visual feedback from the optical flow that expands over their retina
\citep{vanbreugel2012visual,baird2013universal,borst2014fly}.
A defensive coat pattern can interfere with any of these
stages. It can make the animal less noticeable against the background \citep{britten2016zebras},
reduce the contrast that polarotactic flies exploit
\citep{egri2012polarotactic,horvath2010whiteness,horvath2024polarotaxis},
disrupt the figure-ground separation that aids landing \citep{brady1988landing,gibson1992tsetse}, or introduce false motion signals into the optical flow estimator
\citep{how2014motion}.

\subsection{Apposition compound eye}

Biting flies have apposition compound eyes. Each photoreceptor unit, called an ommatidium, is made up of a corneal lens and a crystalline cone.
Each ommatidium also contains a fused or open rhabdom, which is made up of eight retinular cells (R1 to R8)\citep{hawkes2022vision,land1997mosquito,land1999fundamental}. Two geometric factors determine how the eye samples the visual scene. These are the diameter of the lens facet $D$ (typically
\SIrang{15}{30}{\micro\meter}) and the inter-ommatidial angle
$\Delta\varphi$ (typically \SIrang{1}{5}{\degree})
\citep{land1997mosquito,landnilsson2002eyes}. Their product, the
The product of these, called the eye parameter $D\,\Delta\varphi$, has been used by
\citet{kawada2006eye} to classify culicid species along a continuum
from diurnal species, which have a small eye parameter and higher resolution, to nocturnal species, which have a large eye parameter and higher photon catch. For \textit{Aedes aegypti}, a diurnal anthropophilic vector used in this study as a representative of biting flies
$\Delta\varphi$ lies near \SI{2.5}{\degree}. For this species, $\Delta\varphi$ is about \SI{2.5}{\degree}, which corresponds to a Nyquist spatial frequency of
about $0.2~\mathrm{cycles/deg}$ 
\citep{land1997mosquito,land1999fundamental}.
Two consequences are central to the present work. The
ommatidial array constitutes a regular, quasi-hexagonal sampling
lattice. Like any regular sampler, this pattern can cause aliasing of spatial 
frequencies above its Nyquist limit \citep{snyder1979physics}.
In addition, the angular acceptance function of each rhabdom acts as a
low-pass spatial filter \citep{snyder1979physics,warrant1999seeing}.

\subsection{Moir\'e interference}

When two periodic structures with comparable spatial frequencies are
combined, as happens when a striped scene is viewed through the
ommatidial array, the resulting signal contains not only the
original spatial frequencies, but also their sums and differences.
If one of these difference frequencies falls within the eye's
pass-band, it appears in the retinal image as a low-frequency
"parasitic" fringe pattern that does not actually exist on the
zebra's surface. This is known as the Moiré effect in optics, and its impact on the biting fly vision is studied in this publication.
This effect is biologically relevant because it aligns with the experimental observations.
The widths of the zebra stripe span $\sim$\SIrang{0.2}{7.5}{\centi\meter} between the body regions and the species
\citep{egri2012polarotactic}, which translates to spatial periods of
$\sim$\SIrang{0.1}{4}{\degree} at viewing distances of
\SIrang{1}{5}{\meter}, bracketing the ommatidial pitch of glossinids, culicids, and tabanids. Tabanids are much less attracted to stripes
that are narrower than the spacing between their ommatidia \citep{egri2012polarotactic}, and the critical distance found in experiments for
\textit{Glossina pallidipes} estimated at $\sim$\SI{3.5}{\meter} for
\SI{20}{\centi\meter} stripes \citep{gibson1992tsetse}. Recent
high-speed video of tabanids around live zebras and zebra-coated
horses shows that flies approach at similar rates, but fail in
land successfully. Instead, they often abort their landings or simply fly by
\citep{caro2019benefits}, which is behaviour consistent with disruption of
the optic-flow signals that usually help them land
\citep{baird2013universal,vanbreugel2012visual}. 
Although stripe-induced disruption of fly optic flow has long been suspected \citep{how2014motion,caro2019benefits}, and a recent geometric model has shown that an undersampled apposition eye misjudges motion against striped patterns \citep{mouy2025aliasing}, a rigorous optical-physical (Fourier) derivation of the parasitic spectrum, its validation on real zebra images under realistic viewing geometry, and a two-dimensional motion stage that resolves the radial looming cue governing landing have not yet been provided

\section{Optical Model}
The following chapter describes how the diptera vision on Zebra can be modeled in a Fourier optic framework. 
It provides the theoretical basis for the following chapters.

\label{sec:optical_model}
 
\begin{figure}[H]
\centering
\begin{tikzpicture}[
  >={Stealth[length=2.5mm,width=1.8mm]},
  font=\small,
  stage/.style={
      draw=black!70, thick,
      fill=blue!4,
      rounded corners=2pt,
      align=center,
      inner sep=4pt,
      minimum height=11mm,
      minimum width=22mm,
      text width=22mm,
  },
  output/.style={
      draw=black!70, thick,
      fill=orange!12,
      rounded corners=2pt,
      align=center,
      inner sep=4pt,
      minimum height=11mm,
      minimum width=22mm,
      text width=22mm,
  },
  nlstage/.style={
      draw=violet!50!black, thick, dashed,
      fill=violet!6,
      rounded corners=2pt,
      align=center,
      inner sep=4pt,
      minimum height=11mm,
      minimum width=24mm,
      text width=24mm,
  },
  param/.style={
      draw=black!50, thin, dashed,
      fill=black!4,
      rounded corners=1pt,
      align=center,
      font=\footnotesize,
      inner sep=3pt,
  },
  signal/.style={font=\footnotesize, inner sep=1pt, fill=white},
  flow/.style={->, thick, black!70, shorten >=1pt, shorten <=1pt},
  nlflow/.style={->, thick, violet!70!black, dashed,
                 shorten >=1pt, shorten <=1pt},
  param-arrow/.style={->, thin, dashed, black!55, shorten >=2pt},
]
 
\node[stage] (stim)  at (0,    0) {Striped stimulus\\[1pt]$I(x,y)$};
\node[stage] (blur)  at (3.2,  0) {Acceptance blur\\[1pt]$H_\rho(f)$\\[1pt]\scriptsize Eqs.\,\ref{eq:eye_otf}--\ref{eq:airy_otf}};
\node[stage] (samp)  at (6.4,  0) {Spatial sampling\\[1pt]$\mathrm{III}_{f_s}$\\[1pt]\scriptsize Eqs.\,\ref{eq:spatial_sampling}--\ref{eq:freq_replication}};
\node[stage] (recon) at (9.6,  0) {Reconstruction\\[1pt]$H_\mathrm{box}(f)$\\[1pt]\scriptsize Eq.\,\ref{eq:hbox}};
 
\node[nlstage] (nl) at (1.6,  -2.8)
  {Non-linearity \emph{(optional)}\\[1pt]
   \scriptsize Naka--Rushton (Eq.\,\ref{eq:naka_rushton})\\[1pt]
   \scriptsize \emph{or} Volterra (Eq.\,\ref{eq:volterra})};
 
\node[output] (out)  at (9.6, -2.8) {Parasitic content\\[1pt]$F_\mathrm{aliases}$\\[1pt]\scriptsize Eq.\,\ref{eq:F_aliases}};
\node[output] (epar) at (5.0, -2.8) {Relative parasitic energy\\[1pt]$E_\mathrm{par,rel}$\\[1pt]\scriptsize Eq.\,\ref{eq:E_par_rel}};
 
\draw[flow] (stim)  -- node[signal,above]{$F_\mathrm{stim}$}    (blur);
\draw[flow] (blur)  -- node[signal,above]{$F_\mathrm{blurred}$} (samp);
\draw[flow] (samp)  -- node[signal,above]{$F_\mathrm{repl}$}    (recon);
 
\draw[flow] (recon.south) -- node[signal,right]{$F_\mathrm{retina}$} (out.north);
 
\draw[flow] (out.west) -- node[signal,above,align=center]
            {\scriptsize integrate $|F|^{2}$\\[-1pt]\scriptsize $|\mathbf f|\!\le\!\nu_\mathrm{eye}$}
            (epar.east);
 
\draw[nlflow] (blur.south west) -- (nl.north west)
              node[signal,pos=0.5,left=2pt]
                  {$F_\mathrm{blurred}$};
 
\draw[nlflow] (nl.north east) -- (samp.south west)
              node[signal,pos=0.5,right=2pt]
                  {$F_\mathrm{blurred}^{\,\mathrm{NL}}$};
 
\node[param,above=7mm of stim] (dist)
  {Approach distance $d$\\[0.5pt]\scriptsize scales spatial freq.\\[0.5pt]
   \scriptsize Eqs.\,\ref{eq:distance_scaling_real}--\ref{eq:distance_scaling_fourier}};
\draw[param-arrow] (dist.south) -- (stim.north);
 
\node[param,above=7mm of blur] (drho)
  {$\sigma_\rho$\\[0.5pt]\scriptsize acceptance angle};
\draw[param-arrow] (drho.south) -- (blur.north);
 
\node[param,above=7mm of samp] (dphi)
  {$\Delta\varphi,\;f_s\!=\!1/\Delta\varphi$\\[0.5pt]\scriptsize sampling pitch};
\draw[param-arrow] (dphi.south) -- (samp.north);
 
\end{tikzpicture}

\caption{Optical-model scheme, the diptera eye
is modelled as a cascade of four linear operations on the
striped stimulus $I(x,y)$ (blue boxes): acceptance blur
by the angular photoreceptor MTF $H_\rho$, spatial sampling by
the Dirac comb $\mathrm{III}_{f_s}$ at the inter-ommatidial
spacing $\Delta\varphi$, and Voronoi reconstruction by the
box-cell aperture $H_\mathrm{box}$.  Three eye parameters control the chain (dashed boxes on top). The approach
distance $d$ scales the spatial-frequency content of the
stimulus, the acceptance angle $\sigma_\rho$ pins $H_\rho$, and
the inter-ommatidial spacing $\Delta\varphi$ pins the comb pitch
$f_s$. The output stages (orange boxes) extract the
parasitic content $F_\mathrm{aliases}$ 
(Eq.~\ref{eq:F_aliases}) and integrate its power inside the
eye Nyquist disc to obtain the dimensionless metric
$E_\mathrm{par,rel}$ (Eq.~\ref{eq:E_par_rel}). An 
photoreceptor non-linearity (dashed violet branch) replaces $F_\mathrm{blurred}$ with
$F_\mathrm{blurred}^\mathrm{NL}$ before sampling and produces a
parallel parasitic signal $F_\mathrm{aliases}^\mathrm{NL}$ along
the same final path.}
\label{fig:optical_pipeline}
\end{figure}

\subsection{Biting fly vision}
\label{subsec:optical_model}

The spatial vision of the host-approaching biting fly is represented as
a linear, shift-invariant optical system. Additionally, a memoryless photoreceptor
non-linearity is incorporated as described in
\S\ref{subsec:nonlinearity}. This idealization excludes temporally adaptive gain control of the lamina and post-receptor lateral interactions within the optic lobe
\citep{borst2014fly,landnilsson2002eyes}, but retains the
essential features of the phenomenon under investigation: the emergence of new
spatial frequencies resulting from the interaction between a periodic stimulus
and a periodic sampling lattice
\citep{snyder1979physics,land1997mosquito,land1999fundamental}.
The retinal image $I_{\mathrm{retina}}(x,y)$ is the convolution of the
external stimulus $I_{\mathrm{stim}}(x,y)$ with the spatial impulse
response $h_{\mathrm{eye}}(x,y)$ of the compound eye:
\begin{equation}
I_{\mathrm{retina}}(x,y)
\;=\;
I_{\mathrm{stim}}(x,y) * h_{\mathrm{eye}}(x,y).
\label{eq:convolution}
\end{equation}
Spatial coordinates $(x,y)$ are expressed in degrees of visual angle
relative to the optical axis. The corresponding Fourier-domain
relation is the product
\begin{equation}
F_{\mathrm{retina}}(u,v)
\;=\;
F_{\mathrm{stim}}(u,v)\cdot H_{\mathrm{eye}}(u,v),
\label{eq:fourier_product}
\end{equation}
where $(u,v)$ are spatial frequency coordinates in
$\mathrm{cycles\,deg^{-1}}$, $F_{\mathrm{stim}} =
\mathcal{F}\{I_{\mathrm{stim}}\}$, and $H_{\mathrm{eye}}(u,v)$ is the
optical transfer function (OTF) of the compound eye.
The shift-invariance assumption is justified by the local and
quasi-periodic structure of the ommatidial array. 
In the small part of the visual field considered for approaching the target when optical vision takes over, $\Delta\varphi$ changes very little. However, $\Delta\varphi$ changes more
across the whole eye, which is relevant immediately before landing when the field of view is larger. In such a case, the differences in $\Delta\varphi$ between the acute zone
and the periphery\citep{land1997mosquito,land1999fundamental,hawkes2022vision} are handled separately by running the analysis for each region of the eye (\S\ref{subsec:acute_zone_sim}).

\subsection{Distance scaling in Fourier space}
\label{subsec:distance_scaling}

The model accounts for viewing distance by considering the visual angle it creates.
Let $I_{0}(x,y)$ represent the stimulus measured in
degrees of visual angle at a reference distance $d_{0}$, where the
body height is set to a standard value. For a different viewing distance
of $d$, the projected stimulus is uniformly scaled,
\begin{equation}
I_{\mathrm{stim}}^{(d)}(x,y)
\;=\;
I_{0}\!\left(\frac{d}{d_{0}}\,x,\;\frac{d}{d_{0}}\,y\right),
\label{eq:distance_scaling_real}
\end{equation}
where the scale factor $d/d_{0}$ reduces the angular extent of the
animal. The Fourier scaling theorem gives
\begin{equation}
F_{\mathrm{stim}}^{(d)}(u,v)
\;=\;
\left(\frac{d_{0}}{d}\right)^{2}
F_{0}\!\left(\frac{d_{0}}{d}\,u,\;\frac{d_{0}}{d}\,v\right),
\label{eq:distance_scaling_fourier}
\end{equation}
i.e., as the fly moves further from the host, the stripe pattern
shifts towards higher spatial frequencies in the angular
spectrum. Distance therefore acts as a controlled spectral sweep
that determines which components of the stripe pattern fall into
resonance with the fixed sampling lattice of the eye.

\subsection{Compound-eye transfer function}
\label{subsec:eye_otf}
The fly-eye OTF is modeled as the product of two physical factors.
A low-pass envelope, $H_{\rho}$, which represents
the angular acceptance function of a single ommatidium
\citep{snyder1979physics,landnilsson2002eyes} and $H_{\mathrm{samp}}$, which describes the periodic ommatidial lattice \citep{land1997mosquito,land1999fundamental}:
\begin{equation}
H_{\mathrm{eye}}(u,v)
\;=\;
H_{\rho}(u,v)\cdot H_{\mathrm{samp}}(u,v).
\label{eq:eye_otf}
\end{equation}
Treating each ommatidium as a diffraction-limited optical system with
a circular facet aperture of diameter $D$ operating at wavelength
$\lambda$, the angular point-spread function is the Airy pattern:
\begin{equation}
\mathrm{PSF}(\theta)
\;=\;
\left[\frac{2\,J_{1}(\pi D\theta/\lambda)}{\pi D\theta/\lambda}\right]^{2},
\label{eq:airy_psf}
\end{equation}
where, $J_{1}$ refers to the Bessel function of the first kind of order one.
$\theta$ represents the off-axis angle. The corresponding low-pass
envelope $H_{\rho}$ is found by taking the normalized autocorrelation of
the circular pupil. This is the standard diffraction-limited OTF for
incoherent imaging \citep{goodman2005fourier,born1999principles}:
\begin{equation}
H_{\rho}(u,v)
\;=\;
\begin{cases}
\dfrac{2}{\pi}\!\left[
\arccos\!\left(\dfrac{\rho}{\rho_{c}}\right)
-\dfrac{\rho}{\rho_{c}}\sqrt{\,1-\left(\dfrac{\rho}{\rho_{c}}\right)^{2}}
\,\right], & \rho \le \rho_{c}, \\[6pt]
0, & \rho > \rho_{c},
\end{cases}
\label{eq:airy_otf}
\end{equation}
with, $\rho=\sqrt{u^{2}+v^{2}}$ is the radial spatial frequency, and
$\rho_{c}=D/\lambda$ is the diffraction cutoff, measured in cycles per radian.
The aperture-to-wavelength ratio $D/\lambda$ is set by matching the half-power
width from Eq.~\eqref{eq:airy_psf} to the measured ommatidial acceptance
angle. The parameter values used for \textit{Aedes aegypti}
($\Delta\varphi$, $\sigma_{\rho}$, and the resulting $\rho_{c}$) and their
empirical basis are provided in \S\ref{subsec:eye_geometry}. The sampling
factor $H_{\mathrm{samp}}$ is left undefined here. Its exact form is the
subject of \S\ref{subsec:sampling}.

\subsection{Apposition baseline}
\label{par:apposition_baseline}

In the apposition compound eye of \textit{Aedes aegypti}, each
ommatidium has its own dioptric apparatus and a single
isolated rhabdom. The receptor signals are not combined across
neighboring ommatidia in the lamina. This is different from the neural
superposition found in higher Diptera (Calliphoridae,
Drosophilidae), where seven aligned receptors from seven
neighboring facets connect to a single cartridge
\citep{kirschfeld1967general,landnilsson2002eyes,borst2014fly}.
Even in higher Diptera, this convergence is a coordinated summing of
aligned signals. This increases sensitivity, but maintains
angular resolution.
For the current model, this means that the only physical
neighbor coupling in the optical front-end is the partial overlap
of the angular acceptance functions of nearby ommatidia, which is
already included in $H_{\rho}$ (Eq.~\ref{eq:airy_otf}). After this acceptance blur, each
ommatidium works as an independent point sampler.

\subsection{Sampling and reconstruction}
\label{subsec:sampling}

Ommatidial sampling on the lattice
$\Lambda_{\Delta\varphi} = \{(m\Delta\varphi,n\Delta\varphi)\,|\,
m,n\in\mathbb{Z}\}$ is, in the spatial domain, multiplication of
the acceptance-blurred stimulus by a two-dimensional Dirac comb,
\begin{equation}
I_{\mathrm{samples}}(x,y)
\;=\;
I_{\mathrm{blurred}}(x,y) \cdot \mathrm{III}_{\Delta\varphi}(x,y),
\quad
I_{\mathrm{blurred}} = I_{\mathrm{stim}} * h_{\rho},
\label{eq:spatial_sampling}
\end{equation}
where
$\mathrm{III}_{\Delta\varphi}(x,y) = \sum_{m,n}\delta(x{-}m\Delta\varphi)\,\delta(y{-}n\Delta\varphi)$
and $h_{\rho} = \mathcal{F}^{-1}\{H_{\rho}\}$. The dual statement
in the frequency domain is convolution of the blurred-stimulus
spectrum with the dual comb at spacing $f_{s} = 1/\Delta\varphi$,
\begin{equation}
F_{\mathrm{repl}}(u,v)
\;=\;
F_{\mathrm{blurred}}(u,v) \,\circledast\, \mathrm{III}_{f_{s}}(u,v)
\;=\;
\sum_{k,m}
F_{\mathrm{blurred}}\!\left(u-k f_{s},\,v-m f_{s}\right),
\label{eq:freq_replication}
\end{equation}
where $\circledast$ denotes two-dimensional convolution. The
$(k,m)\!=\!(0,0)$ term reproduces the signal the eye would deliver if
sampling were ideal; every other $(k,m)$ term is a spectral replica
of $F_{\mathrm{blurred}}$ shifted by integer multiples of $f_{s}$
along each axis. The spectral content of these replicas that falls within
the eye's Nyquist disc
$|f|\!\leq\!\nu_{\mathrm{eye}}=1/(2\Delta\varphi)$
is called the moiré signal. These are frequencies the brain receives, even though
the original stimulus did not contain them. Each ommatidium represents an angular Voronoi cell (a
$\Delta\varphi\!\times\!\Delta\varphi$ square for a square lattice).
By using the nearest-neighbour reconstruction rule,
the value reported by the closest ommatidium is assigned to every direction, which
spatially means convolving the spike train with a square indicator function of
side $\Delta\varphi$. The Fourier-domain twin is
multiplication by the corresponding box transfer function,
\begin{equation}
H_{\mathrm{box}}(u,v)
\;=\;
\operatorname{sinc}(\Delta\varphi\,u)\,
\operatorname{sinc}(\Delta\varphi\,v),
\label{eq:hbox}
\end{equation}
with $\operatorname{sinc}(x)\!=\!\sin(\pi x)/(\pi x)$. The full linear
retinal spectrum is then
\begin{equation}
F_{\mathrm{retina}}(u,v)
\;=\;
H_{\mathrm{box}}(u,v)\,
\sum_{k,m}
F_{\mathrm{blurred}}\!\left(u-k f_{s},\,v-m f_{s}\right).
\label{eq:F_retina_linear}
\end{equation}
The sinc factors vanish exactly at $u,v\!=\!\pm k f_{s}$ for
$k\!\geq\!1$, so $H_{\mathrm{box}}$ kills replicas at the comb spike
centres. The sinc envelope decays
as $1/u^{2}$ in each direction, so $F_{\mathrm{retina}}$ has finite
total energy as required for a physically realisable image.

\subsection{Photoreceptor non-linearity}
\label{subsec:nonlinearity}

The compressive response of insect photoreceptors is well described
by a Naka--Rushton function
\citep{laughlin1981retina,landnilsson2002eyes},
\begin{equation}
\mathrm{NR}_{\beta,s_{50}}(s)
\;=\;
\frac{s^{\beta}}{s^{\beta}+s_{50}^{\beta}},
\label{eq:naka_rushton}
\end{equation}
with shape exponent $\beta=0.7$ and half-saturation level
$s_{50}=0.5$. Inserted between the acceptance blur and the comb
convolution, this point-wise non-linearity creates harmonic and
intermodulation content
$f_{1}\!\pm\! f_{2}$, $2 f_{1}$, $2 f_{1}\!\pm\! f_{2}$
that the strictly linear chain of \S\ref{subsec:sampling} cannot
generate. To make the contribution of this non-linearity explicit, a non-linear blurred image is defined as:
\begin{equation}
I_{\mathrm{blurred}}^{\mathrm{NL}}(x,y)
\;=\;
(1-\alpha)\,I_{\mathrm{blurred}}(x,y)
+ \alpha\,\mathrm{NR}_{\beta,s_{50}}\!\bigl(I_{\mathrm{blurred}}(x,y)\bigr),
\label{eq:nl_blurred_spatial}
\end{equation}
with mix factor $\alpha\in[0,1]$ interpolating between linear
($\alpha=0$) and fully compressed ($\alpha=1$). The remainder of the
chain is unchanged but uses
$F_{\mathrm{blurred}}^{\mathrm{NL}} = \mathcal{F}\{I_{\mathrm{blurred}}^{\mathrm{NL}}\}$
in place of $F_{\mathrm{blurred}}$, giving
\begin{equation}
F_{\mathrm{aliases}}^{\mathrm{NL}}(u,v)
\;=\;
H_{\mathrm{box}}(u,v) \!\sum_{k,m}\!
F_{\mathrm{blurred}}^{\mathrm{NL}}\!\left(u-k f_{s},\,v-m f_{s}\right)
\;-\;
F_{\mathrm{blurred}}(u,v)\,H_{\mathrm{box}}(u,v).
\label{eq:F_aliases_NL}
\end{equation}
Subtracting the linear baseline
$F_{\mathrm{blurred}}\!\cdot\!H_{\mathrm{box}}$ in
Eq.~\ref{eq:F_aliases_NL} ensures that all content the brain
receives but the stimulus optics alone do not supply is counted as parasitic.

\subsection{Moir\'e isolation}
\label{subsec:moire_isolation}

The signal the eye would have delivered with ideal sampling is
$F_{\mathrm{signal}} = F_{\mathrm{blurred}}\!\cdot\! H_{\mathrm{box}}$.
Subtracting it from the full retinal spectrum
(Eq.~\ref{eq:F_retina_linear}) leaves the parasitic content of every
$(k,m)\!\neq\!(0,0)$ replica multiplied by the same reconstruction
MTF:
\begin{equation}
F_{\mathrm{aliases}}(u,v)
\;=\;
F_{\mathrm{retina}}(u,v) \;-\; F_{\mathrm{blurred}}(u,v)\,H_{\mathrm{box}}(u,v).
\label{eq:F_aliases}
\end{equation}
This is the definition of the parasitic spectrum used
throughout this publication. By construction, it is identically zero at
every $(u,v)$ where the brain receives the same content the
acceptance-blurred stimulus already supplied; what survives are the
genuinely added frequencies introduced by the
sampling-and-reconstruction stage.
The spatial-domain image of the parasitic information is recovered by
inverse Fourier transform,
\begin{equation}
I_{\mathrm{par}}(x,y)
\;=\;
\mathcal{F}^{-1}\{F_{\mathrm{aliases}}(u,v)\},
\label{eq:inverse_fft}
\end{equation}
yields an angular map of the artificial contrast that arises
exclusively from the interaction of the striped coat with
ommatidial sampling. All features seen in $I_{\mathrm{par}}$ are
phantom perceptions. They are patterns of contrast that have no
physical equivalent on the host's coat.

\subsection{Quantifying parasitic content}
\label{subsec:e_par}

The value that represents a parasitic spectrum is its total power measured within the eye's Nyquist disk:
\begin{equation}
E_{\mathrm{par}}(d)
\;=\;
\!\!\!\!\!\sum_{|f|\leq\nu_{\mathrm{eye}}}\!\!\!
|F_{\mathrm{aliases}}(u,v;d)|^{2},
\quad
E_{\mathrm{par}}^{\mathrm{NL}}(d)
\;=\;
\!\!\!\!\!\sum_{|f|\leq\nu_{\mathrm{eye}}}\!\!\!
|F_{\mathrm{aliases}}^{\mathrm{NL}}(u,v;d)|^{2}.
\label{eq:E_par}
\end{equation}
According to Parseval’s theorem, these are equivalent to
$\sum_{x,y}|I_{\mathrm{par}}(x,y)|^{2}$ in the spatial domain.
The restriction $|f|\leq\nu_{\mathrm{eye}}$ is essential because the parasitic
content outside the Nyquist disk is not detected by the fly brain, since the box reconstruction $H_{\mathrm{box}}$ has already filtered it out.
The raw $E_{\mathrm{par}}$ has units of $|F|^{2}$ summed over
a frequency disk, that is, (linear-intensity)$^{2}$ multiplied by a
$N$-dependent FFT normalisation factor. So, it is only
meaningful when comparing cases where the input intensity scale and FFT
gain remain constant. To obtain a scale-invariant scalar that is
intrinsic to the optics of the eye-stimulus system and can be
compared across images, cameras, and runs, the
relative parasitic energy
\begin{equation}
E_{\mathrm{par,rel}}(d)
\;=\;
\frac{E_{\mathrm{par}}(d)}{E_{\mathrm{sig}}(d)},
\qquad
E_{\mathrm{sig}}(d)
\;=\;
\!\!\!\!\!\sum_{|f|\leq\nu_{\mathrm{eye}}}\!\!\!
\bigl|F_{\mathrm{blurred}}(u,v;d)\,H_{\mathrm{box}}(u,v)\bigr|^{2},
\label{eq:E_par_rel}
\end{equation}
where $E_{\mathrm{sig}}$ is the integrated power of the alias-free
in-band signal ($k\!=\!m\!=\!0$ of
Eq.~\ref{eq:F_retina_linear}). This is what the brain would receive
if the sampling were ideal at the same approach distance. The
non-linear analog uses the same denominator:
\begin{equation}
E_{\mathrm{par,rel}}^{\mathrm{NL}}(d)
\;=\;
E_{\mathrm{par}}^{\mathrm{NL}}(d)\,/\,E_{\mathrm{sig}}(d).
\label{eq:E_par_rel_NL}
\end{equation}
Both the numerator and denominator are calculated from the same FFT,
using the same windowed image but over separate frequency ranges. This means that
the effects of intensity, Hann-window, and FFT normalization cancel out
completely in the ratio. As a result, $E_{\mathrm{par,rel}}$
is dimensionless and appears on the y-axis of every Fourier-stage
parasitic-energy figure in this paper. If an idealized eye is considered that samples at a sufficiently high rate, or whose acceptance modulation transfer function (MTF) exhibits strong attenuation
such that no stripe content aliases above its Nyquist limit, nor
the fundamental nor the supra-Nyquist edge harmonics that dominate the
parasitic band (\S\ref{subsec:fourier_dissection}), 
the moir\'e mechanism is absent and $E_{\mathrm{par,rel}}\!\to\!0$.
On the other hand, a fly sampling lattice viewing the close-approach band
where $d\!\in\![1,2.5]\,\si{\meter}$, produces values that
have $E_{\mathrm{par,rel}}\!>\!0$. This is the
biologically relevant regime, which will be discussed in
\S\ref{subsec:reichardt_control_discussion}. Because $E_{\mathrm{par,rel}}$ is a ratio of squared-magnitude spectral sums, it is non-negative by definition. At this
Fourier stage, the moir\'e effect can only add spectral power in
the brain-accessible passband and never subtract from it. This
non-negativity is a property of the optical stage only.
Later stages that produce a signed output, such as the Hassenstein--Reichardt motion detector (see \S\ref{subsec:motion}), do not have this property. Instead, a separate signed metric $E_{\mathrm{HR}}^{\mathrm{moir\acute{e}}}$
(Eq.~\ref{eq:E_HR_moire}) is used for the corresponding
Elementary Motion Detector (EMD) stage comparison in
\S\ref{subsec:reichardt_control_discussion}.

\section{Simulation}
\label{sec:simulation}

This chapter describes how the optical model of \S\ref{sec:optical_model}
is realized numerically. Eq.~\ref{eq:F_aliases} is the analytical
target. The simulation reproduces each operation in this equation
within Fourier space, and validates each Fast Fourier Transform (FFT) stage using numerical controls.
Additionally, the simulation reports the relative parasitic energy as defined by
Equation~\ref{eq:E_par_rel}.

\subsection{Image dataset and preprocessing}
\label{subsec:dataset}

The pipeline is applied to a set of photographs of
\textit{Equus quagga} captured using a Nikon D50 with a 70–300 mm lens, typically at $f_{\mathrm{lens}}\!=\!300\,\mathrm{mm}$ and $f/5.6$,
ISO\,400). The native angular pixel pitch is
$\mathrm{dpp_{cam}}\!=\!\arctan(p_{\mathrm{pix}}/f_{\mathrm{lens}})\!\approx\!5.42\,\mathrm{arcsec/pixel}$,
which is derived from the XMP focal length $f_{\mathrm{lens}}$ and a sensor pixel pitch
$p_{\mathrm{pix}}\!=\!7.88\,\si{\micro\meter}$. Each image is converted to linear
grayscale, assuming an sRGB gamma of $2.2$.
A square region of interest centred on the body is selected 
in each photograph. The resulting body height in camera pixels
$h_{\mathrm{body,pix}}$, together with the focal length, determines the
camera-to-host distance using
\begin{equation}
d_{\mathrm{cam}}
\;=\;
\frac{h_{\mathrm{body,m}}}
{2\tan\!\bigl(h_{\mathrm{body,pix}}\!\cdot\!\mathrm{dpp_{cam}}/2\bigr)},
\qquad h_{\mathrm{body,m}} = \SI{1.4}{\meter}.
\label{eq:cam_distance}
\end{equation}
The same processing chain is then evaluated at a sequence of biting fly approach
distances $d$ using isotropic crop scaling. For each $d$, the script
extracts a square crop of side
$\bigl(d/d_{\mathrm{cam}}\bigr)\,\mathrm{FoV}/\mathrm{dpp_{cam}}$
camera-pixels around the selected center, rescales it to the simulation
grid with $N\!=\!384$ samples per side, and inputs it to the chain described in
\S\ref{subsec:optical_model}--\S\ref{subsec:moire_isolation}. Body height is calibrated to a generic value $h_{\mathrm{body}} = \SI{1.4}{\meter}$. As described above, the error introduced by this calibration results in only a minor shift in spatial frequency in the angular spectrum and does not affect the absolute amount of parasitic energy.

\subsection{Compound-eye and ommatidial sampling}
\label{subsec:eye_geometry}

\begin{figure}[H]
\centering
\includegraphics[width=0.9\textwidth]{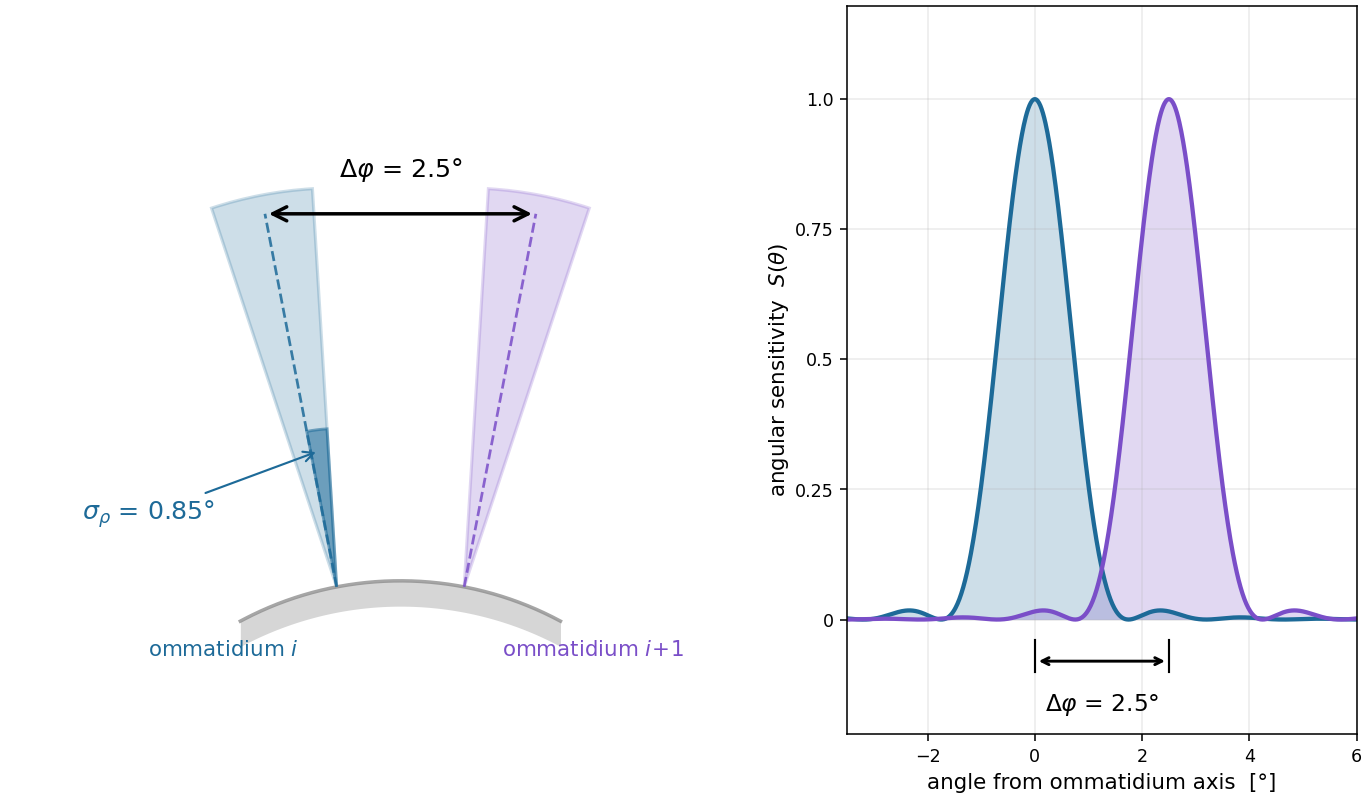}
\caption{Compound-eye sampling geometry assumed in the simulation. The left side sketches two adjacent ommatidia of the apposition compound eye with inter-ommatidial angle and 
$\Delta\varphi$ and angular acceptance half-width
$\sigma_{\rho}$. On the right side, the angular acceptance function $S(\theta)$ of one ommatidium is plotted.}
\label{fig:eye_geometry}
\end{figure}

The two geometric quantities that fully parameterise the sampling
front-end of the model are the inter-ommatidial angle
$\Delta\varphi$ and the angular acceptance half-width
$\sigma_{\rho}$ of a single ommatidium
(Eq.~\ref{eq:airy_otf}--\ref{eq:spatial_sampling}). Both are
empirically constrained in \textit{Aedes aegypti} from goniometric
and electrophysiological measurements
\citep{land1997mosquito,land1999fundamental,landnilsson2002eyes,hawkes2022vision}.
Figure~\ref{fig:eye_geometry} summarizes the geometry and the
parameters adopted throughout the simulations. The left side sketches two adjacent ommatidia of the apposition compound eye, separated by the inter-ommatidial angle $\Delta\varphi\!=\!\SI{2.5}{\degree}$, which is the value for diurnal culicids reported in Refs.~\citep{land1997mosquito,land1999fundamental}.
On the right side, the angular acceptance function $S(\theta)$ of one ommatidium is plotted. $S(\theta)$ is modelled as the diffraction-limited Airy pattern of a circular facet aperture (Eq.~\ref{eq:airy_psf}) \citep{snyder1979physics,goodman2005fourier,born1999principles},
scaled so that its half-width at half maximum matches the empirical
acceptance half-width $\sigma_{\rho}\!\approx\!\SI{0.85}{\degree}$, which is 
equivalent to a full width at half maximum
$\Delta\rho\!\approx\!\SI{2}{\degree}$, as reported for diurnal Culicidae
\citep{land1997mosquito,land1999fundamental,landnilsson2002eyes}.
The two curves correspond to two neighboring ommatidia $i$ and
$i\!+\!1$ whose optical axes are separated by $\Delta\varphi$. Their
acceptance functions overlap only weakly at half-maximum, which
determines the spatial-frequency cutoff of the eye through
Eq.~\ref{eq:airy_otf}: the Airy MTF $H_{\rho}(u,v)$ has a hard
diffraction cutoff at $\rho_{c}\!=\!D/\lambda$, with the
aperture-to-wavelength ratio $D/\lambda$ chosen so that the
half-power width of the angular PSF matches the empirical acceptance
half-width $\sigma_{\rho}$.

For the simulations the standard
\textit{Aedes aegypti} parameter set
$(\Delta\varphi,\sigma_{\rho})\!=\!(\SI{2.5}{\degree},\SI{0.85}{\degree})$ is adopted
\citep{land1997mosquito,land1999fundamental,landnilsson2002eyes},
which yields a sampling rate $f_{s}\!=\!1/\Delta\varphi
\!=\!\SI{0.40}{cyc.deg^{-1}}$ and a corresponding Nyquist limit
$\nu_{\mathrm{eye}}\!=\!f_{s}/2\!=\!\SI{0.20}{cyc.deg^{-1}}$.
Matching the half-power width of the Airy PSF
(Eq.~\ref{eq:airy_psf}) to $\sigma_{\rho}\!=\!\SI{0.85}{\degree}$
sets the aperture-to-wavelength ratio $D/\lambda$ and therefore
the diffraction cutoff of the Airy MTF
(Eq.~\ref{eq:airy_otf}) at
$\rho_{c}\!=\!D/\lambda\!\approx\!\SI{0.61}{cyc.deg^{-1}}$, with a
$-3\,\mathrm{dB}$ amplitude point at
$\rho\!\approx\!0.23\,\rho_{c}\!\approx\!\SI{0.14}{cyc.deg^{-1}}$.
The hard cutoff is about three times higher than
$\nu_{\mathrm{eye}}$, so the spatial-frequency band
$\nu_{\mathrm{eye}}\!<\!\rho\!\leq\!\rho_{c}$ is a window of finite
width that the optics transmits with non-zero MTF but the
ommatidial sampling lattice cannot faithfully represent. This is
precisely the regime in which the moir\'e mechanism of
\S\ref{subsec:moire_isolation} operates: stripe fundamentals above
the Nyquist limit but below the diffraction cutoff are not killed
by the acceptance blur before they are aliased by the sampling
comb. The same parameter set has been used by
\citet{warrant1999seeing} as the canonical reference point for the
apposition eyes of diurnal small-headed Diptera, is consistent
with the comparative review of culicid eye optics by
\citet{kawada2006eye}, and reproduces the antero-ventral
acute-zone values reported in the recent vision survey of
\citet{hawkes2022vision}.

\subsection{Numerical pipeline}
\label{subsec:pipeline}

$N\!=\!384$ samples per side over a fixed angular canvas of
FoV~$=\!30^\circ$ are used, yielding $\Delta x\!=\!\mathrm{FoV}/N\!=\!0.078^\circ$
per pixel. The FFT bin spacing is
$\Delta f\!=\!1/(N\Delta x)\!=\!1/\mathrm{FoV}$, and the comb shift in
FFT-bin units is
\begin{equation}
\frac{f_{s}}{\Delta f}
\;=\;
\frac{N\Delta x}{\Delta\varphi}
\;=\;
\frac{\mathrm{FoV}}{\Delta\varphi}.
\label{eq:comb_shift_bins}
\end{equation}
For $\mathrm{FoV}\!=\!30^\circ$ and $\Delta\varphi\!=\!2.5^\circ$ this
is exactly $12$, so the replica sum of Eq.~\ref{eq:freq_replication}
can be evaluated without sub-pixel interpolation by integer-shifted
summation. When the canvas is fractionally off
$\mathrm{FoV}$ (camera-pixel rounding produces $29.97^\circ$ in
practice), the comb shift becomes $11.99$ bins; the corresponding
sub-pixel spectral shift is implemented via the spatial-domain phase
ramp
$f_{\mathrm{spatial}}(x,y)\,\sum_{k,m}\!\exp\bigl(2\pi i\,(k f_{s} x + m f_{s} y)\bigr)$,
exact by Poisson duality. $|k|,|m|\!\leq\!3$ replica shells
($49$ shifts in total) are summed up; higher-order shells contribute below
floating-point noise inside the displayed window.
The simulated patch contains
$(\mathrm{FoV}/\Delta\varphi)^{2}\!=\!12\!\times\!12\!=\!144$
ommatidia, each separated by $\Delta\varphi\!=\!2.5^\circ$. This is a
focal sub-region of one eye, not the entire eye; a complete
\textit{Aedes aegypti} compound eye contains on the order of $350$
ommatidia
\citep{liu2019general,hawkes2022vision} and covers a substantially
larger solid angle. The choice $\mathrm{FoV}\!=\!30^\circ$ corresponds
to the angular extent over which the assumption of a constant
inter-ommatidial angle is reasonable in the antero-ventral acute zone
\citep{land1997mosquito,land1999fundamental,hawkes2022vision}; the
regional variation away from this zone is treated explicitly in
\S\ref{subsec:acute_zone_sim}.
The simulation crop is not periodic. A rectangular FFT window
introduces sinc-shaped leakage zeros that appear as a regular grid
of dark pixels in the displayed power spectra, with no physical
counterpart in the eye. A separable two-dimensional Hann
window $w(x,y) = w_{1}(x)\,w_{1}(y)$ is applied to the spatial signals before
each FFT, where $w_{1}$ is the standard one-dimensional Hann taper
of length $N$. Because the same window factor multiplies
$F_{\mathrm{stim}}$, $F_{\mathrm{blurred}}$ and $F_{\mathrm{retina}}$,
the moir\'e-isolation difference of Eq.~\ref{eq:F_aliases} is
unaffected up to a constant gain. The Hann window is the only
modification to the standard pipeline of \S\ref{subsec:optical_model}
required by the Fourier-domain construction.
The Naka--Rushton compression of Eq.~\ref{eq:nl_blurred_spatial} is
applied point-wise in the spatial domain on $I_{\mathrm{blurred}}$,
rescaled to zero-mean unit-variance before compression and re-matched
to the original mean and variance after to remove global gain
shifts. The compressed image is then inverse Fourier transformed and used
as $F_{\mathrm{blurred}}^{\mathrm{NL}}$ throughout the
non-linear branch (Eq.~\ref{eq:F_aliases_NL}).

\subsection{Airy MTF}
\label{subsec:airy}

The main analysis uses the acceptance function $H_{\rho}$. This
function is based on the diffraction-limited Airy pattern of an
apposition lens with an effective aperture $D$ that operates at
wavelength $\lambda$ \citep{nilsson1989optics,landnilsson2002eyes},
and has an angular cutoff
\begin{equation}
f_{c} \;=\; D / \lambda \quad [\mathrm{cyc/rad}].
\label{eq:airy_cutoff}
\end{equation}
The closed-form circular-aperture MTF is
\begin{equation}
H_{\mathrm{Airy}}(f)
\;=\;
\frac{2}{\pi}
\Bigl[
   \arccos\!\bigl(f/f_{c}\bigr)
\;-\;
   (f/f_{c})\sqrt{1 - (f/f_{c})^{2}}
\Bigr]
\quad \text{for } f \leq f_{c},
\label{eq:airy_mtf}
\end{equation}
The function is zero above the cutoff. For \textit{Aedes aegypti}
facets, $D \!\approx\! 15$ to $25\,\si{\micro\meter}$, operating
around $\lambda \!\approx\! 500\,\si{\nano\meter}$, the diffraction
cutoff is about $2.4^{\circ}$ FWHM, which is similar to the geometric
acceptance angle $\Delta\rho \!\approx\! 2^{\circ}$
\citep{land1997mosquito}. Diffraction and rhabdom-aperture geometry
contribute about equally to $H_{\rho}$, and the Airy pattern
describes both the central lobe and the sharp high-frequency cutoff,
which a Gaussian envelope (an approximation reviewed in
\S\ref{app:gaussian_otf}) does not have. The simulation works the
same way with either option because the diffraction-limited MTF acts
as a spatial filter that can be replaced and applied in the
frequency domain. If the Airy MTF is replaced with the Gaussian
approximation from \S\ref{app:gaussian_otf}
($D\!=\!\SI{20}{\micro\meter}$, $\lambda\!=\!\SI{500}{\nano\meter}$)
the moir\'e-peak distance stays the same, but the peak
$E_{\mathrm{par,rel}}$ drops by about a factor of 2, because the Airy
MTF has a sharper cutoff than the Gaussian envelope and so lets more
of the stripe-fundamental energy into the alias-generating band. In
this way, the Gaussian approximation is a conservative under-estimate.
The moir\'e-relevant approach distance and the existence of a clearly
defined peak are not affected by the choice of PSF.

\subsection{Regional variation in $\Delta\varphi$}
\label{subsec:acute_zone_sim}

The compound eye of \textit{Aedes aegypti} is regionally
specialized: the inter-ommatidial angle $\Delta\varphi$ varies
throughout the visual field, with the antero-ventral region, which is the
part that looks down and forward at a host during the landing approach, having higher resolution than the dorsal periphery. Studies of biting flies show that this acute zone is closely related to how they fixate on and land on hosts
\citep{land1997mosquito,land1999fundamental}. The ommatidia
sketched in Fig.~\ref{fig:eye_geometry} represent the sampling geometry of the compound eye. The regional difference between the
acute zone and the periphery is a difference in angle 
$\Delta\varphi$ between the adjacent ommatidia. The corresponding acceptance half-width $\sigma_{\rho}$ scales with the diameter of the rhabdom and is empirically smaller in the acute zone\citealp{land1997mosquito,landnilsson2002eyes}.
When $\Delta\varphi$ depends on position, shift-invariance is no longer maintained.
A piecewise-stationary regional analysis helps. By splitting the visual
field into a small number of regions, the existing pipeline can be used
independently in each region with its own $(\Delta\varphi,\,\sigma_{\rho})$. The simulation shows three canonical configurations with regional
values estimated from the lens-diameter gradient documented in the
references~\citep{land1999fundamental,hawkes2022vision} and
propagated to the acceptance angle by the diffraction-limited
optics of \S\ref{subsec:eye_otf}:
\[
\begin{array}{lcc}
\text{Foveal core} &
   \Delta\varphi\!=\!\SI{1.5}{\degree}, &
   \sigma_{\rho}\!=\!\SI{0.6}{\degree}\\[2pt]
\text{Standard (canonical acute zone)} &
   \Delta\varphi\!=\!\SI{2.5}{\degree}, &
   \sigma_{\rho}\!=\!\SI{0.85}{\degree}\\[2pt]
\text{Dorsal periphery} &
   \Delta\varphi\!=\!\SI{3.5}{\degree}, &
   \sigma_{\rho}\!=\!\SI{1.2}{\degree}
\end{array}
\]
The Standard parameters are
used throughout this study and represent a high-resolution diurnal
culicid, comparable to the antero-ventral acute zone of
\textit{Aedes aegypti}.
Eye-averaged \textit{Aedes aegypti} is substantially coarser
($\Delta\varphi\!\approx\!\SI{6.40}{\degree}$,
$p_{\mathrm{eye}}\!\approx\!2.08\,\si{\micro\meter\radian}$ 
\citealp{kawada2005nocturnal}), and applying the same pipeline at
these parameters increases the peak $E_{\mathrm{par,rel}}$ from
$\sim\!\SI{8}{\percent}$ to $>\!\SI{40}{\percent}$ across the
close-approach band. Within the
family of biting Diptera the parasitic-energy fraction is therefore
monotonic in $\Delta\varphi$ over the biologically relevant range of
tabanids, glossinids.  Other culicid taxa with eye parameters
between $p_{\mathrm{eye}}\!\approx\!1$ and
$p_{\mathrm{eye}}\!\approx\!3\,\si{\micro\meter\radian}$ would all show
parasitic-energy fractions in the range \SIrange{8}{40}{\percent}.
The Standard parameter set is in this sense a conservative
lower bound on the moir\'e effect; coarser-eyed biting flies see
more parasitic content, not less.
The Foveal core row of the table is a model probe of a
hypothetically more-resolved sub-region of the acute zone, and the Dorsal periphery row a corresponding probe in the opposite
direction. The moir\'e mechanism is therefore not specific to
culicids, and the quantitative shifts implied by family-level
differences are captured by the same piecewise-stationary regional
construction used here for within-eye variation. The biological
interpretation of how the moir\'e signature varies across these
regions is deferred to \S\ref{subsec:acute_zone_disc}, where the
resulting $E_{\mathrm{par,rel}}(d)$ curves are reported
(Fig.~\ref{fig:acute_chromatic_a}).

\subsection{Fourier model simulation}
\label{subsec:pipeline_output}

\begin{figure}[H]
\centering
\includegraphics[width=\textwidth]{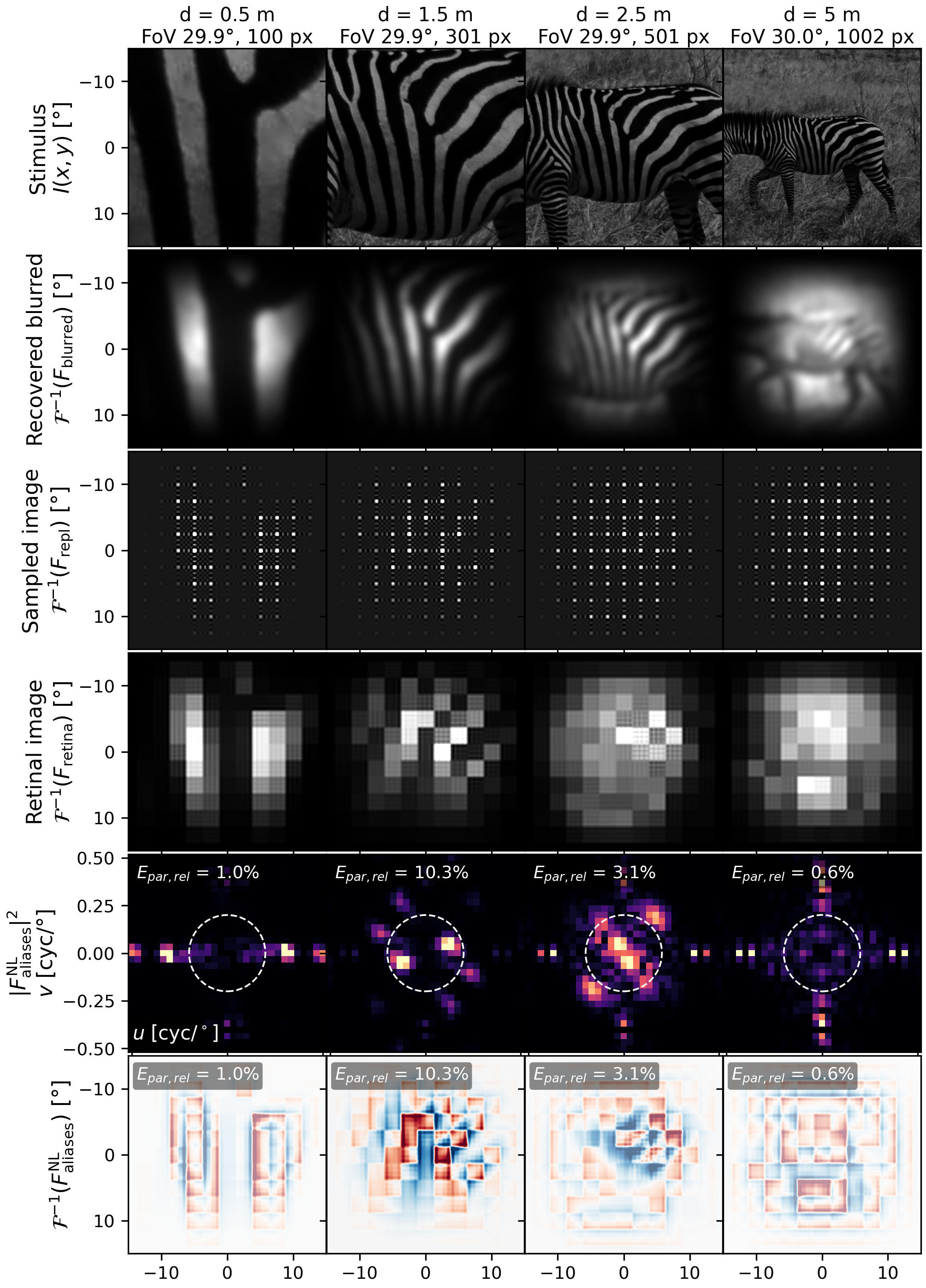}
\caption{Full pipeline output on a single representative photograph
(\textit{DSC\_0084}, \textit{Equus quagga}). It tracks one approach
distance per column ($d \in \{0.5,\,\textit{1.5},\,2.5,\,5\}\,\si{\meter}$)
through the six stages of the cascade developed in
\S\ref{subsec:sampling}--\S\ref{subsec:moire_isolation}, ending at
the non-linear parasitic image
$\mathcal{F}^{-1}\{F_{\mathrm{aliases}}^{\mathrm{NL}}\}$.
Rows~5 and~6 print the relative parasitic energy
$E_{\mathrm{par,rel}}$ (Eq.~\ref{eq:E_par_rel}) of each column.}
\label{fig:fourier_moire}
\end{figure}

The complete simulation pipeline output on a Zebra photograph (DSC\_0084, the same photograph that serves as striped host in Figs.~\ref{fig:zebra_vs_horse_a} and~\ref{fig:reichardt_control_zebra}, \textit{Equus quagga}) through the six stages of the cascade developed in
\S\ref{subsec:sampling}--\S\ref{subsec:moire_isolation}, ending in
the non-linear parasitic image
$\mathcal{F}^{-1}\{F_{\mathrm{aliases}}^{\mathrm{NL}}\}$ is shown in Fig.~\ref{fig:fourier_moire}. Each column represents a different approach distance ($d \in \{0.5,\,1.5,\,2.5,\,5\}\,\si{\meter}$), and each row shows a stage of the
cascade.
Row 1 (Stimulus) shows an angular cut-out at a fixed
field of view ($\mathrm{FoV} = 30^{\circ}$) centered on the body. The
camera-pixel side-length for each column is given in the panel header.
Row 2 (Recovered blurred) shows the inverse FFT of the windowed
spectrum,
$F_{\mathrm{blurred}} = \mathcal{F}\{I_{\mathrm{blurred}}\}$,
after applying the acceptance-MTF low-pass filter and Naka–Rushton compression. This is a
numerical control step in the FFT pipeline (round-trip relative error
$\lesssim 10^{-15}$, see Section \ref{subsec:fft_controls}).
Row 3 (Sampled image) shows the inverse FFT of the
comb-replicated spectrum,
$F_{\mathrm{repl}} = F_{\mathrm{blurred}} \circledast$
$\mathrm{III}_{f_{s}}$ (see Eq.~\ref{eq:freq_replication}). Each
bright dot represents one of the
$(\mathrm{FoV}/\Delta\varphi)^{2} = 144$ ommatidial sample
values.
Row 4 (Retinal image) shows the inverse FFT of the
reconstructed spectrum,
$F_{\mathrm{retina}} = F_{\mathrm{repl}} \cdot H_{\mathrm{box}}$,
which is the pixelated mosaic that the post-sampling stages of the
visual system actually receive.
Row 5 ($|F_{\mathrm{aliases}}^{\mathrm{NL}}|^{2}$) shows the isolated parasitic power spectrum (see Eq.~\ref{eq:F_aliases_NL})
as linear power, normalized in each column to its 99.5th
percentile, so that the alias replicas stand out on a common dark
background. The white dashed circle marks the eye's Nyquist limit,
$\nu_{\mathrm{eye}} = \SI{0.20}{cyc.deg^{-1}}$. The number
printed in each panel is the integral of the displayed spectrum over this
Nyquist disc, normalised by the signal energy inside the same disc,
i.e.\ $E_{\mathrm{par,rel}}$ of Eq.~\ref{eq:E_par_rel} for that column.
The false-colour image therefore shows where the parasitic energy sits
in the frequency plane (brightness is relative within each column. No
absolute scale is implied), and the printed number how much of it there
is. The sinc-zeros of the Voronoi reconstruction MTF at $\pm f_{s}$ lie
on the dark background and are not visible on the linear scale.
Row 6 ($\mathcal{F}^{-1}\{F_{\mathrm{aliases}}^{\mathrm{NL}}\}$)
shows the non-linear phantom percept on a symmetric, normalized red–blue scale, normalized independently in each column. The panel repeats the
column's $E_{\mathrm{par,rel}}$ and encodes it in its opacity,
$\eta = 0.25 + 0.75\sqrt{E_{\mathrm{par,rel}}/E_{\max}}$ with
$E_{\max}$ the largest value among the four columns, so a faint panel
carries little parasitic energy and a fully opaque one the most.
High-amplitude features systematically align with the
black-to-white edges of the underlying coat (see also
Section \ref{subsec:fourier_dissection}). 

\begin{figure}[H]
\centering
\includegraphics[width=\textwidth]{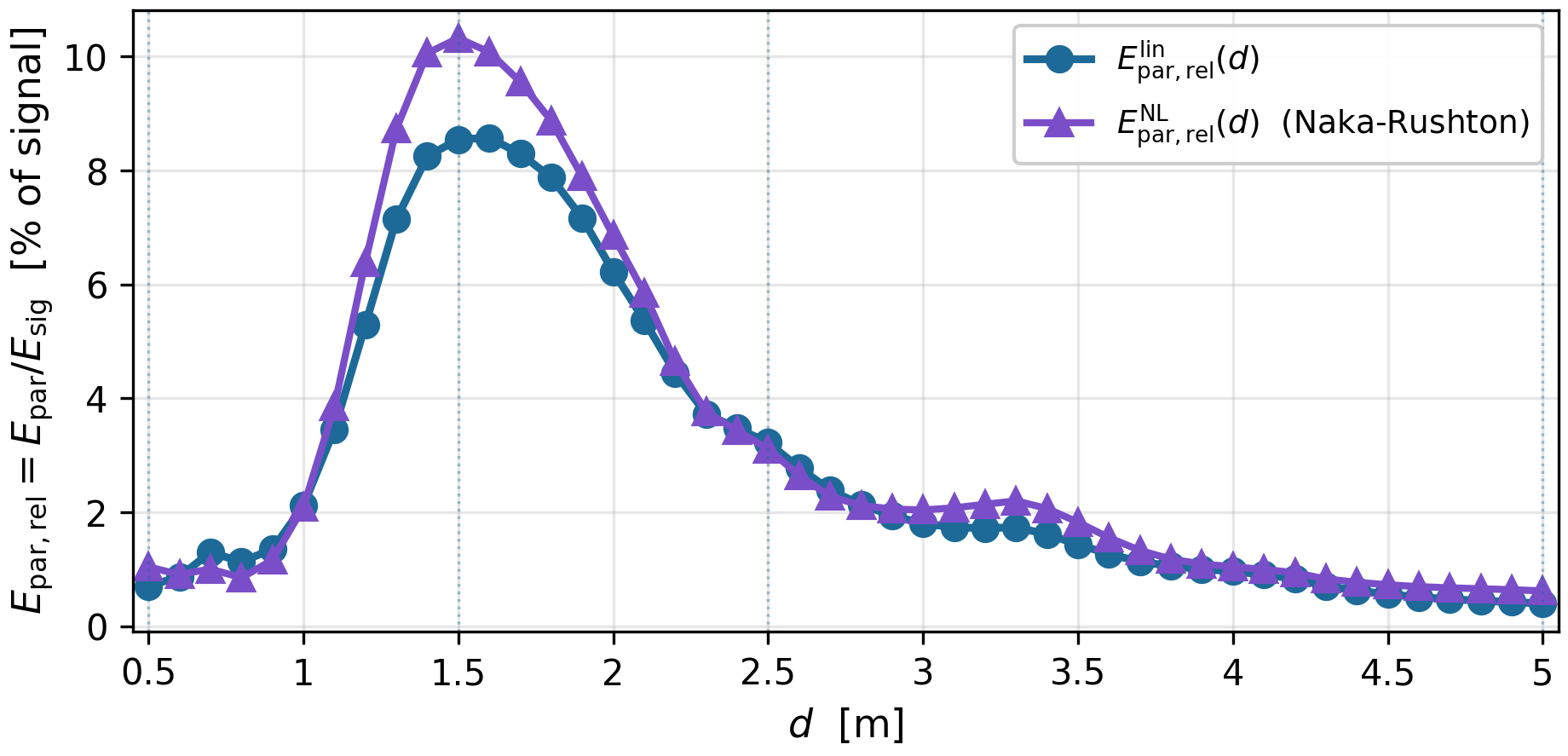}
\caption{Integrated relative parasitic energy $E_{\mathrm{par,rel}}(d)$
(Eq.~\ref{eq:E_par_rel}) as a function of approach distance, for both
the linear branch (Eq.~\ref{eq:F_aliases}) and the non-linear
Naka--Rushton branch (Eq.~\ref{eq:F_aliases_NL}). The two curves are
closely aligned, peak at the same distance \textit{($d\approx\SI{1.5}{\meter}$;
$\SI{10.3}{\percent}$ non-linear, $\SI{8.6}{\percent}$ linear)}, and identify a
moir\'e-relevant window $d\in[1.0,2.5]\,\si{\meter}$. The dashed
vertical guidelines mark the four column distances of Figure~3, whose
printed $E_{\mathrm{par,rel}}$ values are the corresponding points of the
non-linear curve.}
\label{fig:fourier_moire_set}
\end{figure}

Fig.~\ref{fig:fourier_moire_set} shows the relative parasitic energy,
$E_{\mathrm{par,rel}}(d) = E_{\mathrm{par}}(d)/E_{\mathrm{sig}}(d)$,
(see Eq.~\ref{eq:E_par_rel}), as a function of the approach distance $d$ ($\Delta d = 0.1$ m), for the
linear branch (blue circles, see Eq.~\ref{eq:F_aliases}) and the
non-linear Naka–Rushton branch (purple triangles,
see Eq.~\ref{eq:F_aliases_NL}). 
The two curves are closely aligned, peak
at the same approach distance\textit{, $d \approx \SI{1.5}{\meter}$
($E_{\mathrm{par,rel}}^{\mathrm{NL}} \approx \SI{10.3}{\percent}$,
$E_{\mathrm{par,rel}}^{\mathrm{lin}} \approx \SI{8.6}{\percent}$)}, and identify a moiré-relevant
window $d \in [1.0, 2.5]$ meters. The four distances marked in Fig.~\ref{fig:fourier_moire_set} (dashed vertical line) represent the
four columns at their corresponding points on the energy curve. The
values printed in rows~5 and~6 of Fig.~\ref{fig:fourier_moire}
($E_{\mathrm{par,rel}}^{\mathrm{NL}} = 1.0$, $10.3$, $3.1$ and
$\SI{0.6}{\percent}$ at $d = 0.5$, $1.5$, $2.5$ and $\SI{5}{\meter}$)
are these four points of the non-linear curve. The qualitative interpretation of $E_{\mathrm{par,rel}}(d)$ as a function of the approach distance is discussed
in section \ref{subsec:fourier_dissection}.

\subsection{Post-retinal motion processing}
\label{subsec:motion}

To translate the parasitic retinal signal into a behaviorally
meaningful prediction, a model of the fly
motion-detection pathway is fed with it. The Hassenstein--Reichardt elementary
motion detector (EMD)
\citep{hassenstein1956systemtheoretische,borst2010motion}
correlates the temporally low-pass-filtered signal of one ommatidium
with the unfiltered signal of a neighboring ommatidium and
subtracts the mirror-symmetric correlation. For two adjacent
inputs $L(t)$ and $R(t)$ separated by $\Delta\varphi$,
\begin{equation}
\mathrm{HR}(t)
\;=\;
\bigl[ L(t)*\tau_{\mathrm{lp}}(t) \bigr]\,R(t)
\;-\;
L(t)\,\bigl[ R(t)*\tau_{\mathrm{lp}}(t) \bigr],
\label{eq:reichardt}
\end{equation}
where $\tau_{\mathrm{lp}}(t)$ is a first-order low-pass filter with
time constant $\tau_{\mathrm{HR}} \approx \SI{30}{\milli\second}$
\citep{borst2010motion,borst2014fly}. Eq.~\ref{eq:reichardt} is the
opponent form of the detector. Its output is a signed
quantity per detector per frame, positive for motion in one
direction along the $L\!\to\!R$ axis and negative for the mirror
direction. The direction-agnostic motion energy at frame $t$ is
the squared output spatially averaged across the
$N_{\mathrm{det}}\!\times\!N_{\mathrm{det}}$ detector array:
\begin{equation}
E_{\mathrm{HR}}(t)
\;=\;
\bigl\langle \mathrm{HR}_{x}(t)^{2}\bigr\rangle
\;+\;
\bigl\langle \mathrm{HR}_{y}(t)^{2}\bigr\rangle ,
\label{eq:E_HR}
\end{equation}
with detectors applied independently in the horizontal and
vertical directions and $N_{\mathrm{det}}\!=\!12$ ommatidia per side over the
\SI{30}{\degree} simulated patch (one detector pair per
neighbouring ommatidium).
The motion energy of Eq.~\ref{eq:E_HR} is
direction-agnostic. The per-detector squaring discards the sign
that separates expansion from contraction. Landing, however, is
gated by the radial expansion (looming) of the optic-flow field
rather than by its undirected magnitude, so the same detector
array is also read out in a second, signed form. Treating the
horizontal and vertical detectors at each location as the
components of a local motion vector
$\mathbf{HR}(t)\!=\!(\mathrm{HR}_{x},\mathrm{HR}_{y})$, the vector is
projected onto the outward radial direction $\hat{\mathbf{r}}$ from
the focus of expansion and averaged
over the array (see Section \ref{app:reichardt_cancellation}),
\begin{equation}
E_{\mathrm{exp}}(t)
\;=\;
\bigl\langle\, \mathbf{HR}(t)\cdot\hat{\mathbf{r}} \,\bigr\rangle .
\label{eq:E_exp}
\end{equation}
Unlike $E_{\mathrm{HR}}$, the looming signal
$E_{\mathrm{exp}}$ is signed. Positive for a net expanding
(looming) field and negative for a net contracting (receding)
one. The two are complementary projections of the same EMD output. $E_{\mathrm{HR}}$ measures how much coherent motion energy
survives, $E_{\mathrm{exp}}$ measures whether the surviving motion
still points outward, that is, whether the approach is registered
as an approach at all. Evaluated for the full and clean pipelines
below it yields $E_{\mathrm{exp}}^{\mathrm{full}}$ and
$E_{\mathrm{exp}}^{\mathrm{clean}}$, exactly as for the energy.
Because individual detectors are aperture-limited $E_{\mathrm{exp}}$ is informative only as a field average and
only through the perceived-versus-true comparison of
\S\ref{subsec:reichardt_control_discussion}, in which the shared
aperture and sampling geometry cancels.
A close-approach scenario of an attacking biting fly is simulated
by re-cropping the photograph at successive equi-spaced approach
distances $d(t)\!=\!d_{0} - v_{\mathrm{app}}\,t$ with
$d_{0}\!=\!\SI{5.0}{\meter}$, closing speed
$v_{\mathrm{app}}\!=\!\SI{0.5}{\meter\per\second}$, and frame
interval $\Delta t\!=\!\SI{30}{\milli\second}$, which delivers
angular expansion rates inside the
$\SIrange{20}{200}{\degree\per\second}$ range documented for
in-flight tabanids \citep{caro2019benefits}. Each frame of the
approach trajectory is run twice through the optical pipeline of
\S\ref{subsec:pipeline}.

\noindent{Full pipeline:}\,
Acceptance blur ($H_{\rho}$) $\to$ Naka--Rushton ($\to
I_{\mathrm{blurred}}^{\mathrm{NL}}$) $\to$ comb-replication
(Eq.~\ref{eq:freq_replication}) $\to$ box reconstruction
$H_{\mathrm{box}}$. This delivers the actual ommatidial mosaic
and therefore contains the aliased moir\'e of
\S\ref{subsec:moire_isolation}.

\noindent{Clean pipeline:}\,
Identical to the full pipeline except that the comb-replication
step is explicitly nulled ($F_{\mathrm{aliases}}\!=\!0$ of
Eq.~\ref{eq:F_aliases}). This is the same Naka--Rushton-compressed
signal seen through the same box-reconstruction MTF, but without
the spectral replicas that produce the moir\'e.
Because the two pipelines differ only in the comb step,
subtracting their motion energies isolates the EMD-stage
contribution of that step alone,
\begin{equation}
E_{\mathrm{HR}}^{\mathrm{moir\acute{e}}}(t)
\;=\;
E_{\mathrm{HR}}^{\mathrm{full}}(t)
\;-\;
E_{\mathrm{HR}}^{\mathrm{clean}}(t).
\label{eq:E_HR_moire}
\end{equation}
Two properties of Eq.~\ref{eq:E_HR_moire} are essential to
read the empirical results of
\S\ref{subsec:reichardt_control_discussion}. First,
$E_{\mathrm{HR}}^{\mathrm{moir\acute{e}}}$ is the difference of
two mean-squared signed EMD outputs, not the squared
magnitude of their difference, and therefore takes either sign.
The comb step can either constructively augment or destructively
cancel the genuine-motion EMD output before the per-detector
squaring of Eq.~\ref{eq:E_HR}. Second, this is categorically
distinct from the Fourier-stage power ratio
$E_{\mathrm{par,rel}}$ of Eq.~\ref{eq:E_par_rel}, which is
non-negative by construction. A
moir\'e contribution that is a power addition in the
optical stage can still be subtracted from the coherent motion
percept in the EMD stage. The exact algebraic decomposition that
isolates the canceling (cross) and additive (self) components of
$E_{\mathrm{HR}}^{\mathrm{moir\acute{e}}}$ is reported in
\S\ref{app:reichardt_cancellation}.
This motion-detection stage omits
spatially-extended processing of the lobula plate and 
gain control of wide-field cells \citep{borst2014fly}, but
is sufficient to test whether the parasitic spatial frequencies
of \S\ref{sec:optical_model} translate into measurable
contamination of the local optic flow signal that controls
approach and landing in flies
\citep{baird2013universal,vanbreugel2012visual}.
These two readouts carry the two halves of the analysis
in \S\ref{subsec:reichardt_control_discussion}. The energy
difference $E_{\mathrm{HR}}^{\mathrm{moir\acute{e}}}$
(Eq.~\ref{eq:E_HR_moire}) exposes the destructive cancellation of
motion energy, while the signed looming signal
$E_{\mathrm{exp}}$ (Eq.~\ref{eq:E_exp}) exposes the corruption, which is a
close-range suppression and a transient sign reversal of
the radial-expansion cue that gates the landing decision. The
first quantifies how much coherent motion is lost. The second,
whether what remains still signals an approach.
The single-trajectory control above is extended to the full
matched-pair dataset to test whether the optical parasitic energy
predicts the motion-stage degradation across hosts
(Fig.~\ref{fig:epar_vs_emd}). For each of the $n=28$ pairs the optical
chain is evaluated on a fine distance grid and reduced to band-mean
optical scalars (non-linear branch) averaged over
the close-approach band $d\in[1,2.5]\,\si{\meter}$. The isotropic
relative parasitic energy $\langle E_{\mathrm{par,rel}}\rangle$
(Eq.~\ref{eq:E_par_rel}) and, because the looming read-out
$E_{\mathrm{exp}}$ of Eq.~\ref{eq:E_exp} is a radial projection, its
radial and tangential components. With
$I_{\mathrm{par}}=\mathcal{F}^{-1}\{F_{\mathrm{aliases}}^{\mathrm{NL}}\,M\}$
and $I_{\mathrm{sig}}=\mathcal{F}^{-1}\{F_{\mathrm{signal}}\,M\}$ the
parasitic and alias-free signal images restricted to the Nyquist disc
$M$ ($|f|\leq\nu_{\mathrm{eye}}$), and $\hat{\mathbf{r}}$ the outward
radial unit vector from the focus of expansion,
\begin{equation}
E_{\mathrm{par,rad}}
\;=\;
\frac{\sum_{x,y} I_{\mathrm{par}}^{2}\, w_{\mathrm{rad}}[I_{\mathrm{par}}]}
     {\sum_{x,y} I_{\mathrm{sig}}^{2}\, w_{\mathrm{rad}}[I_{\mathrm{sig}}]},
\qquad
w_{\mathrm{rad}}[I] \;=\; \frac{(\nabla I\cdot\hat{\mathbf{r}})^{2}}{|\nabla I|^{2}},
\qquad
w_{\mathrm{tan}} \;=\; 1 - w_{\mathrm{rad}},
\label{eq:E_par_rad}
\end{equation}
and $E_{\mathrm{par,tan}}$ analogously with $w_{\mathrm{tan}}$. The
weight $w_{\mathrm{rad}}$ is the fraction of the local orientation that
is radial, so the two numerators partition the parasitic energy
$E_{\mathrm{par}}$ (Parseval) into the part whose fringes would move
radially, corrupting looming, and the part that would move tangentially.
Each scalar is normalized by the same projection of the alias-free
signal, mirroring the relative looming loss. The Reichardt stage
of \S\ref{subsec:motion} is run on each striped host in its full and
clean variants, and reduced to three band-restricted motion scalars: the
relative energy deficit
$(E_{\mathrm{HR}}^{\mathrm{clean}}-E_{\mathrm{HR}}^{\mathrm{full}})/
E_{\mathrm{HR}}^{\mathrm{clean}}$, the relative looming loss from
$E_{\mathrm{exp}}$, and the band-integrated cross- and self-terms
$\mathrm{CROSS}$ and $\mathrm{SELF}$ of the decomposition in
\S\ref{app:reichardt_cancellation}. \textit{Each optical scalar} is then related to
each motion scalar across the sample by Pearson and Spearman
correlation. The correlations of the radial and tangential
components with the same motion scalar are compared with the test for
dependent correlations of \citet{meng1992comparing}. All geometry (body centre, camera distance, crop) is taken
from the matched-pair calibration of \S\ref{subsec:matched_pair}, so the
optical and motion stages are evaluated on identical crops. The
resulting coupling is reported in \S\ref{subsec:reichardt_control_discussion}.

\subsection{Adaptive gain}
\label{subsec:adaptive_gain}

This paper uses the static receptor model.
Specifically, it applies the Naka-Rushton compression described in Section \ref{subsec:nonlinearity}.
This model captures how insect photoreceptors reach steady-state saturation.
It describes their response to light intensity, which is the key property used
in the main processing step of the Fourier pipeline described in
Section \ref{sec:optical_model}
\citep{laughlin1981retina,laughlin1989role,howard1984temporal}.
In reality, receptors also have a slow membrane time constant
$\tau_{\mathrm{m}} \approx 10\,\si{\milli\second}$, and
an even slower adaptation pool with a characteristic time
of about $100\,\si{\milli\second}$, which shifts the operating point
$s_{50}$ toward the local average luminance over time
\citep{vanhateren1992theoretical,borst2010motion,borst2014fly}.
As a result, the instantaneous gain depends on recent stimulus
history, not just the current luminance. Still, the static model is the best choice for this paper's research question, for two reasons that apply to different parts of the pipeline.
this paper, for two reasons that act in different parts of the simulation
pipeline.
 
First, for the Fourier-stage matched-pair analysis in
Section \ref{subsec:matched_pair}, the choice of model does not matter. The source
images are photographic stills taken from a single fixed distance, so the
receptor is at its steady-state operating point in each frame.
An adaptive cascade would reduce to the static
Naka-Rushton compression already used here. The light and dark history
does not affect the moiré-peak position or amplitude.
The main result, $E_{\mathrm{par,rel}} \approx \SI{8}{\percent}$ across
the $n = 28$ matched-pair dataset, does not change with this
modelling choice.
Second, for the Reichardt motion-energy control in
Section \ref{subsec:reichardt_control_discussion}, the choice is
based on clear reasoning. In this control, each ommatidium receives a real
temporal sequence of luminance values over a 4-second approach
trajectory, with stripes moving across the visual field at
angular speeds between $20$ and $200\,\si{\degree\per\second}$
\citep{caro2019benefits}. In principle, adaptive dynamics could
change the per-frame contrast that goes into Eq.~\ref{eq:reichardt}.
The static Naka-Rushton model is used here because the goal of the
control is to isolate the effect of the comb step at the EMD stage
(Eq.~\ref{eq:freq_replication}) by itself. If the gain varied over time,
it would mix two effects instead of showing just one.
The motion-energy values reported in
Fig.~\ref{fig:reichardt_control_zebra} therefore represent the
static-receptor baseline for the moiré-interference effect. The destructive-interference signature should also remain
even with more realistic receptor dynamics. At 
higher angular speeds, each receptor experiences luminance changes in
the several hundred hertz range, which is much higher than the corner frequency of the adaptation low-pass (about $10\,\si{\hertz}$). The adaptive
gain therefore affects the slow body-silhouette envelope much
more than the fast comb-step contribution.
The main claim in Section \ref{subsec:reichardt_control_discussion} is that the comb
step adds a strictly negative motion-energy contribution to
the optic flow estimate during the close approach. This does not
depend on the adaptive dynamics of the receptor.

\subsection{Unstriped control}
\label{par:unstriped_control}

Applying the same method to a stimulus with a uniform color (here
$I_{\mathrm{stim}}$ equals $c$, plain grey Zebra), $F_{\mathrm{stim}}$ is
limited to a small low-frequency cluster near DC. The acceptance blur
$H_{\rho}$ keeps only the DC component, so $F_{\mathrm{repl}}$ becomes a sparse
set at integer multiples of $f_{s}$. The box reconstruction $H_{\mathrm{box}}$
(Eq.~\ref{eq:hbox}) removes all of these. As a result, $F_{\mathrm{aliases}}$
is nearly zero everywhere except for numerical noise, and $I_{\mathrm{par}}$ is
also nearly zero. The disappearance of $\|I_{\mathrm{par}}\|_{2}$ for uniform
stimuli acts as a sanity check required by the sampling theorem
\citep{snyder1979physics} and is used in the matched-pair test in Section
\ref{sec:discussion}.

\section{Discussion}
\label{sec:discussion}

The simulation process described in Section \ref{sec:simulation} provides, at each approach distance,
two main outputs: the spatial parasitic image
($I_{\mathrm{par}}(x,y)$) and the dimensionless relative
parasitic energy ($E_{\mathrm{par,rel}}(d)$, Eq.~\ref{eq:E_par_rel}). The following discussion explains
these outputs in four steps: (i) the matched-pair test, which
identifies the stripe pattern as the main variable
(see Section \ref{subsec:matched_pair}). (ii) the way parasitic energy changes with distance
across a set of photographs, which shows a peak
at $d\!\in\![1,2.5] \,\si{\meter}$, overlapping the documented
terminal-hesitation window, where biting flies approach but
do not settle on striped hosts (see Section \ref{subsec:matched_pair});
(iii) how robust the mechanism is to two biologically
important changes: regional acute-zone specialisation and
chromatic decomposition, followed by the established
polarisation channel (see Sections \ref{subsec:acute_zone_disc} to \ref{subsec:polarisation}); and
(iv) how this optical finding is used in a basic
Hassenstein--Reichardt motion-detector model (\S\ref{subsec:reichardt_control_discussion}).

\subsection{Parasitic interference}
\label{subsec:fourier_dissection}

Figure~\ref{fig:fourier_moire} of \S\ref{subsec:pipeline_output}
shows that the spatial parasitic image
$\mathcal{F}^{-1}\{F_{\mathrm{aliases}}^{\mathrm{NL}}\}$ reproduces
the geometry of the underlying retinal mosaic remarkably closely.
In every column the high-amplitude red and blue features of the
parasitic image lie along the black-to-white boundaries
of the striped stimulus rather than on the smooth interiors of
either the dark or the bright stripes. This is a consequence of
how sampling-induced moir\'e is generated. A stripe edge contains
a broad range of spatial frequencies (the FFT of a step is
$\propto 1/u$), so the edge populates the band
$\nu_{\mathrm{eye}} \!<\! |f| \!<\! 2\,\nu_{\mathrm{eye}}$ much
more strongly than the smooth stripe interior does. Replicas of
that edge energy, shifted by $\pm f_{s}$
(Eq.~\ref{eq:freq_replication}), fold back into the eye's Nyquist
disc and appear in $F_{\mathrm{aliases}}$ as the dominant
contribution to the parasitic content. In the spatial domain this
manifests as the alternating red and blue ripple localised around
every black-to-white edge of the zebra coat visible in row~6 of
Fig.~\ref{fig:fourier_moire}. The colour scale of row~6 encodes the sign of $\mathcal{F}^{-1}\{F_{\mathrm{aliases}}^{\mathrm{NL}}\}$ on a diverging red and blue scale (red positive, blue negative) and is
normalised independently in each column of $|I_{\mathrm{par}}|$, so colour saturation is
comparable across columns only in terms of relative
within-column spatial structure. The amount of parasitic
energy is instead given in the figure itself: Rows~5 and~6 print $E_{\mathrm{par,rel}}$ (Eq.~\ref{eq:E_par_rel}), the energy of the displayed parasitic spectrum inside the dashed Nyquist disc relative to the signal energy in the same disc. The colour map of row~5 shows only its distribution, and the opacity of row~6 scales with the value ($\eta = 0.25 + 0.75\sqrt{E_{\mathrm{par,rel}}/E_{\max}}$). The full distance dependence is shown in Fig.~\ref{fig:fourier_moire_set}. The plausibility of this reading is
reinforced by the distance dependence: at $d\!=\!\SI{0.5}{\meter}$
the FoV captures essentially one large stripe transition and the
parasitic image is correspondingly sparse; at
$d\!=\!\SI{1.5}{\meter}$, where many stripes are present and edge
density is highest, the parasitic image is densest. At
$d\!=\!\SI{5}{\meter}$, where the stripes are angularly small but
still resolvable, the edge density is lower again. The proximate predictor of moir\'e energy is therefore not
edge density on the body nor fractional stripe area, but the
angular edge density on the retina at the viewing distance $d$.
The dimensionless metric $E_{\mathrm{par,rel}}(d)$
(Eq.~\ref{eq:E_par_rel}) plotted in Fig.~\ref{fig:fourier_moire_set}
quantifies this geometric resonance. The curve has a clear peak at \SI{1.6}{\meter} in
the close-approach range $d\!\in\![1,2.5]\,\si{\meter}$ and falls
off rapidly outside it. The peak location is set by an interplay
between two competing effects. At small $d$ the body subtends a
large angular size and few stripes fit within the FoV. 
At that distance many of the stripe transitions sit outside the analysed patch,
so their contribution to the alias energy is missing. At large $d$
the stripe pattern shrinks angularly until its dominant frequencies
exceed the eye's Nyquist limit. The acceptance-angle MTF
(Eq.~\ref{eq:airy_otf}) attenuates them before they can be sampled,
leaving little to alias. Between these limits there is a window in
which several full stripe periods fit in the FoV and the dominant
stripe frequency lies within or just above the eye Nyquist limit,
so that $f_{s}$-shifted replicas fold back into the eye's
pass-band. For the canonical \textit{Aedes aegypti} parameter set
($\Delta\varphi\!=\!\SI{2.5}{\degree}$,
$\sigma_{\rho}\!=\!\SI{0.85}{\degree}$,
$f_{s}\!=\!\SI{0.40}{cyc.deg^{-1}}$) and a body height
$\sim\!\SI{1.4}{\meter}$, this window centres on
$d\!\approx\!\SI{1.5}{\meter}$ which is well inside the close-approach
range relevant for terminal landing
\citep{waage1981zebra,brady1988landing,gibson1992tsetse,caro2019benefits}.
The linear branch
$E_{\mathrm{par,rel}}^{\mathrm{lin}}$
(Eq.~\ref{eq:F_aliases}) and the Naka--Rushton branch
$E_{\mathrm{par,rel}}^{\mathrm{NL}}$ (Eq.~\ref{eq:F_aliases_NL})
agree to within a few percent at the peak. The dominant
contribution to the parasitic content is therefore geometric
moir\'e, sampling-induced aliasing, respectively, rather than harmonic
distortion injected by the photoreceptor non-linearity. The
Naka--Rushton compression adds a measurable but subordinate amount
of in-band intermodulation
\citep{laughlin1981retina,landnilsson2002eyes}.

\subsection{Stripes vs.\ stripe-removed control}
\label{subsec:matched_pair}

\begin{figure}[H]
\centering
\includegraphics[height=0.9\textheight,keepaspectratio]{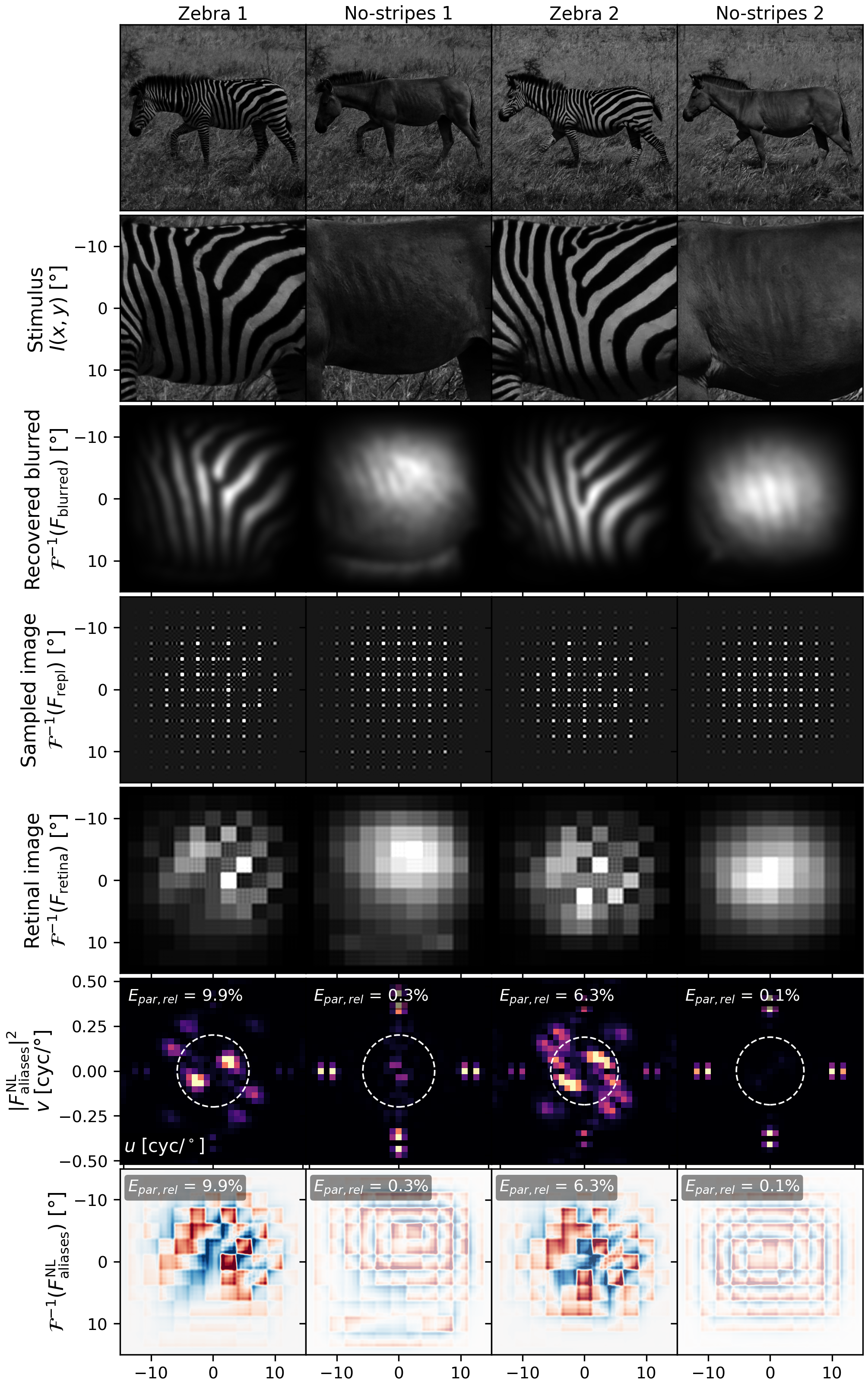}
\caption{Full pipeline output on two matched pairs (\texttt{DSC\_0084} and
\texttt{DSC\_0044}; zebra in columns~1 and~3, stripe-removed control in
columns~2 and~4), each at its moir\'e-peak distance. The seven
rows show the input photograph followed by the six stages of the cascade developed in
\S\ref{subsec:sampling}--\S\ref{subsec:moire_isolation}, ending at
the non-linear parasitic image $\mathcal{F}^{-1}\{F_{\mathrm{aliases}}^{\mathrm{NL}}\}$.}
\label{fig:zebra_vs_horse_a}
\end{figure}

\begin{figure}[H]
\centering
\includegraphics[width=\textwidth]{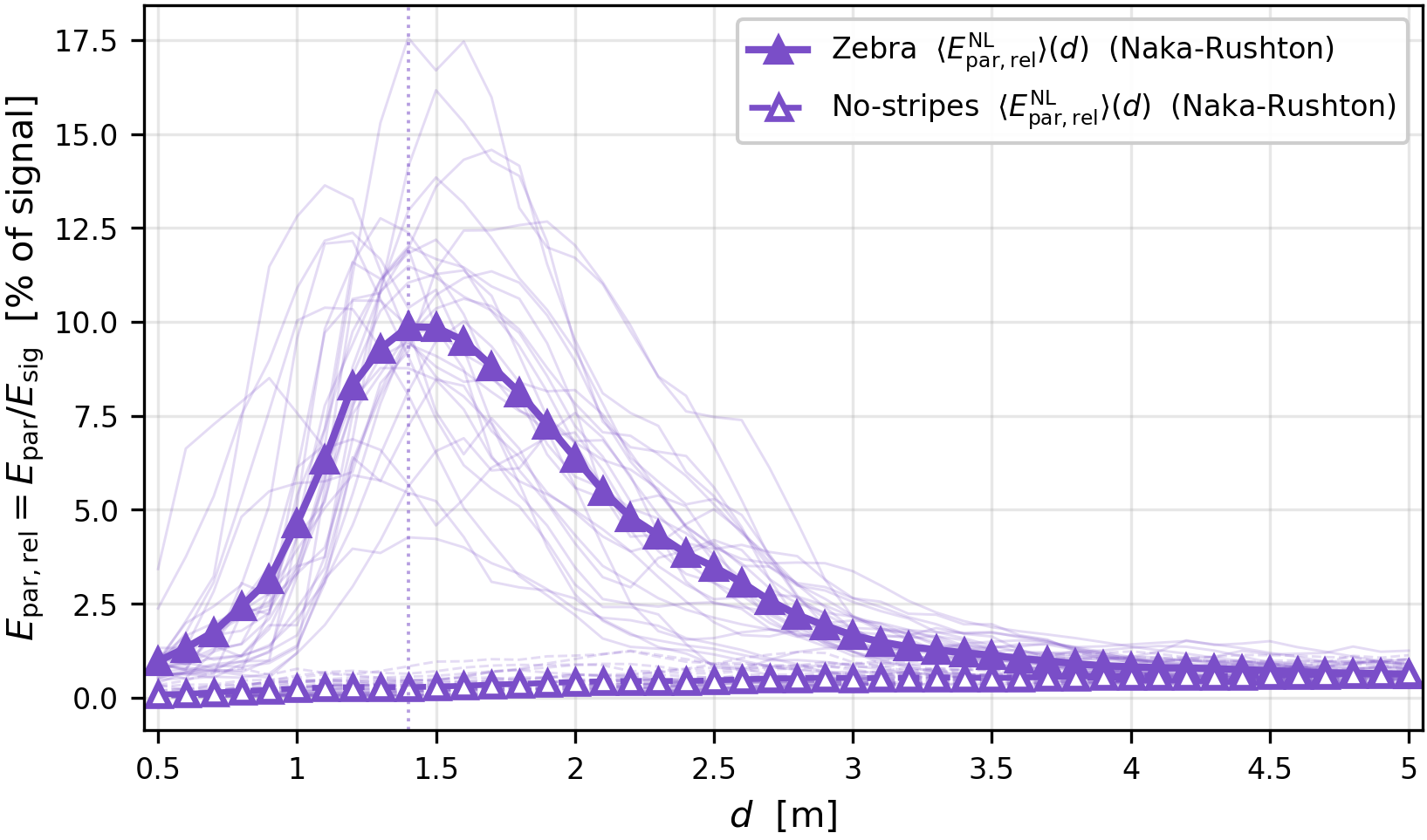}
\caption{Integrated relative parasitic energy $E_{\mathrm{par,rel}}(d)$ (Eq.~\ref{eq:E_par_rel}) of matched pair (zebra, stripe-removed control) as a function of approach distance. Thin background traces: per-pair curves. Solid markers and
solid means: stripes (zebra) condition. Open markers and open means:
stripe-removed control. The vertical dashed guideline marks the cross-pair mean moir\'e-peak
distance $d\!\approx\!\SI{1.4}{\meter}$.}
\label{fig:zebra_vs_horse}
\end{figure}

The most direct empirical test of the moir\'e hypothesis isolates
the role of the stripe pattern itself by removing it from the host
while keeping everything else the same, including silhouette, body size, framing,
lighting, and background. So, all other aspects are identical. 
In accordance with the unstriped-control argument of §3.9, such a test has been constructed. Each striped photograph was digitally re-rendered with the stripes painted out to a uniform grey of matched local luminance. This process produced a stripe-removed twin, referred to as the horse condition.
The same Fourier-moir'e pipeline (see Section \ref{subsec:pipeline}) is run independently on both images in each pair. The picker center and body height are set once on the striped image and then copied exactly to the no-stripe image. The script keeps pixel resolution and focal length the same, so no scaling is needed.
Figure~\ref{fig:zebra_vs_horse_a} shows the result of this
matched-pair test. Stage-by-stage construction of two representative
pairs (\texttt{DSC\_0084} and \texttt{DSC\_0044}, each with its stripe-removed
twin), shown at each pair's moir\'e-peak distance
($d_{\mathrm{peak}}\approx\SI{1.5}{\meter}$, $\mathrm{FoV}\approx30^{\circ}$). The seven rows trace the same
cascade as Fig.~\ref{fig:fourier_moire} side by side for the
two conditions: input photograph, stimulus crop, recovered blurred
image, sampled image, retinal mosaic, parasitic power spectrum
$|F_{\mathrm{aliases}}^{\mathrm{NL}}|^{2}$ (dashed circle marks
$\nu_{\mathrm{eye}}=\SI{0.20}{cyc.deg^{-1}}$), and parasitic
image $\mathcal{F}^{-1}\{F_{\mathrm{aliases}}^{\mathrm{NL}}\}$.
Rows~6 and~7 print the column's $E_{\mathrm{par,rel}}$. The
parasitic spectrum (row~6) is shown as linear power normalised per
panel to its 99.5th percentile, and the opacity of the parasitic image
(row~7) scales with $E_{\mathrm{par,rel}}$ as in
Fig.~\ref{fig:fourier_moire}.
The striped zebra (columns~1 and~3) produces a structured parasitic
spectrum $|F_{\mathrm{aliases}}^{\mathrm{NL}}|^{2}$ that
concentrates inside the Nyquist eye disk, and its inverse Fourier transform image shows the body-aligned moir\'e fringes
discussed in \S\ref{subsec:fourier_dissection}. The stripe-removed
control (columns~2 and~4) is run through the identical pipeline. Its
parasitic spectrum and inverse transform do not show any body-aligned structure, only the silhouette-edge
ringing that any bounded image must generate when sampled on a discrete grid. For the two pairs shown,
$E_{\mathrm{par,rel}}$ at the peak distance is $9.9\,\%$ and $6.3\,\%$
for the striped hosts against $0.3\,\%$ and $0.1\,\%$ for their twins. The qualitative conclusion is that the parasitic
content of the zebra columns is genuinely caused by the stripe
pattern, not by any other property of the scene
(silhouette, mean luminance, framing, lighting), and the
the visual difference between the two conditions makes this
immediate.
Figure~\ref{fig:zebra_vs_horse} quantifies the same statement across $n\!=\!28$ matched
zebra/no-stripes pairs. Each pair consists of a striped zebra photograph (Nikon D50, $f_{\mathrm{lens}}\!=\!300\,\si{\milli\meter}$, $f/5.6$, ISO\,400) and a digitally re-rendered twin in which the stripes
are painted out to a uniform grey of matched local luminance. The Fourier-moir\'e pipeline of
\S\ref{subsec:pipeline} is run independently on every member of
every pair, with diffraction-limited Airy MTF
(Eq.~\ref{eq:airy_mtf}, $D\!=\!\SI{20}{\micro\meter}$,
$\lambda\!=\!\SI{500}{\nano\meter}$) and Naka--Rushton receptor
non-linearity (Eq.~\ref{eq:naka_rushton}, mix
$\alpha\!=\!0.5$). The cross-pair mean $\langle E_{\mathrm{par,rel}}^{\mathrm{NL}}(d)\rangle$ (solid
markers, $\Delta d=0.1 m$) of the striped condition show the characteristic broad
peak centred on $d\!\approx\!\SI{1.4}{\meter}$ with mean amplitude
of order $10$ $\%$ of the in-band signal. The
corresponding curves for the no-stripes condition (open markers)
remain at the floor below $\sim\!1\,\%$ across the full sweep.
The peak ratio between the two conditions, taken on the
cross-pair mean curves, is approximately
$10$--$20\times$ inside the moir\'e-relevant band
$d\!\in\![1,2.5]\,\si{\meter}$. For an individual pair at its
own peak distance the ratio is larger (cf.\ the two pairs of
Fig.~\ref{fig:zebra_vs_horse_a}), because the stripe-removed twin sits
at its floor there. Because $E_{\mathrm{par,rel}}$ is by construction gauge-invariant, the per-image gain factors cancel exactly between
numerator and denominator (\S\ref{subsec:e_par}), the
stripes-vs-no-stripes gap of Fig.~\ref{fig:zebra_vs_horse}
cannot be attributed to per-image differences in exposure,
brightness or framing. It is a gauge-invariant statement about the
optics of the eye-plus-coat system. The thin background traces in
the same panel show the per-pair curves. Every single pair in the
dataset reproduces the same gap, with peak distances scattered
slightly around $\SI{1.6}{\meter}$ in a manner consistent with
different intrinsic angular stripe periods (finer-striped
individuals shift the peak toward smaller $d$, in agreement with
Eq.~\ref{eq:distance_scaling_fourier} and with the
edge-density argument of \S\ref{subsec:fourier_dissection}).

A further small source of horizontal scatter on the same peak is
the calibration uncertainty in the assumed body height
$h_{\mathrm{body}}\!=\!\SI{1.4}{\meter}$. An inter-individual range
of $\SI{1.35}{\meter}\!\leq\!h_{\mathrm{body}}\!\leq\!\SI{1.45}{\meter}$
propagates linearly through the angular projection to a
$\pm\SI{6}{\centi\meter}$ shift in $d_{\mathrm{peak}}$ at
$d_{\mathrm{peak}}\!\approx\!\SI{1.6}{\meter}$, well below the
natural width of the peak. Further information is given in
\S\ref{app:body_height_error}. 
The observation that $E_{\mathrm{par,rel}}$ collapses by more than an order of magnitude when the stripes alone are removed from a real photograph is direct empirical evidence that the moir\'e content predicted by the optical
model (\S\ref{subsec:moire_isolation}) is genuinely caused by the
interaction of the stripe pattern with the ommatidial sampling
lattice. Quantitatively, $\sim\!10\,\%$ of the energy the brain receives
inside its own Nyquist disc, at the distance at which a tabanid or
culicid would commit to landing, is genuinely parasitic moir\'e
content with no physical counterpart on the coat. This is a
non-trivial fraction of the available signal and is exactly the band in which the
optic-flow estimator that controls the final landing manoeuvre
operates
\citep{baird2013universal,vanbreugel2012visual,borst2014fly}.
The moir\'e-relevant distance band identified by
Fig.~\ref{fig:zebra_vs_horse}, $d\!\in\![1,2.5]\,\si{\meter}$,
overlaps exactly with the documented terminal-hesitation phase
during which biting fly approach a striped host but fail to
settle
\citep{waage1981zebra,brady1988landing,gibson1992tsetse,caro2019benefits}.
Beyond $\sim\!\SI{3}{\meter}$ the stripes-vs-no-stripes gap
collapses because the body shrinks below the eye's sampling
resolution and the residual alias content is dominated by the
silhouette edge itself rather than by any internal pattern. At
these larger distances host-finding is in any case dominated by
olfactory cues
\citep{vanbreugel2015mosquitoes,coutinho2022human,mcmeniman2014multimodal}.

\subsection{Regional variation}
\label{subsec:acute_zone_disc}

\begin{figure}
\centering
\includegraphics[width=\textwidth]{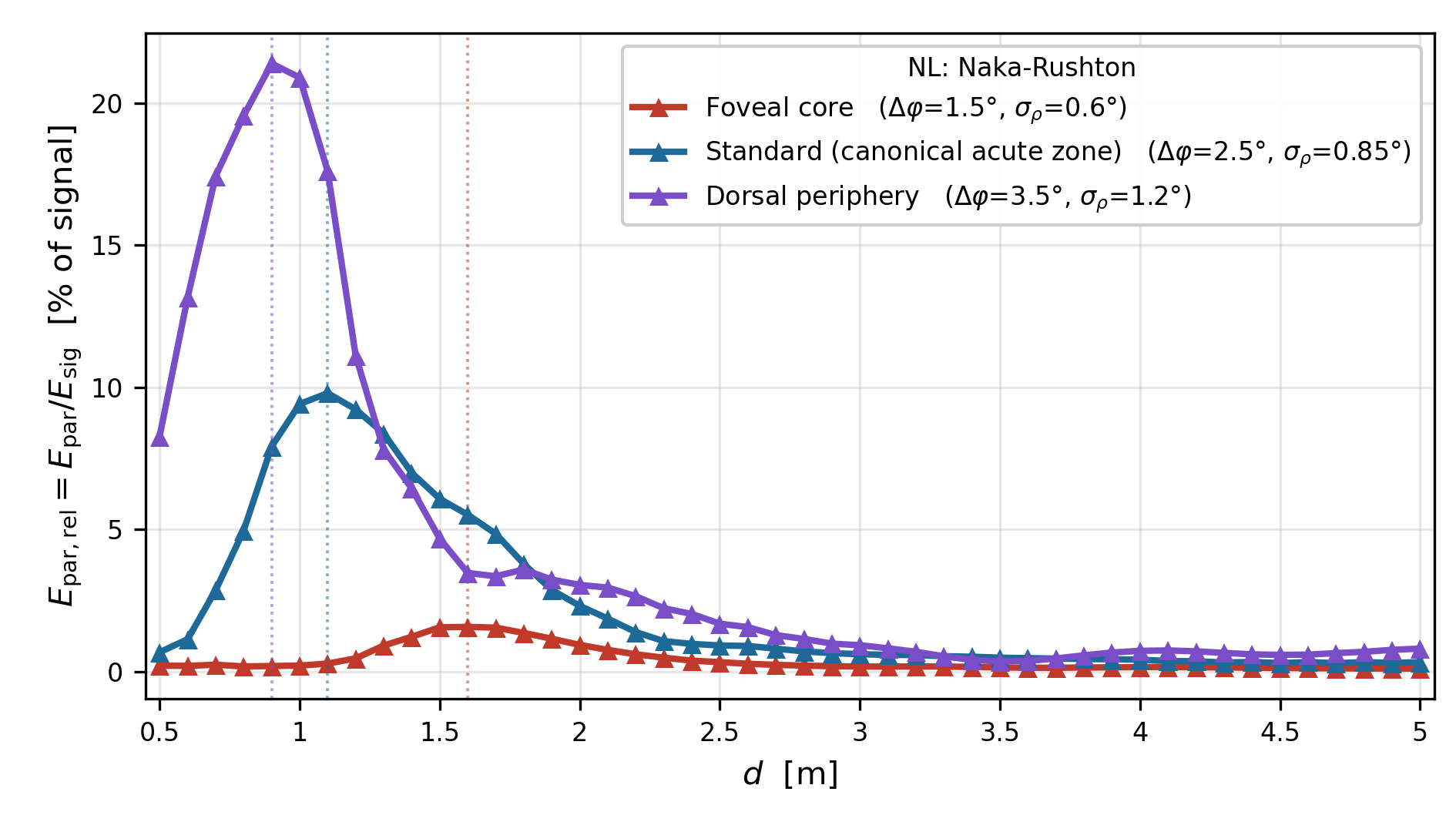}
\caption{Regional sweep across the three canonical eye
configurations of \S\ref{subsec:acute_zone_sim} on photograph
\texttt{DSC\_0158}. Foveal core (red), Standard / canonical acute zone
(blue), Dorsal periphery (purple).}
\label{fig:acute_chromatic_a}
\end{figure}

The simulations of \S\ref{subsec:pipeline_output} use the
Standard (canonical acute zone) parameter set
$(\Delta\varphi,\sigma_{\rho})\!=\!(\SI{2.5}{\degree},\SI{0.85}{\degree})$,
the diurnal-culicid value reported in the comparative-vision
literature
\citep{land1997mosquito,land1999fundamental,landnilsson2002eyes}.
The compound eye is, however, regionally specialised: the
inter-ommatidial angle $\Delta\varphi$ varies systematically
across the visual field, with the antero-ventral region --- the
part that looks down and forward at a host during landing
approach --- having higher resolution than the dorsal periphery
\citep{land1997mosquito,hawkes2022vision}. To probe how robust
the moir\'e mechanism is to this regional specialisation, the
piecewise-stationary construction of \S\ref{subsec:acute_zone_sim}
is run independently in each of the three canonical
configurations: Foveal core
($\Delta\varphi\!=\!\SI{1.5}{\degree}$,
$\sigma_{\rho}\!=\!\SI{0.6}{\degree}$), Standard
($\SI{2.5}{\degree}$, $\SI{0.85}{\degree}$), and
Dorsal periphery
($\SI{3.5}{\degree}$, $\SI{1.2}{\degree}$).
Figure~\ref{fig:acute_chromatic_a} shows
$E_{\mathrm{par,rel}}^{\mathrm{NL}}(d)$ for the three regions on
a representative zebra photograph (\texttt{DSC\_0085.JPG}). All three regions
exhibit a clearly defined moir\'e peak with a comparable shape,
but the peak distance and peak magnitude both shift
systematically with $\Delta\varphi$. The dorsal periphery
($\Delta\varphi\!=\!\SI{3.5}{\degree}$, purple) reaches its
maximum at the smallest distance
($d\!\approx\!\SI{0.9}{\meter}$, purple vertical dashed line)
with the largest peak amplitude ($\sim\!\SI{31}{\percent}$ of
signal). The Standard configuration
($\Delta\varphi\!=\!\SI{2.5}{\degree}$, blue) peaks at
intermediate distance ($d\!\approx\!\SI{1.1}{\meter}$) with
intermediate amplitude ($\sim\!\SI{14}{\percent}$). The
Foveal core ($\Delta\varphi\!=\!\SI{1.5}{\degree}$, red) peaks at
the largest distance ($d\!\approx\!\SI{2.2}{\meter}$, red
vertical dashed line) with the smallest peak amplitude
($\sim\!\SI{2.5}{\percent}$). The ordering follows directly from
the moir\'e-resonance condition between stripe period and
sampling period. A finer sampling lattice has its Nyquist limit
at a higher angular spatial frequency. Therefore the alias-generating
band is reached only when the stripes themselves shrink
angularly and the
resonance lobe is correspondingly weaker because fewer stripe
edges sit close to the Nyquist limit at that distance.
Two points stand out. First, the Foveal core still produces a
clearly identifiable moir\'e peak: the mechanism does not
disappear when the inter-ommatidial angle is reduced by
$\SI{40}{\percent}$ relative to the Standard parameter set.
Second, the Foveal-core peak sits at $d\!\approx\!\SI{1.5}{\meter}$,
which is deeper into the close-approach band than the
Standard or Dorsal-periphery peaks. 
Anatomically and behaviourally, the antero-ventral acute zone, where the Foveal core is located, is the highest-resolution sub-region. This area of the eye is specialized for host fixation and the terminal landing manoeuvre\citep{land1997mosquito,land1999fundamental,hawkes2022vision}.
It is the region whose output is read out by the descending
neurons that drive the legs-out, hover, settle sequence
\citep{borst2014fly,vanbreugel2012visual,baird2013universal}.
Placing the moir\'e peak of the most-resolved sub-region at the
exact distance at which a tabanid or culicid has committed to
landing ($\sim\!1$--$2\,\si{\meter}$) is therefore not a
weakening of the mechanism but a strengthening of its biological
relevance. The parasitic content is injected into the part of
the visual pathway that drives the behaviour the field studies
report disrupted
\citep{waage1981zebra,brady1988landing,caro2019benefits}.
The Dorsal-periphery peak, conversely, sits at
$d\!\approx\!\SI{1.0}{\meter}$. The dorsal eye is not the part
of the visual field used for landing. It serves orientation and
obstacle avoidance
\citep{land1997mosquito,landnilsson2002eyes}. A larger moir\'e
peak in this region is therefore less behaviourally costly,
although a fly performing close-range orientation manoeuvres
around a host body may still receive phantom-motion input from
the dorsal mosaic. The moir\'e mechanism is robust to regional
variation in $\Delta\varphi$, and is strongest where it matters
most.

\subsection{Chromatic channels}
\label{subsec:chromatic_disc}

\begin{figure}
\centering
\includegraphics[width=\textwidth]{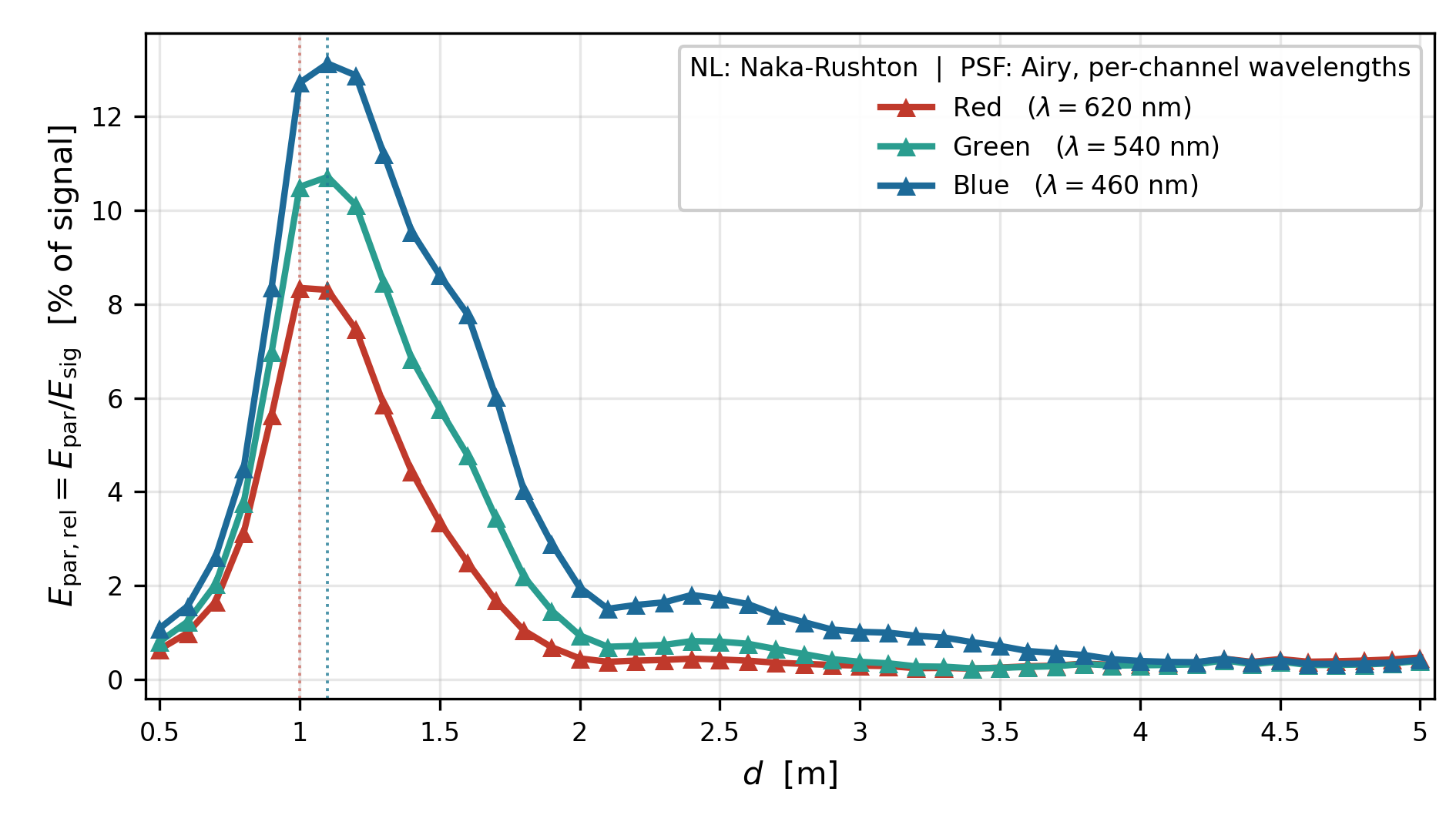}
\caption{Chromatic-channel sweep across the three camera RGB
channels of photograph \texttt{DSC\_0084} with per-channel Airy MTF.}
\label{fig:acute_chromatic_b}
\end{figure}

The simulation runs on a single-channel grayscale stimulus,
which is appropriate for a first-principles demonstration of the
geometric moir\'e mechanism, but ignores the chromatic structure
of the input. \textit{Aedes aegypti} expresses at least three
opsins with peak spectral sensitivities in the UV
($\sim\!\SI{350}{\nano\meter}$), blue
($\sim\!\SI{425}{\nano\meter}$), and green
($\sim\!\SI{525}{\nano\meter}$) bands
\citep{hu2014spectral,zhan2021distinct}. Behavioural work confirms
host attraction to multiple wavelengths
\citep{vanbreugel2015mosquitoes}. A Fourier model that accurately preserves color information
calculates the spectral irradiance of the stimulus for each opsin sensitivity.\citep{kelber2003animal}. The presented study
supports a per-channel sweep (RGB).
Figure~\ref{fig:acute_chromatic_b}\ shows
$E_{\mathrm{par,rel}}^{\mathrm{NL}}(d)$ for the three camera RGB
channels of a representative zebra photograph (DSC\_0084), each run through the pipeline with a diffraction-limited Airy MTF tuned
to the wavelength of the channel
($\lambda_{R}\!=\!\SI{620}{\nano\meter}$,
$\lambda_{G}\!=\!\SI{540}{\nano\meter}$,
$\lambda_{B}\!=\!\SI{460}{\nano\meter}$). The three curves share
the same qualitative shape (a single clear peak in the
close-approach band) but differ in peak amplitude. The blue
channel ($\lambda_{B}\!=\!\SI{460}{\nano\meter}$) produces the
largest moir\'e response, peaking near
$d\!\approx\!\SI{1.1}{\meter}$ with
$E_{\mathrm{par,rel}}^{\mathrm{NL}}\!\approx\!\SI{13.1}{\percent}$.
The green channel ($\lambda_{G}\!=\!\SI{540}{\nano\meter}$) peaks
at the same position ($d\!\approx\!\SI{1.1}{\meter}$) with intermediate
amplitude ($\sim\!\SI{10.8}{\percent}$), and the red channel 
($\lambda_{R}\!=\!\SI{620}{\nano\meter}$) is the weakest of the three ($\sim\!\SI{8.1}{\percent}$) and peaks slightly earlier at $d\!\approx\!\SI{1.0}{\meter}$.
The peak \emph{distance} is essentially channel-invariant. The
peak magnitude scales inversely to the wavelength of the
Airy MTF. The Airy diffraction cutoff (Eq.~\ref{eq:airy_cutoff}) is larger at shorter wavelengths, so the blue channel admits a larger fraction of the stripe-fundamental energy into the alias-generating band
before the diffraction MTF attenuates it. In addition, zebra coats are
 empirically known to exhibit a higher visible contrast in the blue
and UV bands than in the red
\citep{caro2014function,horvath2010whiteness}. Spectroradiometric
characterization of plains zebra \textit{Equus quagga} coat
reflectance shows that white stripes return more UV/blue light
relative to dark stripes than red light, so the per-channel
modulation amplitude entering the FFT is higher at the blue end of
the spectrum.
The photoreceptors most exposed to the moir\'e effect during close approach are the short-wavelength opsins. In \textit{Aedes aegypti} these are the R7 UV-sensitive
photoreceptors and the blue-sensitive R8 photoreceptors, both of
which are known to play a dominant role in host fixation and dark
target detection
\citep{hu2014spectral,zhan2021distinct,vanbreugel2015mosquitoes}.
Therefore, the moir\'e mechanism preferentially corrupts the
chromatic channel on which the diptera itself relies most heavily for
the fixation of the close-range host. The peak distance is unchanged
across the channels because the moir\'e-resonance condition is
geometric rather than chromatic. The chromatic
channels modulate the peak amplitude through a combination of
diffraction MTF and per-channel coat contrast, but not the
distance at which the peak occurs.

\subsection{Polarisation channels}
\label{subsec:polarisation}

The ommatidia of \textit{Aedes aegypti} have two sets of microvilli arranged orthogonal to each other.
These microvilli are found in each rhabdomere and give the photoreceptors sensitivity to polarised light.
This arrangement allows the photoreceptors to detect polarisation \citep{wachowiak1995polarization}.
Polarisation cues are important for the behaviour of host-seeking biting flies.
For example, female tabanids are strongly attracted to linearly polarised light.
They prefer light with a high degree of polarisation,
regardless of its angle of polarisation,
and use this cue
to choose dark hosts that strongly polarise light and to tell them apart from
the dark patches of vegetation in the background, which polarise light weakly
\citep{horvath2010whiteness,egri2012polarotactic,horvath2024polarotaxis}.
At a fixed angle of reflection $\theta_{\mathrm{r}}$ (measured from the surface normal), the darker
the surface, the more polarised the reflected light will be (Umow's rule)
\citep{horvath2024polarotaxis}. However, white
stripes are not just a depolariser compared to a polarising dark
background. Instead, black stripes strongly polarise light and white
stripes do so weakly, so the important factor is the \emph{$\mathcal{P}$-contrast}
between neighbouring black and white stripes. Furthermore, this contrast is
highly dependent on geometry. The polarizing effect is close to zero for near-normal
($\theta_{\mathrm{r}}\!\approx\!0^{\circ}$) and near-grazing
($\theta_{\mathrm{r}}\!\approx\!90^{\circ}$) reflection, and it is highest near the
Brewster angle, where $\mathcal{P}_{\mathrm{black}}$ can reach $75$--$85\,\%$
while $\mathcal{P}_{\mathrm{white}}$ remains below about $25\,\%$
\citep{horvath2024polarotaxis}. If an observer stays in one place and looks in a single
direction, they only sample one point on this
$\mathcal{P}(\theta_{\mathrm{r}})$ curve. In contrast, a host-seeking fly that circles the animal
during its final aerial inspection moves through the entire
$\theta_{\mathrm{r}}$-range, including the high-contrast Brewster band, as it
moves around. This dynamic sampling is the reason why
\citet{horvath2024polarotaxis} argues that reports of only modest,
similar polarisation signals from black and white stripes
\citep{britten2016zebras,caro2023why} are due to measurements taken only
in the low-contrast, near-normal regime, rather than being evidence against a
polarotactic role. In a Fourier-optics framework, polarisation can be incorporated as
three parallel chains operating on the Stokes components $\mathcal{I}$ (total
intensity), $\mathcal{Q}$ (linear horizontal/vertical), and $\mathcal{U}$
(linear $\pm\!45^{\circ}$). The expected impact on the moir\'e mechanism specifically is modest,
but the relationship between the two mechanisms is more entangled than
a clean long-range/close-range split would suggest. Moir\'e is
fundamentally a high-contrast spatial-pattern phenomenon, dominated by
the luminance ($\mathcal{I}$) channel. Polarotaxis has classically been
emphasised as a long-range cue
\citep{horvath2010whiteness,egri2012polarotactic}, governing
whether the fly approaches at all, at distances beyond
$\sim\!\SI{10}{\meter}$ where individual stripes are below the
sampling resolution of the eye. 
However, the dynamic-sampling argument of \citet{horvath2024polarotaxis} indicates that the stripe $\mathcal{P}$-contrast remains behaviourally available during the close-range inspection ($d\!\in\![1, 2.5]\,\si{\meter}$) in which the luminance moir\'e of \S\ref{subsec:matched_pair} operates, so the two mechanisms most plausibly act in concert during terminal approach
rather than in strictly separated phases. Adding polarisation to the moir\'e
pipeline is expected to refine the magnitude predictions and to
sharpen, rather than overturn, the present conclusion.

\subsection{Landing failure on striped hosts}
\label{subsec:synthesis}
\label{subsec:reichardt_control_discussion}

\begin{figure}
\centering
\includegraphics[width=\textwidth]{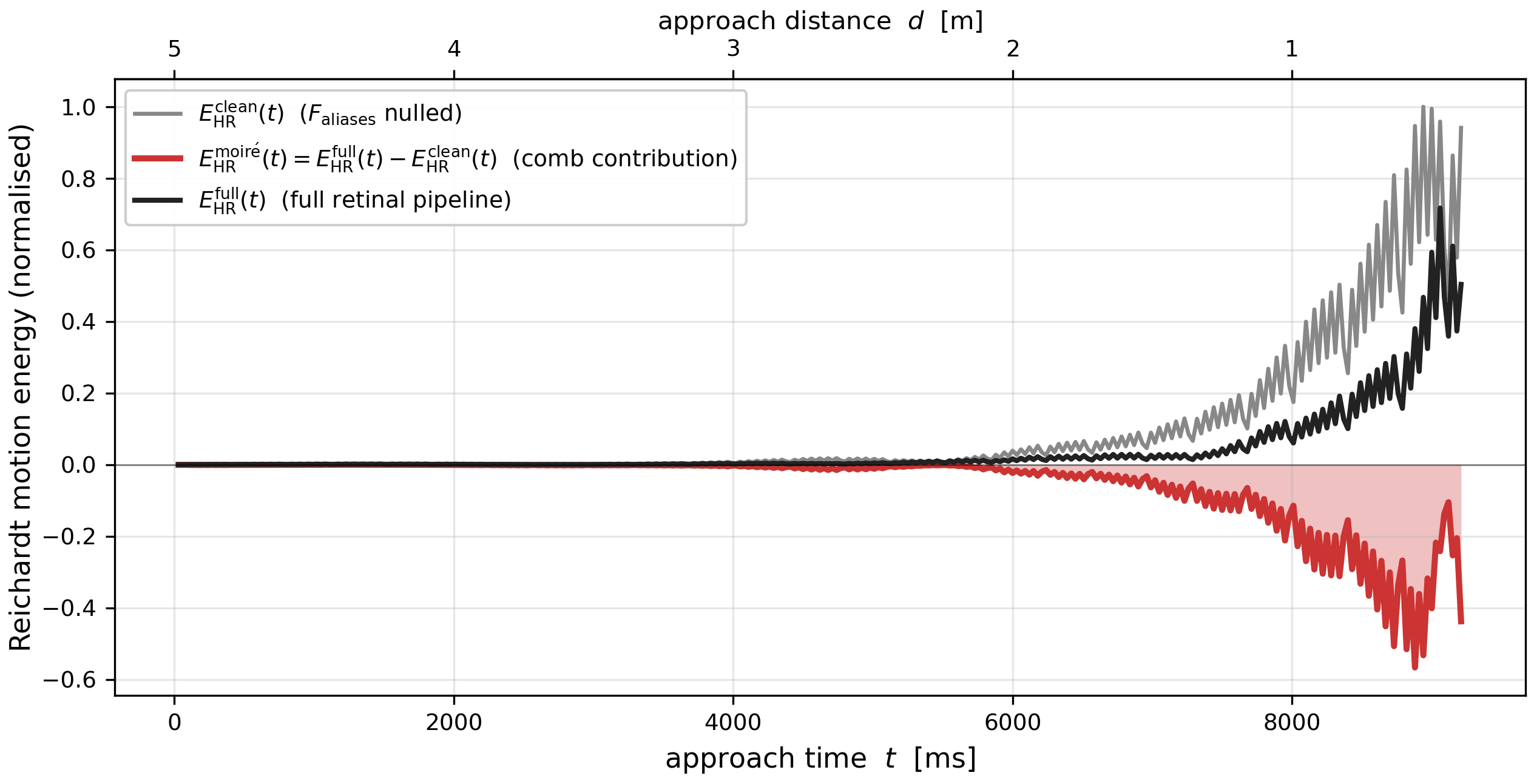}
\caption{Reichardt motion-energy control on a striped host
during a simulated close approach (\texttt{DSC\_0084},
\textit{Equus quagga}; $d_{0}\!=\!\SI{2.5}{\meter}$,
$v_{\mathrm{app}}\!=\!\SI{0.5}{\meter\per\second}$, frame
interval $\Delta t\!=\!\SI{30}{\milli\second}$). Grey:
alias-free EMD output $E_{\mathrm{HR}}^{\mathrm{clean}}(t)$.
Black: full ommatidial EMD output
$E_{\mathrm{HR}}^{\mathrm{full}}(t)$. Red: their signed
difference
$E_{\mathrm{HR}}^{\mathrm{moir\acute{e}}}(t)$
(Eq.~\ref{eq:E_HR_moire}). The moir\'e contribution is negative inside the close-approach band. Aliased drift cancels the genuine motion percept at the EMD output rather than augmenting it. The cancellation is decomposed exactly in
\S\ref{app:reichardt_cancellation}.}
\label{fig:reichardt_control_zebra}
\end{figure}

Taken together, the matched-pair test of
\S\ref{subsec:matched_pair}, the regional acute-zone analysis of
\S\ref{subsec:acute_zone_disc} and the per-channel chromatic
analysis of \S\ref{subsec:chromatic_disc} converge on a single
robust statement: the interaction of a striped zebra coat with the
ommatidial sampling lattice of a high-resolution diurnal culicid compound eye necessarily
generates a parasitic moir\'e signal whose magnitude is of order
$\sim\!10\,\%$ of the in-band brain-accessible signal in the
close-approach band $d\!\in\![1, 2.5]\,\si{\meter}$. This parasitic signal is
stable across receptor non-linearity (linear vs.\ Naka--Rushton),
across eye region (acute zone, standard, periphery), across
chromatic channel (R, G, B), and across a set of $n\!=\!28$ matched photographic pairs. In every case the moir\'e signal disappears
when the stripes alone are removed from the host while every other
property of the scene is preserved, which establishes the stripe
pattern as the proximate cause.
The optical-stage results above establish that a measurable
parasitic spectrum survives into the brain-accessible passband.
They say nothing yet about whether the biting fly's downstream motion
pathway is degraded by it. To close that gap we propagate one
representative approach trajectory (DSC\_0084, ~\ref{fig:zebra_vs_horse}) one
of the $n\!=\!28$ matched pairs of \S\ref{subsec:matched_pair}, 
through the Hassenstein--Reichardt elementary motion detector of
\S\ref{subsec:motion} and report the resulting motion-energy time
series in Fig.~\ref{fig:reichardt_control_zebra}.

\begin{figure}[H]
\centering
\includegraphics[width=\textwidth]{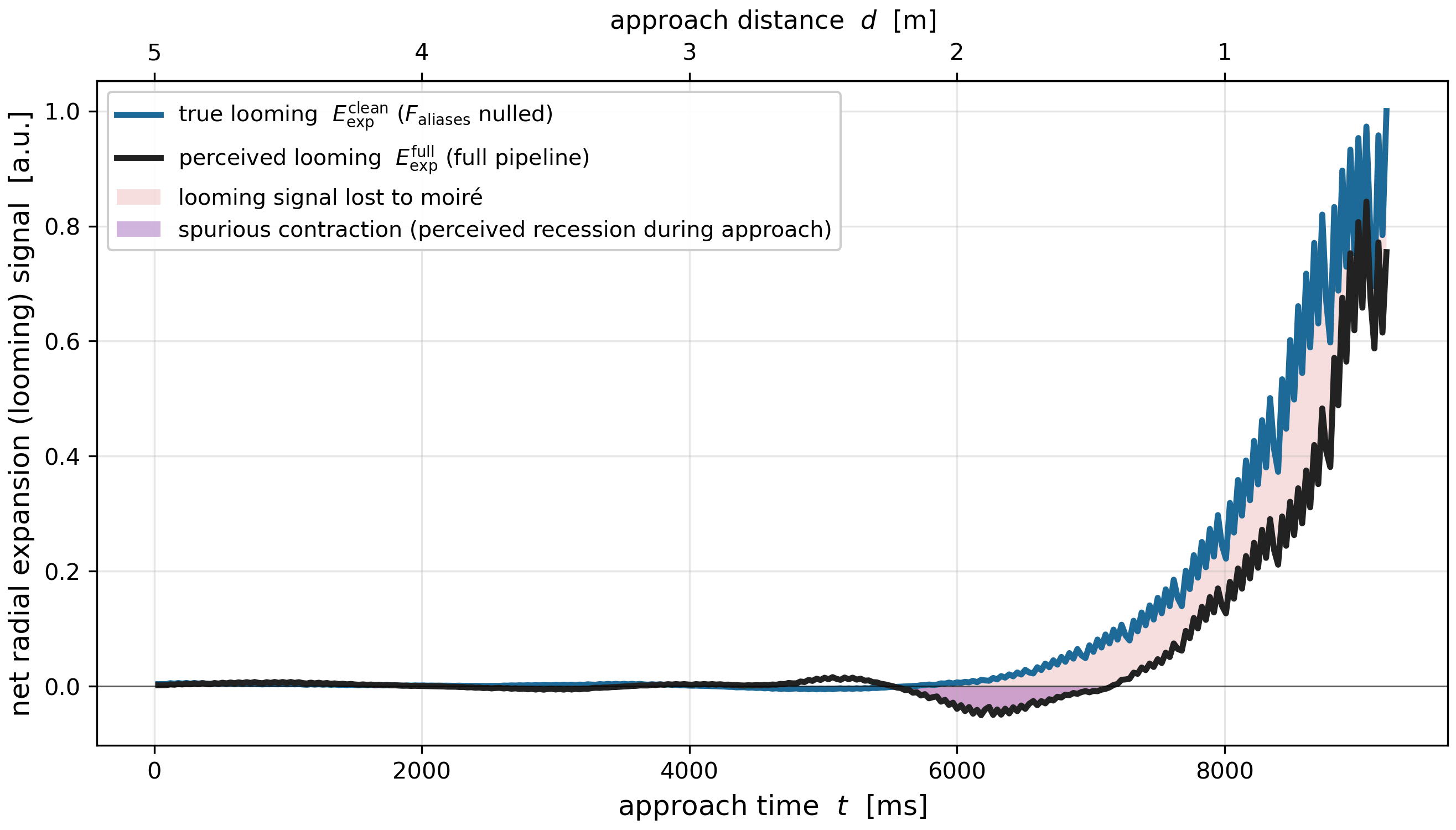}
\caption{Corruption of the radial-expansion (looming) signal during the same simulated approach on a striped host (\texttt{DSC\_0084}, Equus quagga). The Hassenstein--Reichardt vector field is projected onto the radial direction from the focus of expansion (the approached body) and averaged over the mosaic to give the net looming signal that gates landing. Blue: alias-free signal $E_{\mathrm{exp}}^{\mathrm{clean}}(t)$; black: full-pipeline signal $E_{\mathrm{exp}}^{\mathrm{full}}(t)$; red shading: looming signal lost to the moir\'e; purple shading: the moir\'e-resonance window near $d\!\approx\!\SI{1.9}{\meter}$ in which the perceived looming becomes net negative. This means a perceived recession during a genuine approach. The signal is in arbitrary EMD units. Only its sign and its value relative to the alias-free trace are meaningful.}
\label{fig:reichardt_expansion}
\end{figure}

Figure~\ref{fig:reichardt_control_zebra}\ plots three motion-energy time series for the same approach trajectory generated with photograph \texttt{DSC\_0084}. The gray curve $E_{\mathrm{HR}}^{\mathrm{clean}}(t)$ is the motion energy in
the output of the EMD (Elementary Motion Detector) when the comb-replication step
(Eq.~\ref{eq:freq_replication}) is explicitly nulled. This is what
an idealized fly would see if its ommatidial lattice could
sample without aliasing the stripe pattern. The black curve
$E_{\mathrm{HR}}^{\mathrm{full}}(t)$ is the same quantity for
the entire ommatidial pipeline that contains the aliased
moir\'e of \S\ref{subsec:moire_isolation}. The red curve is the
signed difference
$E_{\mathrm{HR}}^{\mathrm{moir\acute{e}}}(t)
= E_{\mathrm{HR}}^{\mathrm{full}}(t)
- E_{\mathrm{HR}}^{\mathrm{clean}}(t)$
of Eq.~\ref{eq:E_HR_moire}, with the red shading indicating
its sign relative to zero. The bottom axis is approach time at
the photoreceptor frame interval $\Delta t\!=\!\SI{30}{\milli\second}$
matched to the membrane time constant
\citep{borst2010motion}, and the top axis is the corresponding
approach distance $d(t)$. The high-frequency ripple visible at
$t\!\gtrsim\!\SI{2.5}{\second}$ is the per-frame discrete
crossing of stripes from one ommatidium to the next as the
body grows in the field of view.
Two features of the curves are the central to the discussion.
First, both the clean and the full motion-energy traces rise
monotonically as the host enters the close-approach band, but
the alias-free trace rises faster. At the end of the
trajectory ($d\!\approx\!\SI{0.4}{\meter}$) the clean signal
exceeds the full signal by a factor of $\sim\!2$. Second,
$E_{\mathrm{HR}}^{\mathrm{moir\acute{e}}}(t)$ is therefore
strictly negative inside the close-approach band
$d\!\in\![1, 2.5]\,\si{\meter}$ and grows progressively more
negative as the host seeking fly closes in. The sign is the opposite of
what would be expected if the moir\'e merely added a spurious
motion signal of arbitrary direction. Instead, the moir\'e is
systematically cancelling coherent motion energy that
the EMD array would otherwise produce. 
A direct algebraic decomposition of $E_{\mathrm{HR}}^{\mathrm{moir\acute{e}}}$ into a cross-term and a self-term, reported in §S.5, attributes the negative sign to a destructive cross-correlation between the moir\'e component and the genuine-motion component of the signed EMD output. 
Aliased moir\'e drift contributes motion signal in
directions inconsistent with the genuine self-motion
translation, and the opponent subtraction of
Eq.~\ref{eq:reichardt} cancels part of the genuine signal
before squaring. The behavioural reading is that the parasitic spectrum
identified at the optical stage is not benign noise that the
downstream brain could integrate out. It is a destructive
contribution to the very motion-energy estimate that controls
approach and landing in biting flies
\citep{baird2013universal,vanbreugel2012visual,borst2014fly}.
An attacking fly approaching a striped host therefore
loses, rather than gains, coherent expansion signal in
exactly the close-approach window
$d\!\in\![1, 2.5]\,\si{\meter}$ that the optical-stage
matched-pair test of \S\ref{subsec:matched_pair} independently
identifies as the moir\'e-peak band. A motion-energy deficit
in this window is a documented cause of stalled landings in
flies \citep{baird2013universal,vanbreugel2012visual}. The
same band coincides with the empirical terminal-hesitation
phase \citep{waage1981zebra,brady1988landing} in which biting
flies approach a striped host but fail to settle. The
optical-mechanism story is therefore not merely that stripes
inject extra signal into the eye, but that the comb-step
geometry of the ommatidial mosaic converts that injected
signal, at the very next processing stage, into a partial
cancellation of the optic-flow estimate the fly needs in order to land.
The motion-energy deficit of Fig.~\ref{fig:reichardt_control_zebra} can be read directly as a corruption of the looming cue that gates landing. Projecting the EMD vector field onto the radial direction from the focus of expansion (the body the fly is approaching) and averaging over the mosaic yields the net radial-expansion (looming) signal, shown in Fig.~\ref{fig:reichardt_expansion}. The alias-free looming signal rises monotonically as the host enters the close-approach band, whereas the full-pipeline signal is suppressed throughout that band and, in the moir\'e-resonance window around $d\!\approx\!\SI{1.9}{\meter}$, turns transiently negative. A negative net looming signal is a perceived radial contraction. At that range the model fly's motion pathway reports the host receding at the very moment it is closing in. The behaviourally weightier error is the suppression at close range ($d\!<\!\SI{1}{\meter}$), where the genuine looming is strongest and the deficit inflates the estimated time-to-contact. The transient sign reversal at intermediate range is a weaker, but it shows unambiguously that the parasitic spectrum corrupts the sign, not merely the magnitude, of the ego-motion estimate. Even if the reversal varies between hosts and situations the feature is robust across the matched-pair set. The key factor for landing behavior is the radial expansion (looming) part of the optic flow described above \citep{baird2013universal,vanbreugel2012visual,borst2014fly}. The main finding is that the moir\'e pattern disrupts this radial ego-motion estimate (Fig.~\ref{fig:reichardt_expansion}). It is probably this disruption of expansion, not any misperception of direction, that most likely explains the stalled and aborted
landings seen on striped hosts \citep{caro2019benefits,brady1988landing}.

\begin{figure}[H]
\centering
\includegraphics[width=\textwidth]{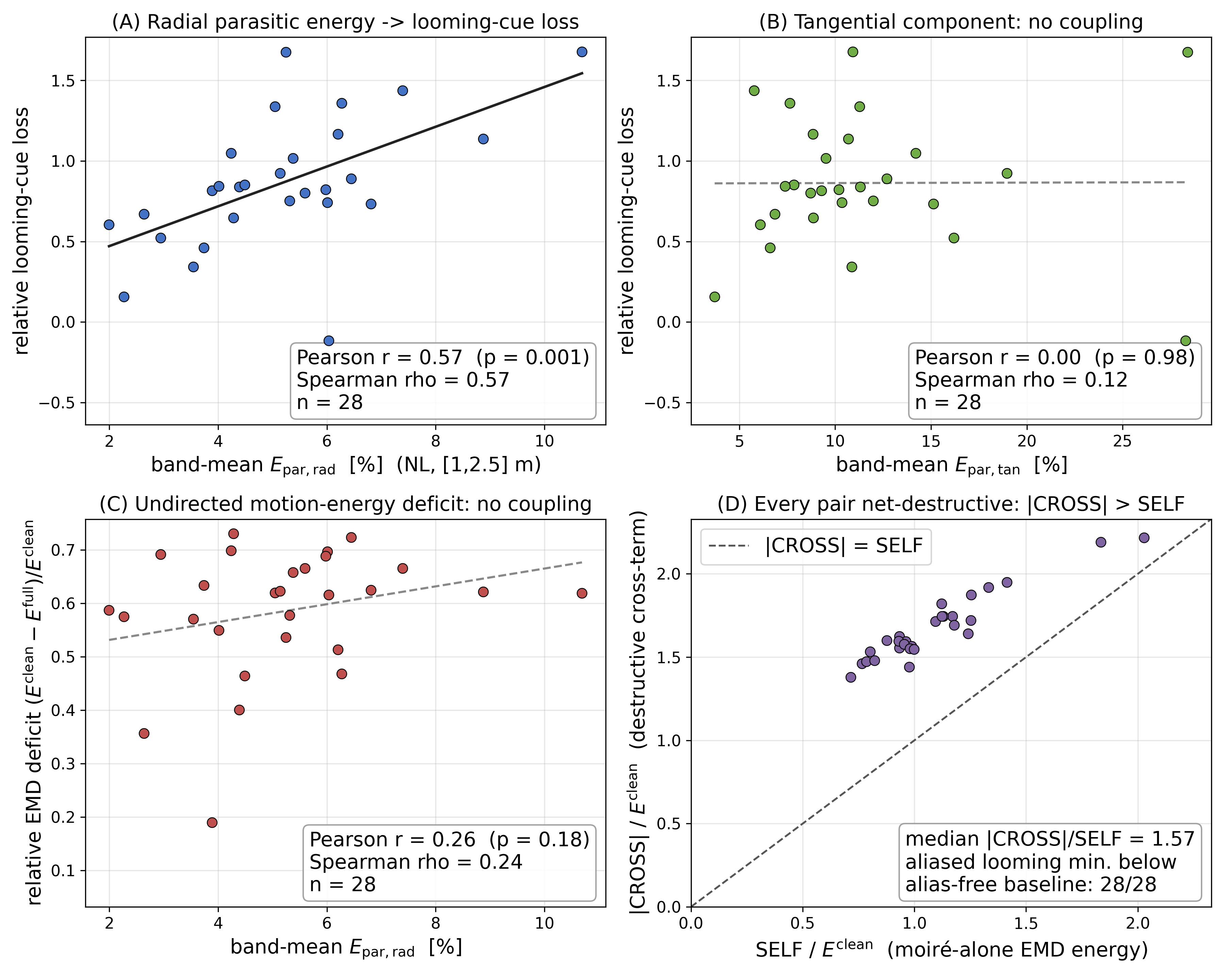}
\caption{Optical parasitic energy versus motion-stage read-outs
across the $n\!=\!28$ matched pairs (hand-picked body centres, band-mean
over $d\!\in\![1,2.5]\,\si{\meter}$, Airy MTF, Naka--Rushton
$\alpha\!=\!0.5$). (A)~The radial component of the parasitic
energy, $E_{\mathrm{par,rad}}$ (Eq.~\ref{eq:E_par_rad}), predicts the
relative looming-cue loss ($r\!=\!0.57$, $p\!=\!0.001$; Spearman
$r_{s}\!=\!0.57$). (B)~The tangential component $E_{\mathrm{par,tan}}$
does not ($r\!=\!0.00$, $p\!=\!0.98$).
(C)~ the undirected EMD energy
deficit is not predicted by $E_{\mathrm{par,rad}}$ ($r\!=\!0.26$, $p\!=\!0.18$). (D)~The destructive cross-term
exceeds the moir\'e-alone self-term in every pair
($|\mathrm{CROSS}|\!>\!\mathrm{SELF}$; median $1.57$), so the comb step
is net-destructive throughout the sample.}
\label{fig:epar_vs_emd}
\end{figure}

The single-trajectory looming result of
Fig.~\ref{fig:reichardt_expansion} holds across the full matched-pair
set, but not in the way a naive energy argument would predict. For each
of the $n\!=\!28$ pairs the optical parasitic energy (band-mean
over the close-approach band, non-linear branch)
was related to two motion-stage read-outs
(see Fig.~\ref{fig:epar_vs_emd}). Because the looming read-out is
a radial projection (Eq.~\ref{eq:E_exp}), the optical scalar is
projected the same way: $E_{\mathrm{par,rad}}$ weights the parasitic
energy inside the Nyquist disc by the fraction of its local orientation
that is radial to the focus of expansion, and $E_{\mathrm{par,tan}}$ by
the tangential fraction (Eq.~\ref{eq:E_par_rad}). Their numerators
partition the parasitic energy $E_{\mathrm{par}}$ of the isotropic
$E_{\mathrm{par,rel}}$. The radial component
carries the coupling to the relative loss of the radial-expansion
(looming) cue across pairs ($r\!=\!0.57$, $p\!=\!0.001$; Spearman
$r_{s}\!=\!0.57$), the tangential component none ($r\!=\!0.00$,
$p\!=\!0.98$; difference of the two dependent correlations
$p\!=\!0.016$), and the isotropic total lies in between
($r\!=\!0.48$, $p\!=\!0.009$). Neither component predicts the
undirected Hassenstein--Reichardt energy
deficit (\textit{$r\!=\!0.26$, $p\!=\!0.18$ for $E_{\mathrm{par,rad}}$}). The energy deficit is the near-cancellation of two large opposing terms, the moir\'e-alone self-energy ($\mathrm{SELF}$)
and its destructive cross-correlation with the genuine-motion signal
($\mathrm{CROSS}$, \S\ref{app:reichardt_cancellation}), whose net is
only weakly tied to the parasitic magnitude. The looming cue, by
contrast, is a signed radial projection onto exactly the axis that gates
landing, so only the radially oriented part of the parasitic drift corrupts it, and it does so in proportion to
$E_{\mathrm{par,rad}}$. The destructive sign is universal across the
sample. Every pair has $|\mathrm{CROSS}|\!>\!\mathrm{SELF}$ (median
ratio $1.57$), and in all $28$ pairs the aliased looming minimum is
driven below the alias-free baseline to a net-negative value, the
alias-free baseline itself sitting at $\approx\!0$ at that minimum.
The causal attribution is complete by construction, since the
full and clean pipelines differ in nothing but the comb step. What
$E_{\mathrm{par,rad}}$ accounts for is a third of the between-host
variation in the size of the loss ($R^{2}\!=\!0.33$). The remainder is
expected from what a power measure cannot carry .The phase of the
parasitic beat relative to the genuine motion, which sets the size of
the destructive cross-term detector by detector
(\S\ref{app:reichardt_cancellation}), and the temporal structure of the
approach, since the loss accumulates along a trajectory through the
temporal filter of the detector whereas the optical scalar is static.
The quantitative coupling between the optical and motion
stages therefore runs through the looming channel specifically, not
through undirected motion energy, which is consistent with landing
being gated by the radial-expansion cue
\citep{baird2013universal,vanbreugel2012visual,borst2014fly}.
The same aliased retinal stimulus is also the basis for the directional motion illusions described by
\citet{how2014motion}. These illusions come from the parasitic moir\'e beat, not from the unresolved stripe carrier. This beat is what the detector array follows, so the local direction reversals counted
in \S\ref{app:reichardt_cancellation} are the wagon-wheel (temporal-aliasing) signature of that beat. An oblique-stripe aperture term, which is not captured by the single-axis control used here,
would produce the barber-pole signature. These directional illusions are therefore
not a separate mechanism, but instead are a companion effect that occurs on the
same moir\'e pattern. They are most relevant to the lateral
and circling phases of approach, rather than the final looming phase. The analytic
details of these relationships and their limits are discussed in \S\ref{app:motion_aliasing}. The model presented here derives the parasitic spectrum
optically and validates it on real zebra coats. Because a looming
estimate requires flow on more than one axis, which the single
facet-line correlator of \citet{mouy2025aliasing} cannot supply, this
corruption of the radial-expansion cue is accessible only to the
two-dimensional detector used here.

\section{Conclusion}
We developed a Fourier-optics model of biting-fly compound-eye sampling
and applied it to photographs of zebra coats in biologically relevant
habitats. When a striped pattern is viewed through the quasi-periodic
ommatidial lattice, the interaction generates parasitic spatial
frequencies that are absent from the coat itself. With published optical
constants for diurnal Culicidae and Tabanidae
\citep{land1997mosquito,land1999fundamental,kawada2006eye,hardie1989compound},
these frequencies fall within the spatial-frequency window that governs
host fixation and landing. Passed through a Hassenstein--Reichardt
elementary motion detector
\citep{hassenstein1956systemtheoretische,borst2010motion}, the
reconstructed parasitic signal $I_{\mathrm{par}}(x,y)$ produces motion
energy unrelated to the fly's true self-motion. Its dominant effect is
to corrupt the radial-expansion (looming) cue that triggers the
legs-out, hover, and settle sequence
\citep{baird2013universal,vanbreugel2012visual,borst2014fly}. The same
aliased signal underlies the wagon-wheel and barber-pole illusions of
\citet{how2014motion}, which ride the moir\'e beat as a subordinate
directional companion. The two-dimensional detector developed here and applied to real zebra coats
shows that within the close-approach band the expansion cue
gating landing is not merely degraded but destructively cancelled, and
can transiently reverse sign. This optical mechanism complements rather than competes with the
polarotactic explanation of \citet{horvath2010whiteness} and
\citet{egri2012polarotactic}. Although polarisation contrast is
classically treated as a long-range attractant, the dynamic-sampling
argument of \citet{horvath2024polarotaxis} indicates that it remains
behaviourally available at the same close range ($1$--$2.5\,\si{\meter}$)
at which the moir\'e operates. The two cues therefore most plausibly act
in concert during terminal approach rather than in strictly separated
phases, and the striped coat may have been selected for both. Incorporating
polarisation explicitly should refine the quantitative predictions and
sharpen rather than overturn this conclusion.
The model represents the optical and first motion-detection stages. It
does not capture higher-level processing in the lamina, medulla, and
lobula plate \citep{borst2014fly}, nor the multimodal interactions with
$\mathrm{CO_2}$ and other olfactory cues that gate visual attraction in
the first place \citep{vanbreugel2015mosquitoes,coutinho2022human,mcmeniman2014multimodal}.
The parasitic frequencies nonetheless arise at the sampling stage and
are an unavoidable retinal perturbation, independent of the downstream
architecture. The Reichardt stage shows that this perturbation
propagates into the motion stream that drives landing. The model contributes
a distinct, luminance-based optical dimension to the anti-parasite
hypothesis of zebra striping
\citep{waage1981zebra,caro2014function,caro2019benefits}.

%

\section*{List of Symbols and Abbreviations}
\addcontentsline{toc}{section}{List of Symbols and Abbreviations}

\newcounter{symtabsave}\setcounter{symtabsave}{\value{table}}

\subsection*{Coordinates and operators}
\begin{longtable}{@{}p{0.18\textwidth} p{0.78\textwidth}@{}}
$(x,y)$            & Spatial coordinates, expressed in degrees of visual angle relative to the optical axis. \\
$(u,v)$            & Spatial-frequency coordinates, $\mathrm{cycles\,deg^{-1}}$. \\
$\rho$             & Radial spatial frequency, $\rho=\sqrt{u^{2}+v^{2}}$. \\
$\theta$           & \textit{Off-axis angle (optics).} \\
$\theta_{\mathrm{r}}$ & \textit{Reflection angle from the surface normal (polarisation); previously also written $\theta$.} \\
$\mathcal{F}$      & Fourier-transform operator; $\mathcal{F}^{-1}$ its inverse. \\
$*$                & One-dimensional convolution. \\
$\circledast$      & Two-dimensional convolution. \\
$\langle\,\cdot\,\rangle$ & Spatial average over the ommatidial/detector array. \\
$\delta(\cdot)$    & Dirac delta function. \\
$\nabla$           & \textit{Spatial gradient operator, $(\partial_{x},\partial_{y})$.} \\
$\operatorname{sinc}(x)$ & Normalised sinc, $\sin(\pi x)/(\pi x)$. \\
$J_{1}$            & Bessel function of the first kind, order one. \\
\end{longtable}

\subsection*{Stimulus and image fields (spatial domain)}
\begin{longtable}{@{}p{0.18\textwidth} p{0.78\textwidth}@{}}
$I(x,y),\ I_{\mathrm{stim}}$        & Striped external stimulus intensity (linear grayscale). \\
$I_{0}(x,y)$                        & Reference stimulus at reference distance $d_{0}$. \\
$I_{\mathrm{retina}}$              & Retinal image after the full eye impulse response. \\
$I_{\mathrm{blurred}}$            & Acceptance-blurred stimulus, $I_{\mathrm{stim}}*h_{\rho}$. \\
$I_{\mathrm{blurred}}^{\mathrm{NL}}$ & Non-linear (Naka--Rushton compressed) blurred image. \\
$I_{\mathrm{samples}}$            & Comb-sampled (ommatidial) image. \\
$I_{\mathrm{par}}(x,y)$           & Parasitic (phantom) image, $\mathcal{F}^{-1}\{F_{\mathrm{aliases}}\}$\textit{; in Eq.~\ref{eq:E_par_rad} the non-linear branch restricted to the Nyquist disc, $\mathcal{F}^{-1}\{F_{\mathrm{aliases}}^{\mathrm{NL}}\,M\}$}. \\
$I_{\mathrm{sig}}(x,y)$           & \textit{Alias-free signal image restricted to the Nyquist disc, $\mathcal{F}^{-1}\{F_{\mathrm{signal}}\,M\}$ (Eq.~\ref{eq:E_par_rad}).} \\
$h_{\mathrm{eye}}(x,y)$           & Spatial impulse response of the compound eye. \\
$h_{\rho}$                         & Acceptance point-spread function, $\mathcal{F}^{-1}\{H_{\rho}\}$. \\
\end{longtable}

\subsection*{Fourier-domain spectra and transfer functions}
\begin{longtable}{@{}p{0.18\textwidth} p{0.78\textwidth}@{}}
$F_{\mathrm{stim}},\ F_{0}$        & Spectrum of the stimulus / reference stimulus. \\
$F_{\mathrm{blurred}}$            & Spectrum of the acceptance-blurred stimulus. \\
$F_{\mathrm{blurred}}^{\mathrm{NL}}$ & Spectrum of the non-linear blurred image. \\
$F_{\mathrm{repl}}$              & Comb-replicated spectrum after sampling. \\
$F_{\mathrm{retina}}$            & Full reconstructed retinal spectrum. \\
$F_{\mathrm{signal}}$            & Alias-free in-band signal, $F_{\mathrm{blurred}}H_{\mathrm{box}}$. \\
$F_{\mathrm{aliases}}$           & Isolated parasitic spectrum (linear branch). \\
$F_{\mathrm{aliases}}^{\mathrm{NL}}$ & Isolated parasitic spectrum (non-linear branch). \\
$H_{\mathrm{eye}}(u,v)$          & Compound-eye optical transfer function (OTF). \\
$H_{\rho}(u,v)$                  & Acceptance (low-pass) modulation transfer function. \\
$H_{\rho}^{\mathrm{G}}$          & Gaussian approximation of $H_{\rho}$ (Supplementary). \\
$H_{\mathrm{Airy}}$              & Diffraction-limited circular-aperture (Airy) MTF. \\
$H_{\mathrm{samp}}$              & Sampling factor of the OTF. \\
$H_{\mathrm{box}}(u,v)$          & Voronoi (box-cell) reconstruction MTF. \\
$\mathrm{III}_{\Delta\varphi},\ \mathrm{III}_{f_{s}}$ & Dirac comb in space / in frequency. \\
$\Lambda_{\Delta\varphi}$        & Ommatidial sampling lattice. \\
\end{longtable}

\subsection*{Eye optics and geometry}
\begin{longtable}{@{}p{0.18\textwidth} p{0.78\textwidth}@{}}
$D$                & Corneal facet (lens) diameter. \\
$\lambda$          & Wavelength of light. \\
$\Delta\varphi$    & Inter-ommatidial angle. \\
$\sigma_{\rho}$    & Angular acceptance half-width (half-width at half-maximum). \\
$\Delta\rho$       & Acceptance full width at half maximum. \\
$S(\theta)$        & Angular acceptance function of one ommatidium. \\
$\mathrm{PSF}(\theta)$ & Angular point-spread function (Airy pattern). \\
$\rho_{c}$         & Diffraction cutoff, $\rho_{c}=D/\lambda$ (cyc.\,rad$^{-1}$ / cyc.\,deg$^{-1}$). \\
$f_{c}$            & Angular cutoff frequency of the Airy MTF. \\
$D\,\Delta\varphi$ & Eye parameter (resolution--sensitivity trade-off). \\
$p_{\mathrm{eye}}$ & \textit{Eye parameter $D\,\Delta\varphi$ in $\si{\micro\meter\radian}$; previously written $p$.} \\
$R1$--$R8$         & The eight retinular (photoreceptor) cells of an ommatidium; $R7$ UV-, $R8$ blue-sensitive. \\
\end{longtable}

\subsection*{Sampling, reconstruction and frequency limits}
\begin{longtable}{@{}p{0.18\textwidth} p{0.78\textwidth}@{}}
$f_{s}$                & Sampling rate, $f_{s}=1/\Delta\varphi$. \\
$\nu_{\mathrm{eye}}$   & Eye Nyquist limit, $\nu_{\mathrm{eye}}=1/(2\Delta\varphi)$. \\
$(k,m)$               & Integer replica indices in the comb sum. \\
$M(u,v)$              & \textit{Nyquist-disc mask, $1$ for $|f|\leq\nu_{\mathrm{eye}}$ and $0$ otherwise.} \\
\end{longtable}

\subsection*{Photoreceptor non-linearity}
\begin{longtable}{@{}p{0.18\textwidth} p{0.78\textwidth}@{}}
$\mathrm{NR}_{\beta,s_{50}}(s)$ & Naka--Rushton compressive response. \\
$\beta$              & \textit{Naka--Rushton shape exponent ($\beta=0.7$); previously written $n$, which is now reserved for the sample size.} \\
$s_{50}$             & Half-saturation level ($s_{50}=0.5$). \\
$\alpha$             & Non-linearity mix factor, $\alpha\in[0,1]$. \\
$\mathrm{Volt}_{a_{2},a_{3}}$ & Truncated Volterra non-linearity (Supplementary). \\
$a_{2},\,a_{3}$      & Volterra quadratic and cubic kernel coefficients. \\
$\mu,\ \sigma$       & Mean and standard deviation of the input image. \\
\end{longtable}

\subsection*{Energy metrics}
\begin{longtable}{@{}p{0.18\textwidth} p{0.78\textwidth}@{}}
$E_{\mathrm{par}}(d)$           & Parasitic energy inside the Nyquist disc (linear). \\
$E_{\mathrm{par}}^{\mathrm{NL}}(d)$ & Parasitic energy (non-linear branch). \\
$E_{\mathrm{sig}}(d)$           & Integrated in-band (alias-free) signal energy. \\
$E_{\mathrm{par,rel}}(d)$       & Relative parasitic energy, $E_{\mathrm{par}}/E_{\mathrm{sig}}$ (dimensionless). \\
$E_{\mathrm{par,rel}}^{\mathrm{lin}},\ E_{\mathrm{par,rel}}^{\mathrm{NL}}$ & Linear / non-linear relative parasitic energy. \\
$\langle E_{\mathrm{par,rel}}\rangle$ & Band-mean relative parasitic energy, averaged over $d\in[1,2.5]\,\si{\meter}$. \\
$E_{\mathrm{par,rad}},\ E_{\mathrm{par,tan}}$ & \textit{Radial / tangential component of the relative parasitic energy (Eq.~\ref{eq:E_par_rad}); band-means over $d\in[1,2.5]\,\si{\meter}$ in Fig.~\ref{fig:epar_vs_emd}.} \\
$w_{\mathrm{rad}}[I],\ w_{\mathrm{tan}}[I]$ & \textit{Radial / tangential fraction of the local orientation of an image $I$, $w_{\mathrm{rad}}=(\nabla I\cdot\hat{\mathbf{r}})^{2}/|\nabla I|^{2}$, $w_{\mathrm{tan}}=1-w_{\mathrm{rad}}$.} \\
$E_{\max}$           & \textit{Largest $E_{\mathrm{par,rel}}$ among the columns of a figure (opacity normalisation in Figs.~\ref{fig:fourier_moire} and~\ref{fig:zebra_vs_horse_a}).} \\
$\eta$               & \textit{Panel opacity of the parasitic image in Figs.~\ref{fig:fourier_moire} and~\ref{fig:zebra_vs_horse_a}, $\eta = 0.25 + 0.75\sqrt{E_{\mathrm{par,rel}}/E_{\max}}$; previously written $\alpha$, which is reserved for the non-linearity mix.} \\
\end{longtable}

\subsection*{Distance, calibration and imaging}
\begin{longtable}{@{}p{0.18\textwidth} p{0.78\textwidth}@{}}
$d$                  & Approach / viewing distance. \\
$d_{0}$              & Reference distance (Fourier scaling) / trajectory start distance. \\
$d_{\mathrm{cam}}$   & Camera-to-host distance. \\
$d_{\mathrm{peak}}$  & Approach distance of the moir\'e-energy peak. \\
$d_{\mathrm{true}}$  & True distance after body-height rescaling. \\
$d(t)$               & Time-dependent approach distance. \\
$h_{\mathrm{body}},\ h_{\mathrm{body,m}}$ & Generic body (shoulder) height, $\SI{1.4}{\meter}$. \\
$h_{\mathrm{body,pix}}$ & Body height in camera pixels. \\
$h_{\mathrm{true}}$  & True per-individual body height. \\
$f_{\mathrm{lens}}$  & \textit{Camera focal length; previously written $f$, which is reserved for spatial frequency.} \\
$p_{\mathrm{pix}}$   & \textit{Sensor pixel pitch, $\SI{7.88}{\micro\meter}$; previously written $p$.} \\
$\mathrm{dpp_{cam}}$ & Native camera angular pixel pitch ($\approx\SI{5.42}{arcsec/pixel}$). \\
\end{longtable}

\subsection*{Numerical grid and windowing}
\begin{longtable}{@{}p{0.18\textwidth} p{0.78\textwidth}@{}}
$N$                  & \textit{Simulation grid samples per side ($N=384$).} \\
$N_{\mathrm{det}}$   & \textit{EMD detectors per side of the simulated patch ($N_{\mathrm{det}}=12$); previously also written $N$.} \\
$\mathrm{FoV}$       & Simulated angular field of view ($30^{\circ}$). \\
$\Delta x$           & Angular pixel size, $\mathrm{FoV}/N$. \\
$\Delta f$           & FFT bin spacing, $1/(N\Delta x)$. \\
$w(x,y),\ w_{1}$     & Separable two-dimensional / one-dimensional Hann window. \\
$\sigma_{f}$         & Gaussian frequency width, $1/(2\pi\sigma_{\rho})$ (Supplementary). \\
\end{longtable}

\subsection*{Post-retinal motion detection (Hassenstein--Reichardt)}
\begin{longtable}{@{}p{0.18\textwidth} p{0.78\textwidth}@{}}
$L(t),\ R(t)$            & Left / right ommatidial input signals to an EMD pair. \\
$\tau_{\mathrm{lp}}(t)$  & First-order low-pass filter kernel. \\
$\tau_{\mathrm{HR}}$     & EMD low-pass time constant ($\approx\SI{30}{\milli\second}$). \\
$\tau_{\mathrm{m}}$      & Receptor membrane time constant ($\approx\SI{10}{\milli\second}$). \\
$\mathrm{HR}(t)$         & Signed EMD (opponent) output. \\
$\mathrm{HR}_{x},\ \mathrm{HR}_{y}$ & Horizontal / vertical EMD outputs. \\
$\mathbf{HR}(t)$         & Local EMD motion vector, $(\mathrm{HR}_{x},\mathrm{HR}_{y})$. \\
$\hat{\mathbf{r}}$       & Outward radial unit vector from the focus of expansion. \\
$E_{\mathrm{HR}}(t)$     & Direction-agnostic motion energy. \\
$E_{\mathrm{exp}}(t)$    & Signed radial-expansion (looming) signal. \\
$E_{\mathrm{HR}}^{\mathrm{full}},\ E_{\mathrm{HR}}^{\mathrm{clean}}$ & Motion energy of the full / alias-free pipeline. \\
$E_{\mathrm{exp}}^{\mathrm{full}},\ E_{\mathrm{exp}}^{\mathrm{clean}}$ & Looming signal of the full / alias-free pipeline. \\
$E_{\mathrm{HR}}^{\mathrm{moir\acute{e}}}$ & Signed moir\'e motion-energy difference, full $-$ clean. \\
$\Delta\mathrm{HR}(t)$   & Moir\'e-induced additive EMD contribution. \\
$\mathrm{CROSS}(t)$      & Cross-term (moir\'e $\times$ genuine motion), signed. \\
$\mathrm{SELF}(t)$       & Self-term (moir\'e-only motion energy), non-negative. \\
$v_{\mathrm{app}}$       & Closing (approach) speed ($\SI{0.5}{\meter\per\second}$). \\
$\Delta t$               & Frame interval ($\SI{30}{\milli\second}$). \\
\end{longtable}

\subsection*{Motion aliasing of the moir\'e beat}
\begin{longtable}{@{}p{0.18\textwidth} p{0.78\textwidth}@{}}
$f_{t}$               & Temporal frequency of a drifting field, $v\,f_{s}$. \\
$f_{\mathrm{M}}$      & Spatial frequency of the aliased moir\'e beat. \\
$v$                   & Retinal drift velocity of the stripe field. \\
$v_{\mathrm{M}}$      & Apparent drift velocity of the moir\'e beat. \\
$\Lambda_{s}$         & Stripe carrier period, $1/f_{s}$. \\
$\Lambda_{\mathrm{M}}$ & Moir\'e beat period, $1/f_{\mathrm{M}}$. \\
$\hat{\mathbf{n}}$    & Stripe normal direction. \\
$\mathbf{v}$          & True motion velocity vector. \\
$\epsilon$            & Aperture-problem angular error, $\angle(\mathbf{v},\hat{\mathbf{n}})$. \\
\end{longtable}

\subsection*{Polarisation}
\begin{longtable}{@{}p{0.18\textwidth} p{0.78\textwidth}@{}}
$\mathcal{I},\ \mathcal{Q},\ \mathcal{U}$ & \textit{Stokes components: total intensity, linear H/V, linear $\pm45^{\circ}$; previously written $I,Q,U$ ($I$ is reserved for the stimulus intensity).} \\
$\mathcal{P}(\theta_{\mathrm{r}})$ & \textit{Degree of linear polarisation as a function of the reflection angle; previously written $d(\theta)$ ($d$ is reserved for the approach distance).} \\
$\mathcal{P}_{\mathrm{black}},\ \mathcal{P}_{\mathrm{white}}$ & \textit{Degree of polarisation of black / white stripes; previously $d_{\mathrm{black}}, d_{\mathrm{white}}$.} \\
\end{longtable}

\subsection*{Statistics (matched-pair analysis)}
\begin{longtable}{@{}p{0.18\textwidth} p{0.78\textwidth}@{}}
$n$                  & \textit{Number of matched zebra / stripe-removed pairs ($n=28$).} \\
$r$                  & \textit{Pearson correlation coefficient.} \\
$r_{s}$              & \textit{Spearman rank correlation coefficient; previously written $\rho$, which is reserved for the radial spatial frequency.} \\
$R^{2}$              & \textit{Coefficient of determination of the linear fit.} \\
$p$                  & \textit{Two-sided $p$-value.} \\
$z$                  & \textit{Test statistic for the comparison of two dependent correlations \citep{meng1992comparing}.} \\
\end{longtable}

\subsection*{Abbreviations}
\begin{longtable}{@{}p{0.18\textwidth} p{0.78\textwidth}@{}}
OTF        & Optical transfer function. \\
MTF        & Modulation transfer function. \\
PSF        & Point-spread function. \\
EMD        & Elementary motion detector. \\
HR         & Hassenstein--Reichardt (motion detector). \\
FFT        & Fast Fourier Transform. \\
FoV        & Field of view. \\
DC         & Direct current (zero spatial frequency). \\
HWHM/FWHM  & Half / full width at half maximum. \\
ULP        & Unit in the last place (floating-point resolution). \\
RGB        & Red--green--blue camera colour channels. \\
sRGB       & Standard RGB colour space. \\
UV         & Ultraviolet. \\
ISO        & Film-speed (sensitivity) rating. \\
XMP        & Extensible Metadata Platform (image metadata). \\
NL         & Non-linear (superscript on Fourier/energy quantities). \\
$\mathrm{CO_{2}}$ & Carbon dioxide. \\
spp.       & Species (plural). \\
\textit{E.\ quagga}, \textit{E.\ zebra}, \textit{E.\ grevyi} & \textit{Equus quagga} (plains), \textit{E.\ zebra} (mountain), \textit{E.\ grevyi} (Grevy's) zebra. \\
\end{longtable}

\setcounter{table}{\value{symtabsave}}

\appendix
\section*{Supplementary Information}
\addcontentsline{toc}{section}{Supplementary Information}
\renewcommand{\thesubsection}{S.\arabic{subsection}}
\setcounter{subsection}{0}

\subsection{Gaussian approximation of the ommatidial acceptance}
\label{app:gaussian_otf}
The compound-eye literature the angular acceptance
function of a single ommatidium can be approximated by a two-dimensional
Gaussian. Under this
approximation the low-pass envelope of (Eq.~\ref{eq:eye_otf}) becomes
\begin{equation}
H_{\rho}^{\mathrm{G}}(u,v)
\;=\;
\exp\!\left(-\frac{u^{2}+v^{2}}{2\sigma_{f}^{2}}\right),
\quad
\sigma_{f} \;=\; \frac{1}{2\pi\,\sigma_{\rho}},
\label{eq:lowpass_gaussian}
\end{equation}
where $\sigma_{\rho}$ is the half-width at half-maximum of the
acceptance angle expressed in degrees. For diurnal culicids
($\Delta\varphi \approx \SI{2}{\degree}$
)\citep{land1997mosquito,land1999fundamental} the canonical choice
is $\sigma_{\rho} = \SI{0.85}{\degree}$. The Gaussian form is
analytically convenient and captures the central lobe of
(Eq.~\ref{eq:airy_psf}) reasonably well, but it lacks a hard cutoff and
underestimates attenuation near $\rho_{c}$. We therefore retain the
Airy-based expression (Eq.~\ref{eq:airy_otf}) in the main analysis and
include (Eq.~\ref{eq:lowpass_gaussian}) here for comparison.

\subsection{FFT-pipeline numerical controls}
\label{subsec:fft_controls}

Two numerical controls are applied on every distance to report relative error.

\paragraph{(i) Round-trip control on $F_{\mathrm{blurred}}$.}
The complex spectrum
$F_{\mathrm{blurred}}=\mathcal{F}\{I_{\mathrm{blurred}}\,w\}$
carries both amplitude and phase. Inverse-transforming it must
recover the windowed blurred image up to floating-point precision,
\begin{equation}
\bigl\|\mathrm{Re}\,\mathcal{F}^{-1}\!\{F_{\mathrm{blurred}}\}
- I_{\mathrm{blurred}}\!\cdot\!w\bigr\|_{\infty}
\;/\;
\bigl\|I_{\mathrm{blurred}}\!\cdot\!w\bigr\|_{\infty}
\;\lesssim\;10^{-15}.
\label{eq:roundtrip_blurred}
\end{equation}

\paragraph{(ii) Round-trip control on $F_{\mathrm{repl}}$.}
By Eq.~\ref{eq:freq_replication} and the convolution theorem,
$\mathcal{F}^{-1}\{F_{\mathrm{repl}}\}$ must equal
$I_{\mathrm{blurred}}\!\cdot\!w$ multiplied by the spatial Dirichlet
approximation of the comb $\mathrm{III}_{\Delta\varphi}$ used in the
construction. The result is real to within
$\bigl|\mathrm{Im}\,\mathcal{F}^{-1}\{F_{\mathrm{repl}}\}\bigr|_{\infty}\!\lesssim\!10^{-15}$,
the floating-point noise level. A non-zero imaginary part above this
threshold would indicate a misordered \verb|fftshift|/\verb|ifftshift|
or an asymmetric replica sum.

A third, indirect Wiener--Khinchin check (Parseval's theorem),
\begin{equation}
\mathcal{F}^{-1}\{|F_{\mathrm{blurred}}|^{2}\}\bigr|_{(0,0)}
\;=\;
\sum_{x,y}\bigl|I_{\mathrm{blurred}}(x,y)\,w(x,y)\bigr|^{2},
\label{eq:parseval}
\end{equation}
holds to relative error $\lesssim\!10^{-16}$,
within a single floating-point unit in the last space. The three checks together verify
the FFT pipeline at every stage of Eqs.~\ref{eq:freq_replication}--%
\ref{eq:F_aliases}.

\subsection{Photoreceptor non-linearity}
\label{subsec:non-linearity}

The Naka--Rushton form (Eq.~\ref{eq:naka_rushton}) is one option for
the photoreceptor non-linearity. The framework also supports a truncated \emph{Volterra} polynomial expanded around the
local mean intensity \citep{volterra1930theory,vanhateren1992theoretical}.
For an input image $I$ of mean $\mu$ and standard deviation
$\sigma$, the Volterra non-linearity is
\begin{equation}
\mathrm{Volt}_{a_{2},\,a_{3}}\!\bigl(I(x,y)\bigr)
\;=\;
\mu \;+\; (I-\mu) \;+\; a_{2}\,(I-\mu)^{2} \;+\; a_{3}\,(I-\mu)^{3},
\label{eq:volterra}
\end{equation}
followed by a passive rescaling that preserves $\mu$ and $\sigma$
so that $a_{2}$, $a_{3}$ parameterise the shape of the
non-linearity rather than its overall gain or DC offset (which
would be absorbed by downstream normalisation anyway). The
quadratic kernel $a_{2}$ generates harmonics at $2 f$ and
intermodulation products $f_{1}\!\pm\! f_{2}$ between any two
stimulus frequencies; the cubic kernel $a_{3}$ generates $3 f$
and $2 f_{1}\!\pm\! f_{2}$. Both kernels are folded into
$F_{\mathrm{blurred}}^{\mathrm{NL}}$ and propagate through the
comb / $H_{\mathrm{box}}$ stages in exactly the same way as the
linear content, so Eq.~\ref{eq:F_aliases_NL} applies unchanged.
The two forms are not in competition: Naka--Rushton is a saturating
function whose local Taylor expansion around the operating
point yields negative quadratic and cubic Volterra coefficients;
the Volterra form makes those coefficients independently tunable,
which is what makes it useful for sensitivity analyses. Empirically the
two forms agree on the location of the moir\'e peak and on its
magnitude to within $\sim 2\,\%$ at the peak distance.  Naka--Rushton is adopted as non-linearity throughout the analysis of this paper because it is
the more biologically motivated of the two forms.

\subsection{Body-height calibration}
\label{app:body_height_error}
 
The simulation pipeline of \S\ref{sec:simulation} converts the
body's pixel count $h_{\mathrm{body,pix}}$ into the camera distance
$d_{\mathrm{cam}}$ through Eq.~\ref{eq:cam_distance}, using a
generic body height $h_{\mathrm{body}}\!=\!\SI{1.4}{\meter}$ for
\textit{Equus quagga}. The simulated approach trajectory is then
generated by isotropic crop scaling around the body centre: at each
simulated distance $d$ the script extracts a crop of side
$(d/d_{\mathrm{cam}})\,\mathrm{FoV}/\mathrm{dpp_{cam}}$ camera
pixels (\S\ref{subsec:dataset}). If the true body height differs from the
generic value, the retinal image labelled "$d$" by the simulation
corresponds to the real-world distance
\begin{equation}
d_{\mathrm{true}}
\;=\;
d \;\cdot\; \frac{h_{\mathrm{true}}}{h_{\mathrm{body}}},
\label{eq:body_height_scaling}
\end{equation}
so the entire $E_{\mathrm{par,rel}}(d)$ curve translates 
on the $d$-axis by the multiplicative factor
$h_{\mathrm{true}}/h_{\mathrm{body}}$. The peak position inherits
exactly this shift,
$\Delta d_{\mathrm{peak}}/d_{\mathrm{peak}}
 \!=\!\Delta h/h_{\mathrm{body}}$,
while the peak amplitude is left invariant. The curve
shape is preserved under the rescaling because every retinal
spatial-frequency content at the new $d_{\mathrm{true}}$ is
identical to that at the old ``$d$'' by construction.
For an inter-individual range
$\SI{1.35}{\meter}\!\leq\!h_{\mathrm{true}}\!\leq\!\SI{1.45}{\meter}$
around $h_{\mathrm{body}}\!=\!\SI{1.4}{\meter}$
(\textit{Equus quagga} shoulder height),
the relative uncertainty is $\pm 3.57\,\%$. Applied to the
empirical peak at $d_{\mathrm{peak}}\!\approx\!\SI{1.4}{\meter}$
(\S\ref{subsec:matched_pair}) this gives
\begin{equation}
\Delta d_{\mathrm{peak}}
\;=\;
d_{\mathrm{peak}} \cdot \frac{\Delta h}{h_{\mathrm{body}}}
\;=\;
\SI{1.4}{\meter} \cdot \frac{\pm\SI{0.05}{\meter}}{\SI{1.4}{\meter}}
\;\approx\;
\pm \SI{5.0}{\centi\meter},
\label{eq:body_height_dpeak}
\end{equation}
so the moir\'e peak lies in the range
$d_{\mathrm{peak}}\!\in\![\SI{1.35}{\meter}, \SI{1.45}{\meter}]$
when the per-individual height is varied across its observed
range.
This uncertainty is one order of magnitude smaller than the
natural width of the peak (the close-approach band
$d\!\in\![1, 2.5]\,\si{\meter}$ identified in
\S\ref{subsec:matched_pair} has a half-width of order
$\SI{0.75}{\meter}$), so the substantive claim of
\S\ref{subsec:reichardt_control_discussion} 
that the moir\'e mechanism peaks inside the documented
terminal-hesitation window is unaffected by the calibration
uncertainty. The peak amplitude $\sim 10\,\%$ reported in
\S\ref{subsec:matched_pair} is unaffected by construction.
For matched-pair comparisons within the dataset, the same scaling
applies independently to each pair. Per-individual height variation
shifts the per-pair peak distance by at most $\pm\SI{6}{\centi\meter}$
relative to the cohort mean, which sets the floor for the
horizontal scatter visible in the thin per-pair traces of
Fig.~\ref{fig:zebra_vs_horse}.

\subsection{Diagnostic for the Reichardt motion-energy control}
\label{app:reichardt_cancellation}

The matched full / clean Reichardt control of
\S\ref{subsec:reichardt_control_discussion} reports a strictly
negative moir\'e contribution
$E_{\mathrm{HR}}^{\mathrm{moir\acute{e}}}(t) =
E_{\mathrm{HR}}^{\mathrm{full}}(t) - E_{\mathrm{HR}}^{\mathrm{clean}}(t)$
(Eq.~\ref{eq:E_HR_moire}) inside the close-approach band.
Because $E_{\mathrm{HR}}$ is the difference of two mean-squared
signed EMD outputs the negative sign admits two mutually exclusive
mechanistic readings. To
distinguish them, $E_{\mathrm{HR}}^{\mathrm{moir\acute{e}}}(t)$ is
decomposed exactly.
The full-pipeline EMD output can be written as
$\mathrm{HR}^{\mathrm{full}}(t) =
 \mathrm{HR}^{\mathrm{clean}}(t) + \Delta\mathrm{HR}(t)$,
where
$\Delta\mathrm{HR}(t) = \mathrm{HR}^{\mathrm{full}}(t)
- \mathrm{HR}^{\mathrm{clean}}(t)$
is the moir\'e-induced additive contribution to the signed EMD
output. Squaring and averaging across the detector array gives
the algebraic identity
\begin{equation}
E_{\mathrm{HR}}^{\mathrm{moir\acute{e}}}(t)
\;=\;
\underbrace{
2\,\bigl\langle \mathrm{HR}^{\mathrm{clean}}(t)\,
\Delta\mathrm{HR}(t)\bigr\rangle
}_{\mathrm{CROSS}(t)}
\;+\;
\underbrace{
\bigl\langle \Delta\mathrm{HR}(t)^{2}\bigr\rangle
}_{\mathrm{SELF}(t)} ,
\label{eq:reichardt_decomposition}
\end{equation}
where $\langle\cdot\rangle$ denotes the spatial average over the
ommatidial array, summed over the horizontal and vertical EMD
components. $\mathrm{SELF}(t)$ is the motion energy that the
moir\'e contribution would carry on its own and is non-negative
by construction. $\mathrm{CROSS}(t)$ is twice the inner product
of the moir\'e-only EMD signal with the genuine-motion EMD signal
and is signed. $\mathrm{CROSS}(t)$ is positive if the two are spatially co-oriented
(constructive interference, moir\'e drift aligned with the
genuine optic flow) and negative if they are anti-oriented
(destructive interference, moir\'e drift opposed to the genuine
optic flow). The identity is algebraically exact and is verified
numerically frame-by-frame to a residual $\lesssim\!10^{-20}$ (a
single ULP of the floating-point sum).
Across the close-approach band $\mathrm{CROSS}(t)$ is negative
and dominates $\mathrm{SELF}(t)$ in magnitude. The time integral
$\bigl|\!\int\!\mathrm{CROSS}(t)\,dt\bigr|$ exceeds
$\int\!\mathrm{SELF}(t)\,dt$ by a factor of $\sim\!2$, so the
integrated $E_{\mathrm{HR}}^{\mathrm{moir\acute{e}}}$ inherits the
negative sign of $\mathrm{CROSS}$. This single-trajectory
time-integral ratio is consistent with the per-pair, band-integrated
ratio reported in Fig.~\ref{fig:epar_vs_emd}, whose cross-sample median
is $|\mathrm{CROSS}|/\mathrm{SELF}\approx1.57$. The two values are
different reductions of the same quantity (one trajectory integrated in
time versus the matched-pair median integrated over the close-approach
band) and both exceed unity, so the cross-term dominates in every
case. Since $\mathrm{SELF}\!\geq\!0$
can never drive the sum negative, the negative sign is necessarily
carried by the cross-term. The moir\'e contribution to the
EMD output is therefore not additive noise. It is correlated with
the genuine-motion signal in the orientation-opposed sense, which
is by definition destructive interference at the signed-output
level. This is the cross-term / self-term decomposition referred
to in \S\ref{subsec:reichardt_control_discussion}.
The anti-orientation has a direct optical origin. Stripe
content at spatial frequencies near and above the eye-Nyquist
boundary is folded by the sampling stage
(Eq.~\ref{eq:freq_replication}) into a low-frequency moir\'e whose
effective drift runs opposite to the underlying stripe motion (the folded-band signature made explicit in
\S\ref{app:motion_aliasing}). The moir\'e thus injects a motion
component pointed against the genuine optic flow, the very
configuration that drives $\mathrm{CROSS}$ negative.

\begin{figure}[H]
\centering
\includegraphics[width=\textwidth]{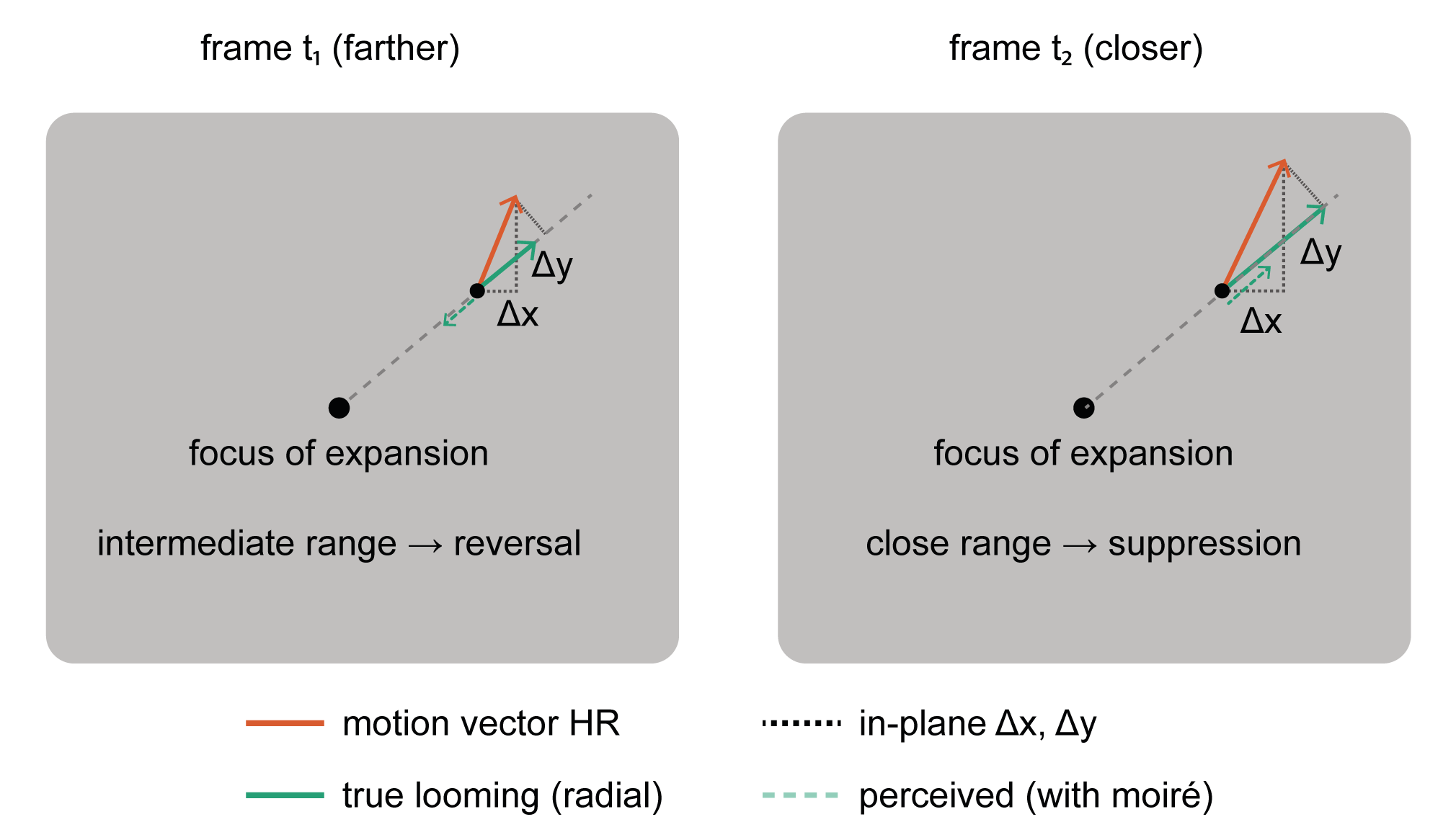}
\caption{Schematic of the radial (looming) projection and
its corruption at two approach distances. At each frame the signed
Hassenstein--Reichardt motion vector at a sample location splits into in-plane components $\Delta x, \Delta y$ and is
projected onto the outward radial direction from the focus of
expansion to give the looming contribution (green arrow). The green
dashed arrow is the moir\'e-corrupted (perceived) looming. At
intermediate range (left) the genuine looming is weak and the
destructive moir\'e component overturns it. The perceived radial
component points inward, which is a perceived recession during a genuine approach. At close range (right) the genuine looming is
strong. So the perceived component keeps the correct outward sign but
is suppressed.}
\label{fig:radial_projection_sketch}
\end{figure}

Figure~\ref{fig:radial_projection_sketch} gives the
intuition for how this destructive cross-term becomes the sign
flip of the looming percept reported in
\S\ref{subsec:reichardt_control_discussion}. The looming cue is the
array average of the EMD vector projected onto the outward radial
direction. The genuine (alias-free) projection points outward and
grows as the fly closes in. In this case adds the moir\'e a component which points against it. Where the genuine looming is still weak (the
intermediate-range resonance window) this opposed component is large enough to overturn the net radial sign. Thus the model fly reads a recession at the moment it is in fact closing in. Where the
genuine looming is strong (at close range resonance window) the opposed component can no longer flip the sign, but it still subtracts from
the magnitude, so the looming is read out suppressed. The same
destructive cross-correlation that drives $\mathrm{CROSS}$ negative
in the energy decomposition is what reverses the looming sign
at intermediate range and suppresses it at close range.

\subsection{Motion aliasing of the moir\'e beat}
\label{app:motion_aliasing}

The Reichardt control of \S\ref{subsec:reichardt_control_discussion} and its
decomposition in \S\ref{app:reichardt_cancellation} show numerically that the
parasitic moir\'e adds a signed motion component that opposes the
genuine optic flow. This appendix sets the result in the context of the motion illusions studied by \citet{how2014motion}. The wagon-wheel reversal and the barber-pole offset ride on the moir\'e beat. They are corollaries of the optical mechanism of \S\ref{sec:optical_model} rather than separate effects.
By the decomposition of \S\ref{app:reichardt_cancellation}
the additive false-motion term (SELF) is smaller than the destructive
cancellation term (CROSS), so these directional signatures are the weaker
companion of the dominant loss of radial ego-motion signal and matter most
during the lateral and circling phases of approach, not the terminal looming
phase.
In the close-approach geometry the fine stripe carrier, at angular frequency
$f_{s}$, sits at or beyond the eye-Nyquist limit $\nu_{\mathrm{eye}}$. So the
acceptance MTF (Eq.~\ref{eq:airy_otf}) attenuates it and the detector array
cannot reliably assign it a direction. The parasitic beat folded into the
pass-band by the sampling stage (Eq.~\ref{eq:freq_replication}), at
$f_{\mathrm{M}}\!<\!\nu_{\mathrm{eye}}$, is fully resolved. The EMD array of
Eq.~\ref{eq:reichardt} tracks the coarse beat, not the stripes. Sampling folds
the spatial frequency from $f_{s}$ to $f_{\mathrm{M}}$ but leaves the temporal
frequency $f_{t}\!=\!v\,f_{s}$ of a field drifting at retinal velocity $v$
unchanged, so the beat appears to drift at
\begin{equation}
v_{\mathrm{M}}
\;=\;
\frac{f_{t}}{f_{\mathrm{M}}}
\;=\;
v\,\frac{f_{s}}{f_{\mathrm{M}}}
\;=\;
v\,\frac{\Lambda_{\mathrm{M}}}{\Lambda_{s}}
\;\gg\; v ,
\label{eq:velmag}
\end{equation}
with $\Lambda_{s}\!=\!1/f_{s}$ and $\Lambda_{\mathrm{M}}\!=\!1/f_{\mathrm{M}}$. 
A slowly translating stripe field is read out as a coarse fringe moving much
faster than the stimulus, and near resonance in the opposite direction. This
is the analytic counterpart of the sign reversal measured in
\S\ref{app:reichardt_cancellation}.
The reversal itself is a temporal-aliasing (wagon-wheel) effect. A two-frame
correlator inverts the sign of the recovered drift once the tracked pattern
shifts by more than half its period between frames,
$v_{\mathrm{M}}\,\Delta t\!>\!\Lambda_{\mathrm{M}}/2$. Substituting
Eq.~\ref{eq:velmag} the beat period cancels and the condition reduces to
\begin{equation}
v\,\Delta t \;>\; \tfrac{1}{2}\,\Lambda_{s},
\label{eq:reversal}
\end{equation}
so the onset is set by the true stripe half-period, independent of the
beat, and matches the per-frame displacement at which \citet{how2014motion}
report wagon-wheel reversal. The moir\'e does not move this onset. It relocates
the spurious signal into a resolved, high-amplitude channel
(Eq.~\ref{eq:velmag}), which is what makes it salient at the detector output.
The barber-pole offset has a second, geometric origin. The relations above
concern the near-vertical flank stripes, whose drift is one-dimensional. The
obliquely oriented rump stripes add the aperture problem. A one-dimensional
periodic pattern fixes only the velocity component along its normal
$\hat{\mathbf n}$, so the EMD recovers the projected rather than the true
direction, with error
\begin{equation}
\epsilon \;=\; \angle\!\bigl(\mathbf v,\,\hat{\mathbf n}\bigr),
\label{eq:aperture}
\end{equation}
the aperture origin of the barber-pole illusion. For the rump geometry it
gives the $50$--$60^{\circ}$ offset reported by \citet{how2014motion}. This
term is not contained in the single-axis translational control of
\S\ref{subsec:reichardt_control_discussion}. Capturing it would require the
per-region stripe-orientation map and is left as an extension. Both named illusions therefore follow from the one aliasing mechanism. The
wagon-wheel reversal is the temporal aliasing of the velocity-magnified beat
(Eqs.~\ref{eq:velmag}--\ref{eq:reversal}), and the barber-pole offset is the
aperture projection of the oblique stripes onto that beat
(Eq.~\ref{eq:aperture}). Three caveats bound the reading. The effect requires
the carrier to survive the acceptance blur before it is undersampled. If the
acceptance half-width $\sigma_{\rho}$ suppresses the stripe contrast below the
level needed to form a beat, no carrier remains to alias, so the mechanism
occupies a bounded window in $(\Delta\varphi,\sigma_{\rho},f_{s})$ rather than
holding universally. Eqs.~\ref{eq:velmag}--\ref{eq:reversal} follow from the
sampling relations and confirm the numerical control rather than extending it. Only the aperture term of Eq.~\ref{eq:aperture} adds a genuinely new dynamical
degree of freedom.

\section*{Acknowledgements}
\addcontentsline{toc}{section}{Acknowledgements}
%

The author thanks Tim Caro for providing the zebra images used for the analysis in this article.


\section*{Competing Interests}
\addcontentsline{toc}{section}{Competing Interests}

The author declares no competing interests.


\section*{Author Contributions}
\addcontentsline{toc}{section}{Author Contributions}

K.M.\ Dettlaff is the sole author and is responsible for all aspects of the presented work.


\section*{Funding}
\addcontentsline{toc}{section}{Funding}

This research received no external funding.

%

\section*{Data Availability}
\addcontentsline{toc}{section}{Data Availability}

The simulation code used to produce all results in this work is
available from the author upon reasonable request.

\bibliographystyle{unsrtnat}
\bibliography{references}

%

\section*{Figure Legends}
\addcontentsline{toc}{section}{Figure Legends}

\noindent\textbf{Figure 1. Optical-model scheme.}
The diptera eye is modelled as a cascade of four linear operations on
the striped stimulus $I(x,y)$ (blue boxes): acceptance blur by the
angular photoreceptor MTF $H_{\rho}$, spatial sampling by the Dirac comb
$\mathrm{III}_{f_{s}}$ at the inter-ommatidial spacing $\Delta\varphi$,
and Voronoi reconstruction by the box-cell aperture $H_{\mathrm{box}}$.
Three eye parameters control the chain (dashed boxes on top): the
approach distance $d$ scales the spatial-frequency content of the
stimulus, the acceptance angle $\sigma_{\rho}$ pins $H_{\rho}$, and the
inter-ommatidial spacing $\Delta\varphi$ pins the comb pitch $f_{s}$.
The output stages (orange boxes) extract the parasitic content
$F_{\mathrm{aliases}}$ (Eq.~\ref{eq:F_aliases}) and integrate its power
inside the eye Nyquist disc to obtain the dimensionless metric
$E_{\mathrm{par,rel}}$ (Eq.~\ref{eq:E_par_rel}). An optional
photoreceptor non-linearity (dashed violet branch) replaces
$F_{\mathrm{blurred}}$ with $F_{\mathrm{blurred}}^{\mathrm{NL}}$ before
sampling and produces a parallel parasitic signal
$F_{\mathrm{aliases}}^{\mathrm{NL}}$ along the same final path.

\medskip\noindent\textbf{Figure 2. Compound-eye sampling geometry.}
The sampling geometry assumed in the simulation. The left side sketches
two adjacent ommatidia of the apposition compound eye with
inter-ommatidial angle $\Delta\varphi$ and angular acceptance half-width
$\sigma_{\rho}$. On the right, the angular acceptance function
$S(\theta)$ of one ommatidium is plotted.

\medskip\noindent\textbf{Figure 3. Full pipeline output on a single
representative photograph.}
Full pipeline output on a single representative photograph
(\textit{DSC\_0084}, \textit{Equus quagga}). It tracks one approach
distance per column ($d \in \{0.5,\,\textit{1.5},\,2.5,\,5\}\,\si{\meter}$)
through the six stages of the cascade developed in
\S\ref{subsec:sampling}--\S\ref{subsec:moire_isolation}, ending at
the non-linear parasitic image
$\mathcal{F}^{-1}\{F_{\mathrm{aliases}}^{\mathrm{NL}}\}$.
Rows~5 and~6 print the relative parasitic energy
$E_{\mathrm{par,rel}}$ (Eq.~\ref{eq:E_par_rel}) of each column.

\medskip\noindent\textbf{Figure 4. Relative parasitic energy versus
approach distance.}
Integrated relative parasitic energy $E_{\mathrm{par,rel}}(d)$
(Eq.~\ref{eq:E_par_rel}) as a function of approach distance, for both
the linear branch (Eq.~\ref{eq:F_aliases}) and the non-linear
Naka--Rushton branch (Eq.~\ref{eq:F_aliases_NL}). The two curves are
closely aligned, peak at the same distance \textit{($d\approx\SI{1.5}{\meter}$;
$\SI{10.3}{\percent}$ non-linear, $\SI{8.6}{\percent}$ linear)}, and identify a
moir\'e-relevant window $d\in[1.0,2.5]\,\si{\meter}$. The dashed
vertical guidelines mark the four column distances of Figure~3, whose
printed $E_{\mathrm{par,rel}}$ values are the corresponding points of the
non-linear curve.

\medskip\noindent\textbf{Figure 5. Matched-pair pipeline output (zebra
versus stripe-removed control).}
Full pipeline output on two matched pairs (\texttt{DSC\_0084} and
\texttt{DSC\_0044}; zebra in columns~1 and~3, stripe-removed control in
columns~2 and~4), each at its moir\'e-peak distance. The seven
rows show the input photograph followed by the six stages of the cascade developed in
\S\ref{subsec:sampling}--\S\ref{subsec:moire_isolation}, ending at
the non-linear parasitic image $\mathcal{F}^{-1}\{F_{\mathrm{aliases}}^{\mathrm{NL}}\}$.

\medskip\noindent\textbf{Figure 6. Relative parasitic energy across
matched pairs.}
Integrated relative parasitic energy $E_{\mathrm{par,rel}}(d)$
(Eq.~\ref{eq:E_par_rel}) for matched zebra / stripe-removed pairs as a
function of approach distance. Thin background traces: per-pair curves.
Solid markers and solid mean: striped (zebra) condition; open markers
and open mean: stripe-removed control. The vertical dashed guideline
marks the cross-pair mean moir\'e-peak distance
$d\approx\SI{1.4}{\meter}$.

\medskip\noindent\textbf{Figure 7. Regional sweep across eye
configurations.}
Regional sweep across the three canonical eye configurations of
\S\ref{subsec:acute_zone_sim} on photograph DSC\_0158: Foveal core
(red), Standard / canonical acute zone (blue), and Dorsal periphery
(purple). $E_{\mathrm{par,rel}}^{\mathrm{NL}}(d)$ is shown for each.

\medskip\noindent\textbf{Figure 8. Chromatic-channel sweep.}
Chromatic-channel sweep across the three camera RGB channels of
photograph DSC\_0084, each run through the pipeline with a per-channel
Airy MTF ($\lambda_{R}=\SI{620}{\nano\meter}$,
$\lambda_{G}=\SI{540}{\nano\meter}$,
$\lambda_{B}=\SI{460}{\nano\meter}$).

\medskip\noindent\textbf{Figure 9. Reichardt motion-energy control.}
Reichardt motion-energy control on a striped host during a simulated
close approach (DSC\_0084, \textit{Equus quagga};
$d_{0}=\SI{2.5}{\meter}$, $v_{\mathrm{app}}=\SI{0.5}{\meter\per\second}$,
frame interval $\Delta t=\SI{30}{\milli\second}$). Grey: alias-free EMD
output $E_{\mathrm{HR}}^{\mathrm{clean}}(t)$; black: full ommatidial EMD
output $E_{\mathrm{HR}}^{\mathrm{full}}(t)$; red: their signed difference
$E_{\mathrm{HR}}^{\mathrm{moir\acute{e}}}(t)$ (Eq.~\ref{eq:E_HR_moire}).
The moir\'e contribution is negative inside the close-approach band:
aliased drift cancels the genuine motion percept at the EMD output
rather than augmenting it.

\medskip\noindent\textbf{Figure 10. Corruption of the radial-expansion
(looming) signal.}
The same simulated approach on a striped host (DSC\_0084, \textit{Equus
quagga}). The Hassenstein--Reichardt vector field is projected onto the
radial direction from the focus of expansion and averaged over the
mosaic to give the net looming signal that gates landing. Blue:
alias-free signal $E_{\mathrm{exp}}^{\mathrm{clean}}(t)$; black:
full-pipeline signal $E_{\mathrm{exp}}^{\mathrm{full}}(t)$; red shading:
looming signal lost to the moir\'e; purple shading: the
moir\'e-resonance window near $d\approx\SI{1.9}{\meter}$ in which the
perceived looming becomes net negative --- a perceived recession during
a genuine approach. The signal is in arbitrary EMD units; only its sign
and its value relative to the alias-free trace are meaningful.

\medskip\noindent\textbf{Figure 11. Optical parasitic energy
versus motion-stage read-outs across the $n=28$ matched pairs.}
Optical parasitic energy versus motion-stage read-outs
across the $n\!=\!28$ matched pairs (hand-picked body centres, band-mean
over $d\!\in\![1,2.5]\,\si{\meter}$, Airy MTF, Naka--Rushton
$\alpha\!=\!0.5$). (A)~The radial component of the parasitic
energy, $E_{\mathrm{par,rad}}$ (Eq.~\ref{eq:E_par_rad}), predicts the
relative looming-cue loss ($r\!=\!0.57$, $p\!=\!0.001$; Spearman
$r_{s}\!=\!0.57$). (B)~The tangential component $E_{\mathrm{par,tan}}$
does not ($r\!=\!0.00$, $p\!=\!0.98$).
(C)~ the undirected EMD energy
deficit is not predicted by $E_{\mathrm{par,rad}}$ ($r\!=\!0.26$, $p\!=\!0.18$). (D)~The destructive cross-term
exceeds the moir\'e-alone self-term in every pair
($|\mathrm{CROSS}|\!>\!\mathrm{SELF}$; median $1.57$), so the comb step
is net-destructive throughout the sample.

\medskip\noindent\textbf{Figure 12. Schematic of the radial (looming)
projection and its corruption.}
At each frame the signed Hassenstein--Reichardt motion vector at a sample
location splits into in-plane components $\Delta x,\Delta y$ and is
projected onto the outward radial direction from the focus of expansion
to give the looming contribution (green arrow). The green dashed arrow is
the moir\'e-corrupted (perceived) looming. At intermediate range (left)
the genuine looming is weak and the destructive moir\'e component
overturns it: the perceived radial component points inward, a perceived
recession during a genuine approach. At close range (right) the genuine
looming is strong, so the perceived component keeps the correct outward
sign but is suppressed.

\section*{Tables}
\addcontentsline{toc}{section}{Tables}

The manuscript presents its quantitative content primarily through
figures; the only tabulated material is the set of regional eye
configurations and the fixed simulation parameters, given below.

\begin{table}[H]
\centering
\caption{Canonical regional eye configurations used in the
piecewise-stationary analysis of \S\ref{subsec:acute_zone_sim}.
The Standard (canonical acute zone) row is used throughout the main
analysis. the Foveal core and Dorsal periphery rows are model probes of
the high- and low-resolution extremes of the visual field.}
\label{tab:regional_config}
\begin{tabular}{lcc}
\hline
Region & $\Delta\varphi$ & $\sigma_{\rho}$ \\
\hline
Foveal core                       & \SI{1.5}{\degree} & \SI{0.6}{\degree}  \\
Standard (canonical acute zone)   & \SI{2.5}{\degree} & \SI{0.85}{\degree} \\
Dorsal periphery                  & \SI{3.5}{\degree} & \SI{1.2}{\degree}  \\
\hline
\end{tabular}
\end{table}

\begin{table}[H]
\centering
\caption{Parameters of the Fourier-optics simulation pipeline
(\S\ref{sec:simulation}). Eye-optics values correspond to the Standard
diurnal-culicid configuration of Table~\ref{tab:regional_config}.}
\label{tab:sim_params}
\begin{tabular}{lcc}
\hline
Quantity & Symbol & Value \\
\hline
Inter-ommatidial angle      & $\Delta\varphi$        & \SI{2.5}{\degree} \\
Acceptance half-width       & $\sigma_{\rho}$        & \SI{0.85}{\degree} \\
Sampling rate               & $f_{s}$                & \SI{0.40}{cyc.deg^{-1}} \\
Eye Nyquist limit           & $\nu_{\mathrm{eye}}$   & \SI{0.20}{cyc.deg^{-1}} \\
Diffraction cutoff          & $\rho_{c}$             & $\approx\SI{0.61}{cyc.deg^{-1}}$ \\
Naka--Rushton exponent      & $\beta$                & $0.7$ \\
Half-saturation level       & $s_{50}$               & $0.5$ \\
Non-linearity mix factor    & $\alpha$               & $0.5$ \\
Grid samples per side       & $N$                    & $384$ \\
Field of view               & $\mathrm{FoV}$         & \SI{30}{\degree} \\
Generic body height         & $h_{\mathrm{body}}$    & \SI{1.4}{\meter} \\
Approach (closing) speed    & $v_{\mathrm{app}}$     & \SI{0.5}{\meter\per\second} \\
Frame interval              & $\Delta t$             & \SI{30}{\milli\second} \\
EMD time constant           & $\tau_{\mathrm{HR}}$   & $\approx\SI{30}{\milli\second}$ \\
\hline
\end{tabular}
\end{table}

\end{document}